\documentclass[usenatbib]{mn2e} 
\usepackage{psfig}
\usepackage{lscape}
\usepackage{rotating}
\usepackage{psfig}
\usepackage{amsmath}
\usepackage{amssymb}
\title[Outflows in compact radio sources]{Fast outflows in compact
  radio sources: 
  evidence for AGN-induced feedback in the early stages of radio
  source evolution}
\author[J. Holt et al.]{J. Holt$^{1}$\thanks{E-mail:
jholt@strw.leidenuniv.nl\newline$\dagger$ Present address: Leiden
Observatory, PO Box 9513, 2300 RA Leiden, The Netherlands.}$\dagger$, C. N. Tadhunter$^{1}$ and R. Morganti$^{2}$\\
$^{1}$Department of Physics and Astronomy, University of Sheffield,
Sheffield,  S3 7RH, UK.\\
$^{2}$Netherlands Foundation for Research in Astronomy, Postbus 2,
7990 AA Dwingeloo, The Netherlands}

\newcommand{\cmc}{cm$^{-3}$}
\newcommand{\kms}{km s$^{-1}$}
\newcommand{\ha}{H$\alpha$}
\newcommand{\hb}{H$\beta$}

\newcommand{\ebv}{E(B-V)}
\newcommand{\lala}{$\lambda\lambda$}

\begin{document}
\maketitle
\begin{abstract}
We present intermediate resolution, wide wavelength coverage
spectra for a complete sample of 14 compact radio sources taken with
the aim of investigating the impact of the nuclear activity on the
circumnuclear interstellar medium (ISM) in the early stages of radio
source evolution. We observe spatially extended line emission (up to
$\sim$20 kpc) in the majority of sources which is consistent with a
quiescent halo. 
In the nuclear apertures we observe broad, highly complex
emission line profiles. Multiple Gaussian modelling of the {[O
    III]}$\lambda$5007   
line  reveals between 2 and 4 components which can have velocity widths (FWHM)
and blueshifts relative to the halo of up to $\sim$2000\kms. 
We interpret these broad, blueshifted components as material in outflow and
discuss the kinematical evidence for jet-driven outflows as previously
proposed for 
PKS 1549-79 and PKS 1345+12. 
Comparisons with  samples in the literature show that compact
radio sources harbour more extreme nuclear kinematics than their
extended counterparts,  
a trend seen within our sample with larger velocities in the smaller sources. 
The observed velocities are also likely to be influenced by source orientation
with respect to the observer's line of sight. 
Nine sources have associated HI absorption. In common with the optical
emission line gas,  
the HI profiles are often highly complex with the majority of the
detected components significantly blueshifted, tracing outflows in the
neutral gas. The  sample has been tested for stratification in the ISM
(FWHM/ionisation potential/critical density) as suggested by
\citet{holt03} for PKS 1345+12 but we find no significant trends
within the sample using a Spearman Rank analysis. This study supports
the idea that compact radio sources are young radio loud AGN observed
during the 
early stages of their evolution and currently shedding their natal
cocoons through extreme circumnuclear outflows. 
\end{abstract}


\section{Introduction}
We currently know little about the early evolution of powerful
extragalactic radio sources, in particular, how the presence of a
radio-loud AGN influences the evolution of the host galaxy, and how
the ISM affects the expansion of the radio jets. Recent developments
in both observation and theory have shown the importance of AGN
feedback in galaxy evolution. For example, there exist  close
correlations between the mass of the central black hole and the properties of
the galaxy bulge
(e.g. \citealt{ferrarese00,gebhardt00,tremaine02,marconi03}). Theoretical 
analyses by, for example, \citet{silk98} and \citet{fabian99}, and
more recently, 
simulations by, for example,  \citet{dimatteo05} and
\citet{hopkins05}, describe how, after  
relocating to the centre of the galaxy remnant after a merger
\citep{heckman86}, the  
supermassive black hole grows through accretion, becomes active (a
proto-quasar) and then sheds its enshrouding cocoon (deposited by the
merger) through outflows driven by powerful quasar winds
\citep{balsara93}. With time, the 
central regions will be cleared of fuel, starving both the central AGN
and any star forming regions in the the bulge and activity will
cease. 

Clearly, understanding AGN-induced feedback is of vital importance for
understanding the evolution of both  AGN and their host
galaxies. However, there remain many uncertainties in the models due
to a lack of observational evidence, and feedback is often inserted as
a `black box'. Whilst the models described above provide a good theoretical
description of the shedding of the natal cocoon during the early evolutionary
stages,  these scenarios often {\it assume} that the
feedback  is dominated by quasar-induced winds. Whilst this may
be true for radio-quiet AGN, in radio-loud AGN, the expanding radio
jets may also provide a significant contribution to the overall feedback of
the AGN through jet-induced outflows (e.g. \citealt{bicknell97}), 
particularly during the early stages of the radio source evolution
when the radio jets are still on the same scales as the natal
cocoon. Such a contribution  on small
scales should not be surprising given the increasing support for 
significant feedback from extended radio sources, for example, in the
quenching of  cooling flows in  massive haloes
(e.g. \citealt{best06,croton06,bower06}).

With this in mind, compact radio sources, comprising the 
Gigahertz-Peaked Spectrum radio sources (GPS: D $<$ 1 kpc) and the
larger  Compact Steep Spectrum radio sources (CSS: D $<$ 15 kpc),
  form a large proportion  of the bright
  (centimetre-wavelength-selected) radio source
  population ($\sim$40\%; e.g. \citealt{odea98})  particularly suited  to studying AGN-induced
feedback. First, GPS and CSS
 sources are currently believed to be compact 
due to evolutionary stage 
\citep{fanti95} with estimates of dynamical and radio spectral ages of
$t_{dyn} \sim 10^2 - 10^3$ yr (e.g. \citealt{owsianik98,tschager00})
and $t_{sp} < 10^4$ yr (e.g. \citealt{murgia99}) respectively. This is
in preference to the frustration scenario in which the ISM is so
dense that the radio jets cannot escape, and the radio source remains
confined and frustrated for its entire lifetime \citep{vanbreugel84}. 
Second,  the small-scale radio jets will be on the same scales as the
natal cocoon of gas and dust and so will readily interact with it. Finally, 
radio-loud sources    contain all potential outflow driving mechanisms
  (quasar-induced winds, radio-jets and starburst-driven superwinds)
  and so are the only objects in which the relative importance of the
  different effects can be assessed in individual objects
   (e.g. \citealt{batcheldor07}). 

The most direct way to probe the kinematics and physical conditions of
the nuclear gas is through high quality optical spectroscopy. However,
to date, much of the work on compact radio sources has either
concentrated on the radio wavelength region or has relied on low
signal to noise  optical spectra. Despite
this, an early optical spectroscopic study of a large sample of
compact radio sources 
suggested  evidence for non-gravitational motions
(indicative of flows of gas) in the form of highly broadened emission
line profiles which are often asymmetric with blue wings
\citep{gelderman94}. However, the potential differences between the
spectra of compact and extended sources were not quantified in this
early study.

With new, deeper spectra, it becomes clear that the 
unusually broad nuclear emission lines are due 
to highly complex emission line profiles requiring multiple Gaussians
to model them.
The first concrete evidence for fast outflows in the optical emission line gas
in a  compact radio source was reported by \citet{tadhunter01} in the
southern compact flat spectrum radio source PKS 1549-79. 
 From their low resolution optical
spectra, the high ionisation emission lines (e.g. {[O III]}) were both
broader (FWHM $\sim$ 1350 \kms~compared to $\sim$ 650 \kms) and {\it
  blueshifted} by $\sim$ 600 \kms~with respect to the low ionisation
lines (e.g. {[O II]}).  Recent higher resolution follow up
  on this source has shown that the outflowing component is present in
  all lines with a velocity of 679 $\pm$ 20 \kms~(\citealt{holt06};
  hereafter H06). 
Similarly, \citet{holt03} (hereafter H03) reported  a much more
extreme outflow (up to 2000\kms) 
in the GPS source PKS 1345+12. In both sources, the outflows are
believed to be driven by the small scale radio jets expanding through
   dense circumnuclear cocoons of gas and dust (see e.g. Figure 2 in
  \citealt{tadhunter01} and Figure 4 in H03). A recent 
high-resolution optical and radio imaging study (HST and
  VLBI) was able to confidently rule out large-scale starbursts driven
  winds, but  failed to distinguish between jet-driven outflows and
  AGN-induced winds \citep{batcheldor07}.  However,  previous HST imaging
  studies of compact radio sources have suggested evidence for an
  alignment effect on small scales in some CSS sources
  \citep{devries97,devries99,axon00} which would support the
  jet-driven outflows scenario.  For PKS
  1345+12, due to the differences in velocity width between the
  emission line components in different lines, H03
  suggested a stratified ISM which may be the result of gradients in
  density and/or ionisation potential across the circumnuclear ISM.

Hence, we have obtained intermediate resolution 
(4-6\AA\footnote{This corresponds to $\sim$200-300\kms, dependent on
  redshift, for {[O 
    III]}5007\AA~for the objects in our sample.}) optical spectra with
good 
signal-to-noise over a  
large spectral range, with the WHT, NTT and VLT,
in order to search for such outflows in a complete
  sample of 14 compact radio 
sources. In this paper, the first of two, we present the kinematical data
for the whole sample in order to search for the signatures of outflows in the
nuclear emission line gas. In sections 2 and 3, we describe the sample
selection, observations and data reduction procedure. In section 4, we
model the emission lines, both in the extended halo (to determine the
rest frame of each system) and in the nucleus (to search for
outflows). Our data are of sufficient resolution to model the lines
with multiple Gaussian components. In addition, in section 5 we have
modelled the 
lines with single Gaussians to allow comparisons between the sample
presented here and other samples (compact and extended) in the
literature, and 
to investigate the possibility of density and/or ionisation gradients
in the ISM in some sources. In this paper we concentrate on the
emission line kinematics and the ionisation of the emission line gas
will be discussed in a second paper.

Throughout this paper, we assume the following cosmology: H$_{0}$ = 71
\kms, $\Omega_{\rmn 0}$ = 0.27 and  $\Omega_{\Lambda}$ = 0.73.

\section{Sample selection}
\begin{center}
\begin{table*}
\begin{center}
{\footnotesize
\caption[Properties of the sample.]
{\footnotesize Properties of the sample: $(a)$ radio source; $(b)$
  optical ID where G = galaxy, Q = QSO; $(c)$ radio 
  source type where CSS = Compact Steep Spectrum, GPS =
  Gigahertz-Peaked Spectrum, CF = Compact Flat spectrum and CC =
  Compact Core; $(d)$ 
  redshift $^{\star}$ redshifts from the literature, for the accurate
  redshifts derived in this study, see Table
  \protect\ref{tab:lineparam}; $(e)$ radio spectral index where 
  F$_{\nu} \propto \nu^{-\alpha}$ and for $\alpha_{\rm 2.7 GHz}^{\rm 5
    GHz}$ (for all but 3C 213.1,  
3C 277.1 and 3C 303.1, these are taken 
from \protect\cite{wall85}; for the remaining 
three sources $\alpha$ is calculated from the flux densities
presented in \protect\citet{kellermann69});   $(f)$ radio luminosity
at 5GHz; 
$(g)$ HI 21cm absorption velocity (\kms); 
$(h)$ HI 21cm absorption FWHM (\kms); $(i)$ Angular
  size of radio source in milli arcseconds; $(j)$ Projected linear
  size of radio source in parsecs; $(k)$ Radio axis position
  angle. For references, see 
the detailed biographies of each sources in the text.
$\dagger$ tentative HI detection with optical depth, $\tau$ = 0.003
(Morganti 2004, private communication) where the central velocity is
assumed to be the midpoint of the entire absorption profile.
$^a$ taken from the literature and may be inconsistent with the sizes
in column $(j)$,  
$^b$ compact core component in 3C 213.1 and 3C 459, $^c$ extended
radio source component  in 3C 213.1 and 3C 459. For
3C 459 the different HI results are: $^{\rmn m}$
\protect\citet{morganti01}, $^{\rmn v}$ 
\protect\citet{vermeulen03} and $^{\rmn g}$
\protect\citet{gupta06}. }
\begin{tabular}{lccccccccrrr} \\ \hline\hline
 \\
\multicolumn{1}{c}{Object} &
\multicolumn{1}{c}{ID} & 
\multicolumn{1}{c}{Type$^a$} &
\multicolumn{1}{c}{z$^{\star}$} &
\multicolumn{1}{c}{$\alpha$} 
&\multicolumn{1}{c}{Radio}&
\multicolumn{2}{c}{HI 21cm absorption} &
\multicolumn{1}{c}{Angular} & \multicolumn{1}{c}{Linear} &
\multicolumn{1}{c}{Radio} \\
\multicolumn{1}{c}{ } & 
\multicolumn{1}{c}{ } &
\multicolumn{1}{c}{ } & \multicolumn{1}{c}{ } &
\multicolumn{1}{c}{ } &
\multicolumn{1}{c}{luminosity}&
\multicolumn{1}{c}{V$_{helio}$} &
\multicolumn{1}{c}{FWHM} &
\multicolumn{1}{c}{size} & \multicolumn{1}{c}{size} &
\multicolumn{1}{c}{PA} \\
\multicolumn{1}{c}{} & \multicolumn{1}{c}{} &
\multicolumn{1}{c}{} & \multicolumn{1}{c}{} &
\multicolumn{1}{c}{} &
\multicolumn{1}{c}{log P$_{\rmn 5 GHz}$ (W Hz$^{-1}$)}&
\multicolumn{1}{c}{(\kms)} &
\multicolumn{1}{c}{(\kms)} &
\multicolumn{1}{c}{(arcsec)} & \multicolumn{1}{c}{(kpc)} &
\multicolumn{1}{c}{($^{\circ}$)} \\
\multicolumn{1}{c}{$(a)$}&
$(b)$&$(c)$&$(d)$&$(e)$&$(f)$&$(g)$&$(h)$&${(i)}$&\multicolumn{1}{c}{$(j)$}&\multicolumn{1}{c}{$(k)$}&
\\
\hline 
3C 213.1    %& 0858+292    
& G & CC  & 0.195 &  0.30  & 26.41        & 0.19395  & 115 & 6.0$^{b}$& 19.1$^{b}$& -61? \\%& 1\\
 & & & & & &  & & 40$^{c}$ & 128$^{c}$&-20\\
3C 268.3    %& 1203+645    
& G & CSS & 0.371 & 0.88  & 27.04       & 0.37186 & 101 & 1.36 & 6.9 & -15\\%& 1,8\\
            &             &      &         &       &                  & 0.37227 & 19  & \\
3C 277.1    %& 1250+568    
& Q & CSS & 0.320 &0.61  & 26.86         & --     & --  & 1.67 & 7.7 & -49 \\%& 1,8\\
4C 32.44    %& 1323+321    
& G & GPS& 0.369 & 0.60  & 27.30         & 0.36843 & 229 & 0.06   & 0.31  & -50 \\%& 10,11\\
           % & DA 344 \\
PKS 1345+12 %& 4C 12.50    
& G & GPS & 0.122 & 0.44  & 26.69         & 0.12184  &150  & 0.15   & 0.33  & 160 \\%& 2,4,8\\
3C 303.1    %& 1443+77     
& G & CSS & 0.267 & 1.08  & 26.43         & --     & --  & 1.80 & 7.3 & -47 \\%& 1,8\\
\\
PKS 0023-26 %& OB 238      
& G & CSS & 0.322 &   0.7  & 27.43         & 0.31890  & 126 & 0.68 & 3.2 & -34            \\%&1,5,6\\
            &             &   &     &     &      &         0.32142  & 39  \\
PKS 0252-71 %&             
& G & CSS & 0.566 &  1.14 & 27.55          & --     & --  & 0.24 & 1.6  & 7              \\%&5\\
PKS 1306-09 %&             
& G & CSS & 0.464 &   0.65 & 27.39          & 0.46492$\dagger$    & 350$\dagger$ & 0.46 & 15.7 & -41\\%&raf,5,6\\
PKS 1549-79 %&             
& G & CF  & 0.152 &   0.18 & 27.00          & 0.15224  &     & 0.12    & 0.3  & 90               \\%&3,\\
PKS 1814-63 %&             
& G & CSS$^a$ & 0.063 &  0.91 & 26.54          & 0.06450  &     & 0.41 & 0.5  & -20            \\%&3,5\\
PKS 1934-63 %&             
& G& GPS$^a$& 0.183 &  0.88  & 27.31          & 0.18282  & 18  & 0.70  & 2.1  & 89             \\%&5\\
PKS 2135-20 %&             
& G & CSS & 0.635 &  0.82  & 27.58          & --     & --  & 0.25 & 1.7 & 52             \\%&1,5,6\\
PKS 2314+03 %& 3C 459      
& G & CC  & 0.220 &  0.97  & 27.65          & 0.21850$^{\rmn m}$  &$\sim$400  & 0.20$^b$  & 0.7$^b$& 95     \\%&1,6,??\\
               & &     &    &      &       &     0.21914$^{\rmn v}$  & 130 &
             10.00$^c$     &   35.2$^c$     \\ 
& & & & & & 0.21837$^{\rmn g}$ & 71 \\
& & & & & & 0.21839$^{\rmn g}$ & 121 \\
& & & & & & 0.21963$^{\rmn g}$ & 164 \\\hline\hline
\end{tabular}
}
\end{center}
\label{tab:sample}
\end{table*}
\end{center}

Our complete sample  comprises 14 compact radio sources split across
the northern and southern hemispheres. In the north, we have observed 
a  complete sample of five sources (3 CSS, 2 GPS
with  $z$ $<$ 0.4, 12$^{h}$ $<$ RA $<$ 18$^{h}$ and $\delta >$
15$^{\circ}$) 
from the subset of compact radio sources in the 3rd and 4th
Cambridge Radio Catalogues 
(3C and 4C respectively, at 178 MHz) listed in Table 1 in the \cite{odea98}
review, which is a 
master list of the CSS sample from \cite{fanti90} and the GPS sample
from \cite{stanghellini90a,stanghellini96}. A further seven sources 
(5 CSS, 1 GPS, 1 compact flat-spectrum) form
a complete  redshift ($z$ $<$ 0.7) and RA (12$^h <$ 
RA $< 5^h$)  limited subset of the  2
Jy sample of  \cite{tadhunter93},  which in turn is a 
subset ($z <$ 0.7, $\delta <$ +10$^{\circ}$) of the 2Jy \cite{wall85}
radio survey at 2.7 GHz. We have also observed a further two radio
sources (3C 213.1 and 3C 459) which are classified as compact core
radio sources. Such sources have both  extended radio structure
(jets and lobes which are often disconnected from the radio core)
 and a second, compact radio structure (complete with lobes), which may be
mis-aligned from the large-scale source. Whilst the larger-scale
structures are consistent with old, evolved radio sources, the inner,
compact structures may represent a younger, more recently triggered
active phase, and so these sources can be studied in conjunction with a
sample of compact radio sources.

 The whole sample has a radio power range of
26 $<$ log P$_{\rm 
  5 GHz}$ (W Hz$^{-1}$) $<$ 28. Details of the sample are presented in
Table {\ref{tab:sample}}. 
Hence, the total sample includes 8 CSS, 3 GPS, 1 compact flat-spectrum
source and 2 compact  core radio sources. The
sample also includes both galaxy 
(13) and quasar (1) host morphologies. 

The  subsamples were chosen to fulfill the following criteria: 
i) redshifts low enough to include {[O III]}$\lambda$5007 in the
spectra and similar for the subsamples; 
ii) RA and Dec ranges to be observed during the allocated runs and;
iii) the samples were derived from well-studied samples (2Jy, 3C \& 4C)
to enable good comparisons between compact and extended sources.
 \vspace*{0.5cm}\\
\section{Observations, data reduction and analysis techniques} 
\begin{center}
\begin{table*}
\begin{minipage}{170mm}
\begin{center} 
\caption[Log of observations.]
{\small Log of observations. Column $(c)$ denotes the arm/wavelength
  range of the spectrograph. $(h)$ is the instrumental seeing ($\dagger$
  indicates seeing measured using the object continuum -- for all
  other sources, a star on the chip was used). Note, the values quoted
  in column $(h)$ represent upper limits on the true seeing as 1)
  where the seeing was measured using the target galaxy, these
  measurements may be affected by the extended galaxy haloes i.e. they
  are not stellar; and 2) the measurements based on the profiles of
  stars along the slit only provide a good seeing estimate if the slit
  passes through the centres of the stellar image, otherwise the FWHM
  (and therefore seeing) will be over-estimated. $(i)$ gives the width
  of the extracted  
nuclear aperture in arcseconds, centred on the centroid of the nuclear
continuum emission  (quoted once per object in the table but the same
aperture was used for both red and blue spectral ranges)
and  $(j)$ denotes the photometric
  conditions  (p = photometric, v = variable transparency). $\star$
  denotes spectra taken aligned within 10 degrees of the radio source PA.  }
{\footnotesize
\begin{tabular}{llclrcccccccc}\\  \hline\hline
 \\
\multicolumn{1}{c}{Date}&\multicolumn{1}{c}{Object} &
\multicolumn{1}{c}{`Arm'} &\multicolumn{1}{c}{Exposure}& 
\multicolumn{2}{c}{Slit} 
& \multicolumn{1}{c}{$\lambda$ range}
& \multicolumn{1}{c}{Seeing} & \multicolumn{1}{c}{Nuclear}&
\multicolumn{1}{c}{Notes} \\ 
 & &  &  &  \multicolumn{1}{c}{PA}&
\multicolumn{1}{c}{width} &  & 
 & \multicolumn{1}{c}{aperture}& \\ 
 & &  & \multicolumn{1}{c}{(s)} &  \multicolumn{1}{c}{($^{\circ}$)}&
\multicolumn{1}{c}{(arcsec)} & \multicolumn{1}{c}{(\AA)} & 
\multicolumn{1}{c}{(arcsec)} & \multicolumn{1}{c}{(arcsec)}& \\ 
\multicolumn{1}{c}{$(a)$} & \multicolumn{1}{c}{$(b)$}&
\multicolumn{1}{c}{$(c)$}&\multicolumn{1}{c}{$(d)$}& $(e)$&$(f)$
&$(g)$ &$(h)$&\multicolumn{1}{c}{$(i)$} &\multicolumn{1}{c}{$(j)$}
 \\\hline 
\multicolumn{8}{l}{\bf WHT/ISIS observations}\\
14/05/2001 &3C 213.1 &R & 3*1200 &  80$\star$  &  1.3 & 7700-9200  & 1.5 $\pm$ 0.1$\dagger$ & 2.5&v\\
 && B & 3*1200 &  80$\star$   & 1.3  & 3300-6800  &  1.8 $\pm$ 0.1$\dagger$ & &v\\
13/05/2001 &3C 268.3 & R & 2*1200 &155$\star$  & 1.3 & 7850-9350 & 1.4 $\pm$ 0.3$\dagger$ & 2.2&p\\
 && R & 3*1200 &  155$\star$  &1.3 & 6200-7700 & 1.9 $\pm$ 0.2$\dagger$ & &p\\
 && B & 5*1200 & 155$\star$  &1.3 & 3700-6800 & 2.3 $\pm$ 0.2$\dagger$ & &p\\  
13/05/2001 &3C 277.1 & R & 1*1200 &129$\star$  &1.3 & 7850-9350 & 0.9 $\pm$ 0.1$\dagger$  &1.8& p \\
 &&R & 2*1200 & 129$\star$  & 1.3 & 6200-7700& 0.9 $\pm$ 0.1$\dagger$ & &p\\
 && B & 3*1200 & 129$\star$  & 1.3 & 3700-6800 &1.4 $\pm$ 0.1$\dagger$ &&p\\ 
14/05/2001 &4C 32.44 & R & 3*1200 & 105 & 1.3 &7700-9200& 1.5 $\pm$
0.2$\dagger$ & 2.5&v\\ 
&& R & 3*1200 & 105 & 1.3 &6200-7700& 1.4 $\pm$ 0.2$\dagger$ &&v\\
14/05/2001 &&B & 3*1200 &105 &  1.3 &3300-6800& 1.9 $\pm$ 0.2$\dagger$ & &v\\ 
12/05/2001 &PKS 1345+12  & R & 1*900 & 104 & 1.3 &6200-7700& 1.3 $\pm$
0.2 & 2.2&p \\ 
 &(4C 12.50)&  B & 1*900 & 104&1.3 &3300-6800&  1.3 $\pm$ 0.2 &&p\\
 &&R & 3*1200 &160$\star$  &  1.3 &6200-7700 & 1.3 $\pm$ 0.2 &&p\\
 &&B & 3*1200 & 160$\star$  & 1.3& 3300-6800 & 1.3 $\pm$ 0.2&&p\\
14/05/2001 &PKS 1345+12&R & 3*1200 &230 & 1.3 &6200-7700& 1.7 $\pm$ 0.2 &2.2&v\\
 && B & 3*1200 & 230 &  1.3 &3300-6800& 1.7 $\pm$ 0.2 &&v \\  
12/05/2001 &3C 303.1 & R & 5*1200 & 130$\star$ &  1.3 & 7700-9200&   1.8 $\pm$ 0.1$\dagger$ &  2.5&p\\
 && B & 5*1200 &130$\star$  &  1.3 &  3300-6800 &1.8 $\pm$ 0.1$\dagger$ &&p\\   
\\
\multicolumn{8}{l}{\bf NTT/EMMI observations} \\
13/07/2002 &PKS 0023-26 & B & 3*1200 & -105 &1.5 & 3700-7050 & 2.5
$\pm$ 0.2$\dagger$ &  2.0&p\\ 
 && R & 2*1200 & -105 &1.5 & 4400-11400 & 2.3 $\pm$ 0.2$\dagger$&  &p\\ 
13/07/2002 &PKS 0252-71 & R & 2*1200 & 135$\star$  & 1.5 &5700-8700 & 1.6 $\pm$ 0.1          &  1.7&p\\
 && B & 2*1200 & 135$\star$  &1.5 & 3700-7050 & 1.5 $\pm$ 0.1          && p\\ 
13/07/2002 &PKS 1306-09 & R & 3*1200 & 135 &1.0 & 5700-8700 &  1.7 $\pm$ 0.2$\dagger$ &  2.3&p\\
&& B & 3*1200 & 135 &1.0 & 3700-7050 & 1.8 $\pm$ 0.2$\dagger$ &  &p\\ 
12/07/2002 &PKS 1549-79 & R & 3*1200 & -5  &1.0 & 5700-8700 & 1.9
$\pm$ 0.1          & &p \\ 
 &&B & 3*1200 & -5  &1.0 & 3700-7050 & 2.2 $\pm$ 0.1          & &p \\
&&R & 2*1200 & 25  & 1.5 & 5700-8700 & 0.8 $\pm$ 0.1          & & p\\
 &&B & 2*1200 & 25  & 1.5 &3700-7050 &  0.8 $\pm$ 0.1          & & p \\
12/07/2002 &PKS 1814-63 & R & 1*1200 & -72 &1.0 & 5700-8700 &  1.1
$\pm$ 0.2$\dagger$ &  1.0&p\\ 
 &&R & 1*1200 & 65  & 1.0 & 5700-8700 &  1.3 $\pm$ 0.1          && p\\
 && B & 1*1200 & 65  & 1.0 & 3700-7050 &  1.2 $\pm$ 0.1          &&  p\\ 
12/07/2002 &PKS 1934-63 & R & 3*1200 & -20 &1.5 & 5700-8700 &  1.2
$\pm$ 0.1          & 2.0&p\\ 
 && B & 3*1200 &-20 & 1.5 & 3700-7050 &  1.2 $\pm$ 0.1          &  &p\\ 
13/07/2002 &PKS 2135-20 & R & 3*1200 & -115&1.5 & 5700-8700 & 1.4 $\pm$ 0.2$\dagger$ & 1.7&p\\
&&B & 3*1200 & -115&1.5 & 3700-7050 &  1.4 $\pm$ 0.2$\dagger$ &   &p\\ 
12/07/2002 &3C 459 & R & 3*1200 &-175& 1.5 & 5700-8700 &  1.5 $\pm$ 0.2$\dagger$ & 1.7&p\\
 &(PKS 2314+03)& B & 3*1200 &-175& 1.5 & 3700-7050 &  1.5 $\pm$ 0.2$\dagger$ && p\\
 && R & 1*1200 & 95$\star$   &1.5 & 5700-8700 & 1.5 $\pm$ 0.2$\dagger$ & &p\\
\\
\multicolumn{8}{l}{\bf VLT/FORS2 observations}\\
24/09/2003 &PKS 1549-79 & R & 3*1200 &75  & 1.3 & 4950-8250 &  2.0 $\pm$ 0.1          & 1.5&p\\
24/09/2003 &&B & 3*600  & 75  & 1.3 & 3050-6000 &  2.1 $\pm$ 0.1          & &p \\\hline\hline 
\end{tabular}
}
\end{center}
\label{tab:obs}
\end{minipage}
\end{table*}
\end{center}
Long-slit optical spectroscopic observations of the full sample were
obtained during two observing runs.  The northern sample was observed
on 12-14 May 2001 with ISIS, the dual arm spectrograph on the 4.2m
William Herschel Telescope on La Palma.  In the red, the TEK4 CCD was
used with the R316R grating with 1x1 binning and two central
wavelengths ($\sim$6950\AA~and 
$\sim$8450\AA, dependent on object). The resulting  wavelength calibration
accuracies, calculated using the standard error on the mean deviation
of the night sky emission line wavelengths from published values
\citealt{osterbrock96}, 
 were 0.06-0.11\AA~($\sim$6950\AA) and 0.10-0.53\AA~($\sim$8450\AA). 
The spectral resolutions, calculated using the widths of the night sky
emission 
lines, were 3.3-3.7 $\pm$ 0.1\AA~($\sim$6950\AA) and 3.8--4.1
$\pm$ 0.1\AA~($\sim$8450\AA). In the blue, the EEV12 CCD was used with the
R300B grating with 2x2 binning. The  wavelength calibration
accuracy was 0.1-0.2\AA~with a spectral resolution of 4.3--4.8 $\pm$
0.2\AA.  The spatial scale was 0.36 arcsec/pixel. Further details,
including the instrumental setups and the 
useful wavelength ranges of the spectra are summarised in Table 2.

The southern sample was
observed on 12-13 July 2002 using the EMMI spectrograph on the ESO New
Technology Telescope (NTT) on La Silla, Chile in RILD mode. The MIT/LL
CCD was used with grisms \#4, \#5 and \#6 to obtain spectra with
central wavelengths 7935\AA, 5385\AA~and 7223\AA~with 2x2 binning. The
 wavelength calibration accuracies were 0.24\AA~(7935\AA),
0.06--0.15\AA~(5385\AA) and 0.10--0.17\AA~(7223\AA)
with spectral resolutions of  14 $\pm$ 1\AA~(7935\AA), 5.6 $\pm$
0.1\AA~(5385\AA) and 4.3-6.7 $\pm$ 0.1\AA~(7223\AA). The spatial scale
was 0.33 arcsec/pixel. Again, further
details are presented in Table 2.

PKS 1549-79 was also observed with the FORS2
spectrograph on the ESO Very
Large Telescope (VLT) on Cerro Paranal, Chile, in September 2003 to
improve the S/N of the fainter features and resolve a seeing-slit
width matching problem in the NTT data. The detailed analysis of the
data for PKS 1345+12 and PKS 
1549-79 have already been presented in H03, H06
and \citet{javi07}. The spatial scale for the VLT observations was
0.25 arcsec/pixel.

The aim of this project is to search for outflows in the circumnuclear
regions using optical spectroscopy with sufficient spectral resolution
and signal to noise to accurately separate and 
model the different components of the highly broadened and complex
emission lines. In order to include all of the outflowing regions in
the slit, and to ensure the spectra were of sufficient resolution,
all spectra were taken with a 1.0-1.5 arcsec slit. To reduce the
effects of differential atmospheric refraction, all exposures were taken at low
airmass (sec $z <$ 1.1) and/or with the slit aligned along the parallactic
angle. Due to various observational constraints, we have only aligned
the slit along the radio axis for approximately half of the sources, and
the PAs are listed in Table 2. However, as this study is concerned
with compact rather than extended radio sources, the feedback effects
we are interested in will be confined to the nuclear regions and the
mis-alignment of the slit with the radio axis will not affect our results.
\vspace*{0.5cm}\\
\subsection{Data reduction and analysis}
\begin{table*}
\caption[Continuum modelling results.]{Summary of the continuum
  modelling results (see also Section 3.2) for the nuclear aperture of all galaxies in the
  sample. The columns are: 
  $(a)$ object, $(b)$ best fitting (lowest $\chi_{\rm reduced}^{2}$) model
  -- OSP (12.5 Gyr) plus YSP 
   (Gyr) +  power-law components, $(c)$ percentage (flux) contribution
  of OSP, $(d)$ percentage (flux) contribution of YSP, $(e)$ \ebv value (if any) used to redden the
  YSP, $(f)$ percentage (flux) contribution of the power
  law component, $(g)$ spectral index of the power-law component where
  F$_{\lambda} \propto \lambda^{\alpha}$ , $(h)$ minimum reduced $\chi^{2}$ of
  the model for the whole SED. The nuclear continuum was not modelled
  for three galaxies: $\star$ 
  3C 277.1:  pure QSO spectrum; $\diamond$ PKS 1814-63: continuum is
  strongly contaminated by a nearby Galactic star which drowns out any
  continuum information from the galaxy;  $\dagger$ 3C 303.1:
  mis-matching between the blue and red arms as the blue arm was
  observed without a dichroic (due to the position of {[O
  III]}\lala4959,5007).  For these three galaxies, see the text for a
  discussion of alternative techniques. More detailed studies of the
  stellar populations in some of the radio sources in this sample are
  presented elsewhere and we refer readers to the following papers:
PKS 1345+12 \protect\citep{tadhunter05,javi07}, PKS 1549-79
\protect\citep{holt06}, PKS 0023-26 and PKS 2135-20
\protect\citet{holt07} and 3C 213.1 and 3C 459 \protect\citet{wills07}.}
\begin{center}
\begin{tabular}{llrrcrrc} \hline\hline
\\
\multicolumn{1}{c}{Object} & \multicolumn{1}{c}{Best fitting model} &
\multicolumn{6}{c}{Model parameters} \\
 & & \multicolumn{1}{c}{12.5 Gyr} & \multicolumn{1}{c}{YSP} &
\multicolumn{1}{c}{\ebv} &
\multicolumn{1}{c}{power-law} & \multicolumn{1}{c}{$\alpha$} &
\multicolumn{1}{c}{$\chi_{\rm min. red.}^{2}$} \\
\multicolumn{1}{c}{$(a)$} &\multicolumn{1}{c}{$(b)$} &
\multicolumn{1}{c}{$(c)$} &\multicolumn{1}{c}{$(d)$} &
\multicolumn{1}{c}{$(e)$} &\multicolumn{1}{c}{$(f)$} &
\multicolumn{1}{c}{$(g)$} &\multicolumn{1}{c}{$(h)$} \\ \hline\hline
3C 213.1    & 12.5 + 1.0  & 44 $\pm$ 2  & 61 $\pm$ 2 & 0.0& -- & -- & 0.62 \\
3C 268.3    & 12.5 + 1.0 + $\alpha$ & 6$_{-6}^{+14}$ & 3$_{-3}^{+14}$& 0.0 & 84$_{-9}^{+7}$ & 1.8 $\pm$ 0.2 & 1.22 \\ 
3C 277.1$\star$    & -- & -- & -- & -- & -- & -- & --\\ 
PKS 1345+12 & 12.5 + 0.1 + $\alpha$ & 48$_{-7}^{+8}$ & 19$_{-19}^{+12}$ & 0.0 & 19$_{-17}^{+22}$ & 2.86$_{-1.40}^{+4.08}$ & 1.07 \\
4C 32.44    & 12.5 + 0.5 & 95 $\pm$ 3 & 8 $\pm$ 2 & 0.0 & -- & -- & 1.34 \\
3C 303.1$\dagger$    & -- & -- & -- & -- & -- & -- & --\\  %\multicolumn{7}{c}{no fitting -- see caption}\\
%\\
PKS 0023-26 & 12.5 + 0.1 & 46 $\pm$ 16 & 51 $\pm$ 18 & 1.1 & -- & -- & 1.46 \\
PKS 0252-71 & 12.5 + 2.0 + $\alpha$ & 2$_{-2}^{+35}$ & 57 $_{-39}^{+17}$ & 0.1 & 39$_{-16}^{+15}$ & 1.14$_{-0.75}^{+0.39}$ & 1.23 \\
PKS 1306-09 & 12.5 + 1.0 + $\alpha$ & 52$_{-21}^{+20}$ & 20 $\pm$ 20 & 0.4 & 31$_{-13}^{+11}$ & 0.85$_{-1.08}^{+0.50}$ & 1.11 \\
PKS 1549-79 & 12.5 + 0.1 + $\alpha$ & 34$_{-14}^{+18}$ & 39$_{-9}^{+5}$ & 0.1 & 22$_{-16}^{+18}$ & 2.98$_{-1.02}^{+2.42}$ & 0.49 \\
PKS 1814-63$\diamond$ & -- & -- & -- & -- & -- & -- & --\\  
PKS 1934-63 & 12.5 + $\alpha$ & 85 $\pm$ 6 & -- & -- & 16 $\pm$ 4 & 1.56$_{-0.33}^{+0.28}$ & 0.59 \\
PKS 2135-20 & 12.5 + 0.1 + $\alpha$ & 16$_{-16}^{+21}$ & 49$_{-12}^{+11}$ & 0.0 & 37$_{-15}^{+16}$ & 0.66 $\pm$ 1.20 & 0.59\\
3C 459      & 12.5 + 0.05 & 34 $\pm$ 1 & 67 $\pm$ 1 & 0.0 & -- & -- & 0.43 
\\\hline\hline
\end{tabular}
\label{tab:cont-model}
\end{center}
\end{table*}
The data were reduced in the usual way (bias subtraction, flat
fielding, cosmic ray removal, wavelength calibration, flux
calibration) using the standard packages in {\sc iraf}. The
two-dimensional spectra were also corrected for spatial distortions of 
the CCD. To reduce wavelength calibration errors due to flexure of the
telescope and instrument (ISIS observations), separate arcs were taken
at each position on the sky. Such calibrations were not made for EMMI
or FORS2 as the reported flexure is small. The calculated wavelength
calibration accuracies and spectral resolutions are given above.

Comparison of several spectrophotometric standard stars taken with a
wide slit (5 arcsec) throughout each run gave relative flux
calibrations accurate to $\pm$5 per cent. This accuracy was confirmed
by good matching in the flux between the red and blue spectra. The
only exception was 3C 303.1 and whilst this will pose problems for
line ratios, the kinematic information is unaffected.
In order to correct the spectra for atmospheric
absorption features (e.g. A and B bands at $\sim$7600 and $\sim$6800
\AA, respectively), standard stars were observed with a narrow slit,
matched to the slit width used to observe the objects. For the ISIS
observations, these were observed at random during the run and the
normalised spectra 
had to be scaled to match the depth of the absorption features. For
the NTT and VLT observations, a separate calibration star was observed
for each object, close in both position and time to the target
observations and so no scaling was required. 
The spectra of all
sources were corrected for Galactic extinction using the \ebv~values
from \citet{schlegel98} and the \citet{seaton79} extinction law.

The spectra were extracted and analysed using the {\sc starlink}
packages {\sc figaro} and {\sc dipso}.

\subsection{Continuum subtraction}

Prior to emission line modelling of the nuclear apertures, the
continuum was modelled and subtracted for most of the sources. 
Initially, a nebular continuum 
was generated and subtracted following \citet{dickson95} taking
full account of reddening following  H03. The remaining
continuum was then modelled  using a customised {\sc idl}
minimum $\chi^2$ fitting programme (see
e.g. \citealt{robinson01,tadhunter05,holt07} for details). The modelling
program allows up to three separate continuum components
 --  Old (OSP, here set to 12.5 Gyr with
 no reddening) and  Young Stellar Populations (YSP, reddened with 
0 $<$ \ebv~$<$ 1.6 using \citealt{seaton79} with age 0.01-5.0 Gyr)
taken from \citet{bruzual93} and an AGN/power-law component. We define
the best fitting model 
as that with the least number of components required to adequately
model both the overall SED and discrete stellar absorption features
(e.g. Ca H+K, Balmer series).

 Continuum modelling and
subtraction was necessary for the detection and accurate modelling of
the broader emission line components. For completeness, details of the
continuum models used for subtraction prior to emission line
modelling are given in Table {\ref{tab:cont-model}}. Note, the continua
of several galaxies in this sample have been more recently accurately
modelled and we refer readers to the corresponding papers: 
PKS 1345+12 \citep{tadhunter05,javi07}; PKS 1549-79 (H06);
PKS 0023-26, PKS 1549-79 and PKS 2135-20 \citep{holt07};  
3C 213.1 and 3C 459 \citep{wills07}.

The continua of three sources were not modelled. 3C 277.1 has a pure
quasar  continuum (there is no evidence for an underlying stellar
continuum), but it could not be reproduced by a simple 
power-law, and so we have not modelled or subtracted the
continuum. However, when modelling the emission lines, care was taken
to remove the BLR contamination.  
The radio source PKS 1814-63 lies close to a bright
foreground Galactic star and so the spectra are strongly contaminated
by light from this star.
Fortunately our observations were made during
good seeing conditions\footnote{Table 2 gives the seeing measured from
  a star on the 2-dimensional spectra whereas the seeing measured
  using several stars on   the acquisition image was 0.8 arcsec.} and
we clearly resolve PKS 1814-63  
from the star. Whilst it was not possible to model the continuum, for
accurate modelling of the broader emission line components, we have
attempted to remove the stellar spectrum.  Because the star is
point-like, its spectrum  will have no 
significant spatial variation and the star itself could be used for
subtraction. Hence, each frame was copied, inverted and then shifted in
the spatial direction, so the centroid of the continuum was
aligned. The inverted spectrum was then  subtracted from the
original spectrum to remove the contamination from the continuum
emission. Finally, no corrections were made for 3C 303.1 due to the
mis-matching of the blue and red arm spectra.

\subsection{Kinematic component definitions}
The emission lines and their components vary significantly in width
(FWHM). Hence, for our analysis,  we  define the
following kinematical components for the Narrow Line Region (NLR) as
used in H03:  
\begin{itemize}
\item narrow: FWHM $<$ 600 \kms;
\item intermediate: 600 $<$ FWHM $<$ 1400 \kms;
\item broad: 1400 $<$ FWHM $<$ 2000 \kms;
\item very broad: FWHM $>$ 2000 \kms.
\end{itemize}

\section{Results}
In this section we discuss the kinematic results derived from
two-dimensional optical spectra. Initially, the extended line emission
is used to establish the systemic velocities followed by detailed
modelling of the emission lines in the nuclear apertures.

\subsection{Kinematics of the extended gaseous halo}
\label{sect:halo}
\begin{figure*}
\begin{minipage}{15cm}
\begin{center}
\begin{tabular}{ccc}
3C 213.1 & 3C 268.3 \\
\hspace*{-0.6cm}\psfig{file=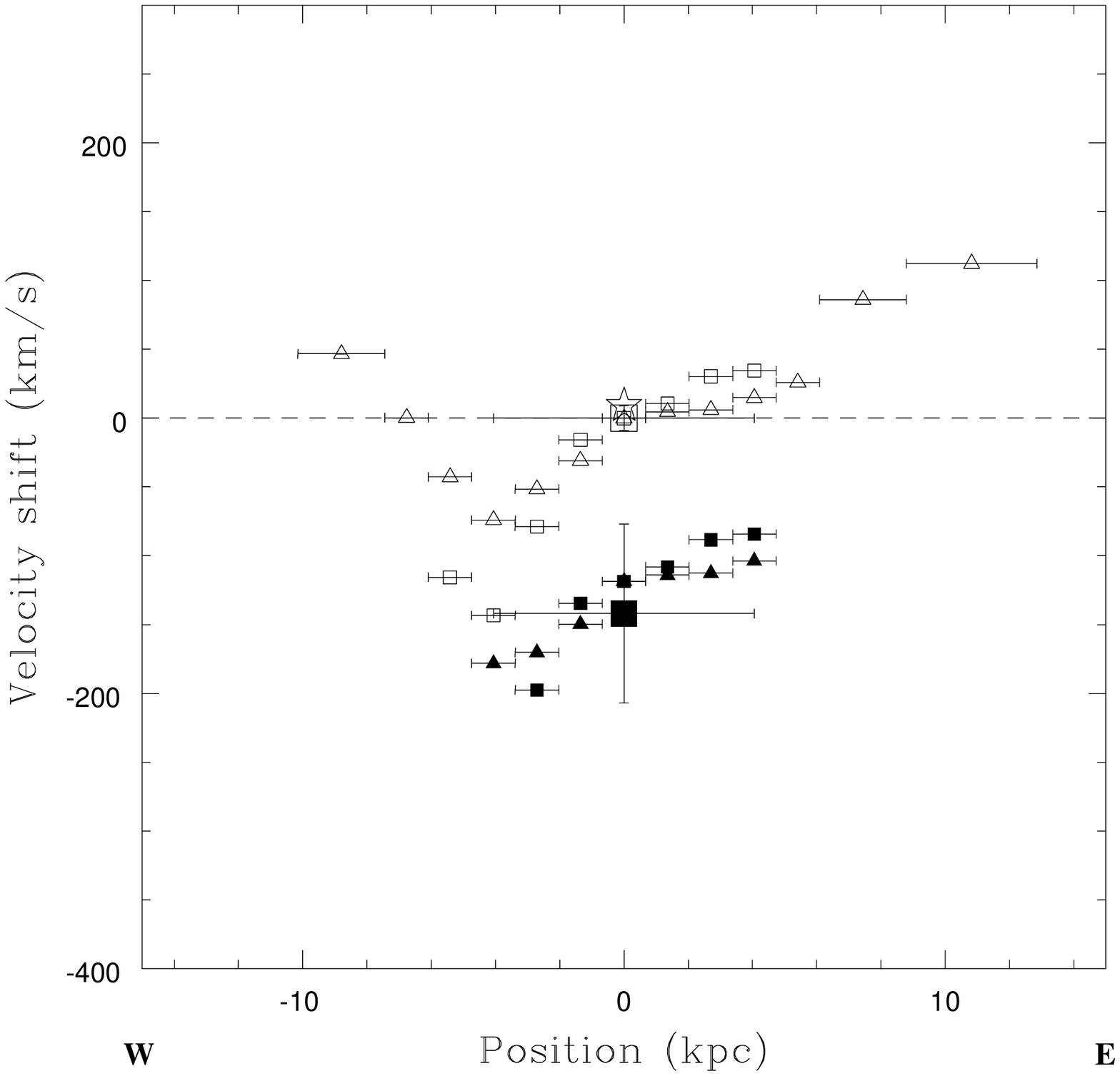,width=7.8cm,angle=0.}&
\psfig{file=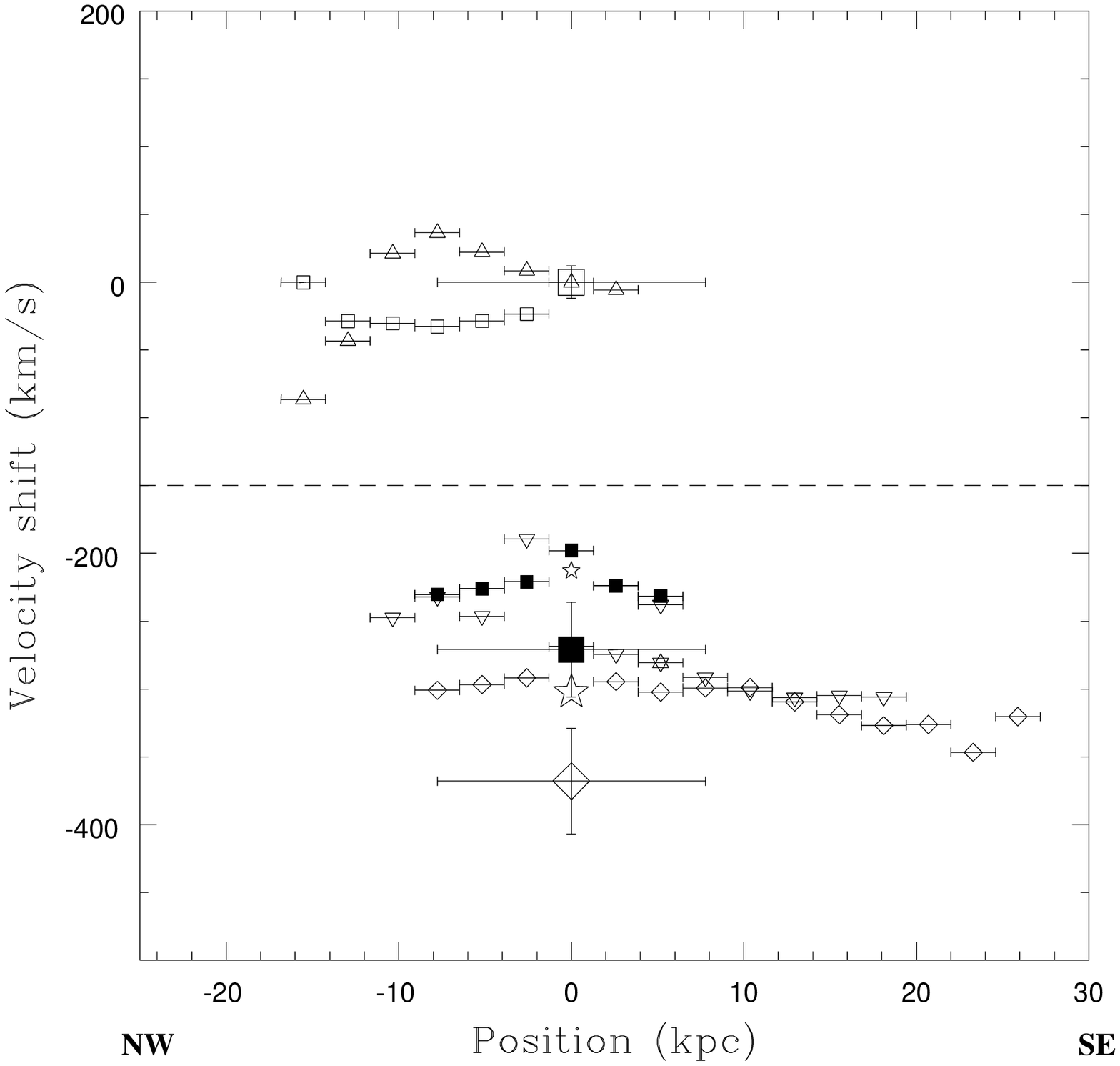,width=7.8cm,angle=0.}\\
\\
3C 277.1 & 4C 32.44 \\
\hspace*{-0.6cm}\psfig{file=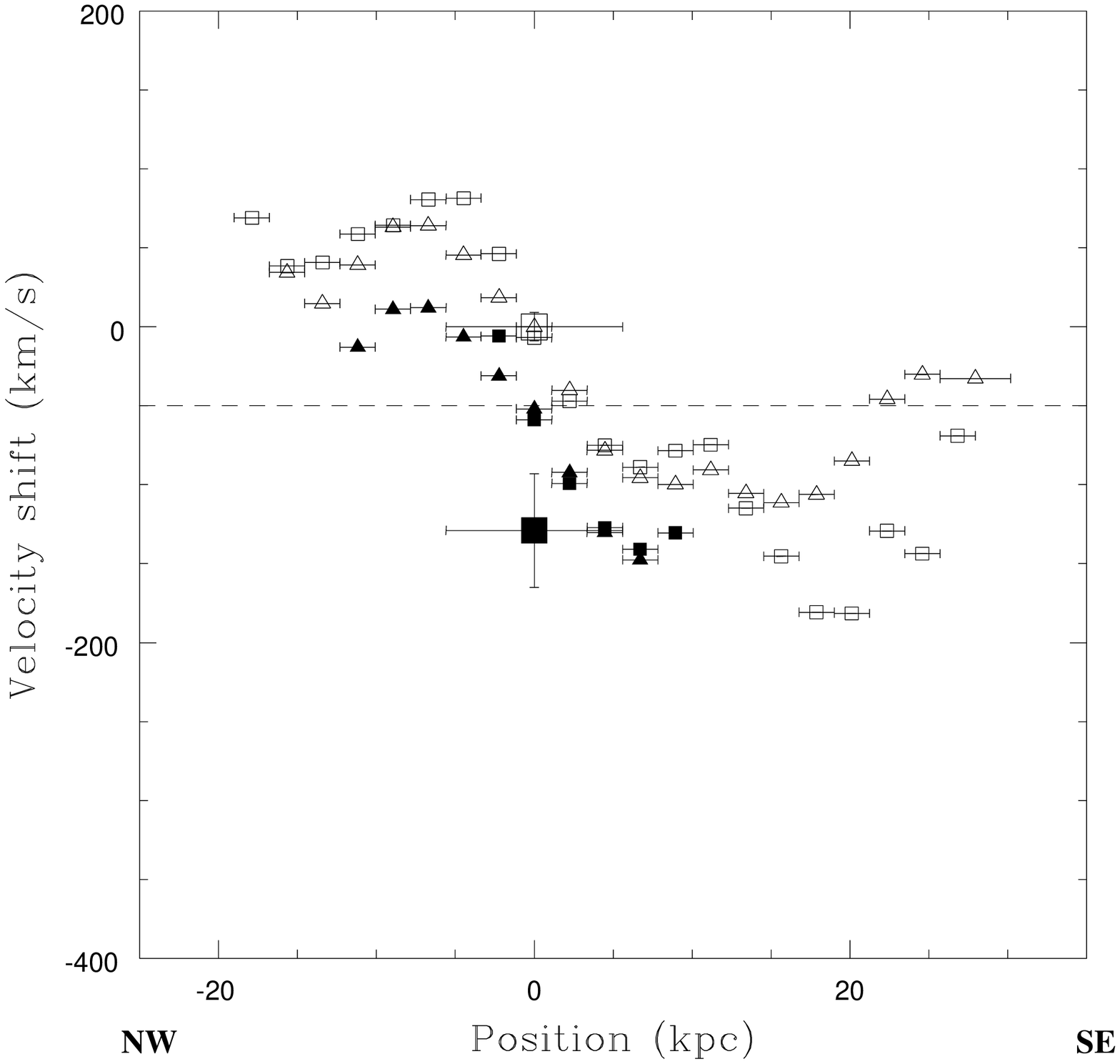,width=7.8cm,angle=0.}&
\psfig{file=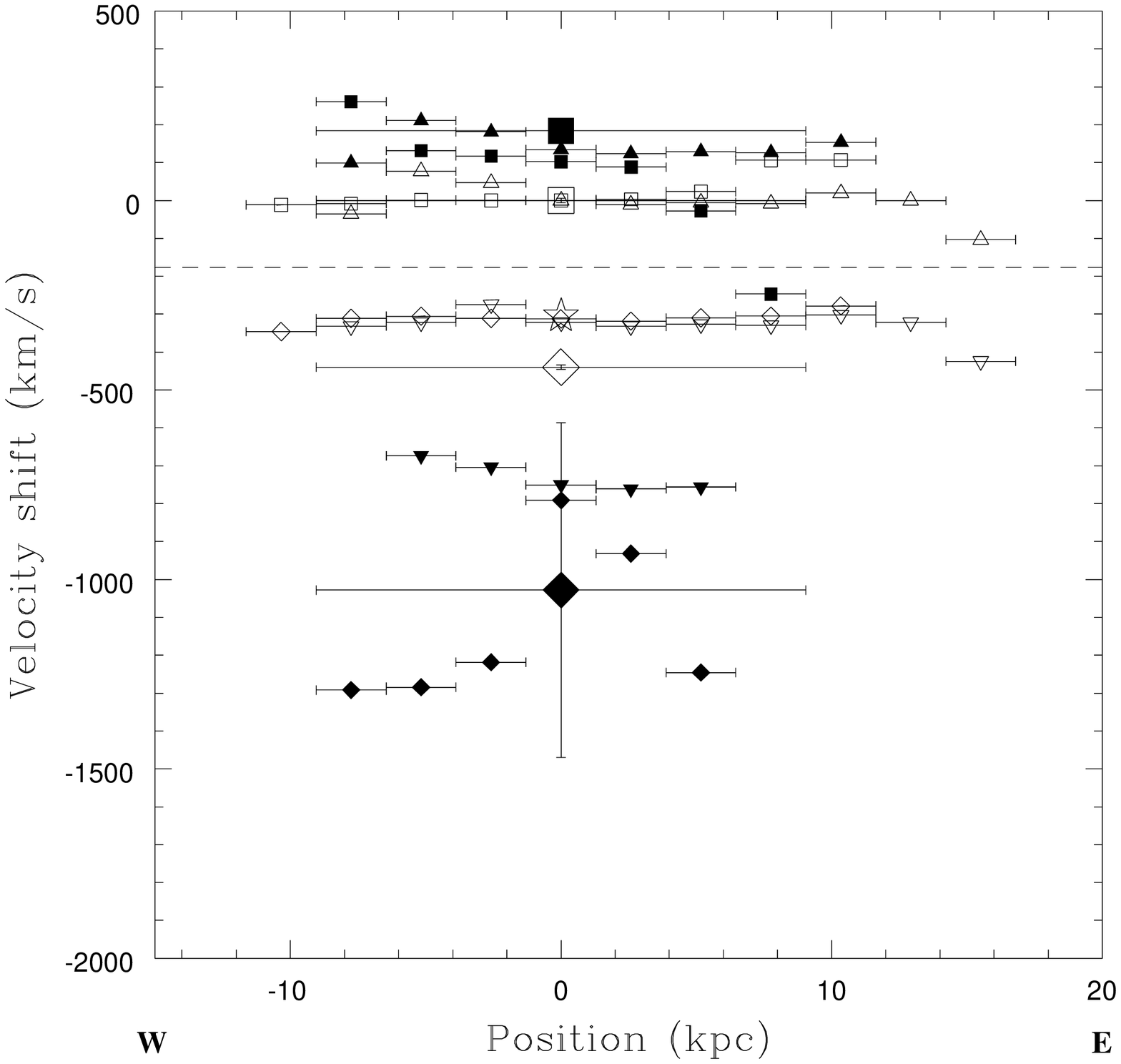,width=7.8cm,angle=0.}\\
\end{tabular}
\end{center}
\caption[]{Rest frame velocity profiles for all sources in the sample. The
  numerous small points represent the pixel-by-pixel fitting where:
triangles ($\triangle$,$\nabla$) are {[O II]}\lala3727 and
  squares/diamonds ($\Box$,$\Diamond$) are {[O III]}\lala4959,5007. 
   In general, open symbols ($\triangle$,$\nabla$,$\Box$,$\Diamond$)
  are narrow components (where two narrow components are present, the
  second is traced by  inverted triangles and diamonds), filled
  symbols represent the intermediate ({\large
    $\blacktriangle$},$\blacksquare$) 
  and broad ({\large $\blacktriangledown$},$\blacklozenge$) components.
  Exceptions are: 4C 32.44 ({\large $\blacktriangle$},$\blacksquare$: broad
  component; {\large $\blacktriangledown$},$\blacklozenge$: very broad
  component),  
 PKS 1306-09  (open and filled points are narrow components),
PKS 1814-63 ({[O III]} and \ha~emission plotted where
  $\Box$,$\blacksquare$: {[O III]} and $\Diamond$,$\blacklozenge$:
  \ha), PKS 1934-63 \& 3C 459: (open: narrow,  filled: broad), 
PKS  2135-20 (open: intermediate, filled:   broad). 
Stars (where shown) mark the  velocity of the detected HI
  absorption. Sources to note are:  PKS 0023-26 (two components, 
small: narrow, large: broad), PKS
  1306-09 (vertical dashed line marks the range of velocities over
  which HI absorption has 
been tentatively  detected, Morganti, private communication), PKS
  1814-63 (in addition to the deep HI absorption line (star), the
  range of velocities covered by the broad shallow absorption feature
  are marked by the vertical dashed line) and 3C 459 (large star:
  \protect\citet{morganti01} detection, large inverted star:
  \protect\citet{vermeulen03} detection, small stars: three components
  detected by \protect\citet{gupta06}).
  The large symbols
  ($\Box$,$\Diamond$,$\blacksquare$,$\blacklozenge$) represent the 
components of {[O III]}\lala4959,5007 detected in the nuclear aperture
where the symbols (open/filled etc) are consistent with the
smaller points in the same plot. The large symbols and their horizontal
  error bars show the position and 
  size of the nuclear aperture. Finally, for 3C 459 (PA 95), 
two further points mark an extended region of line splitting
 ({[O III]}; {\LARGE $\circ$}: narrow, {\LARGE $\bullet$}: broad) 
 All extended apertures will be discussed in a future
  paper. The horizontal dashed line in each plot 
indicates the assumed systemic velocity (derived from {[O III]}). 
Note, for PKS 1345+12, this figure
includes all three PAs observed and is taken directly from H03. For
PKS 1549-79, we show the velocity profile along PA -5 only, again,
taken directly from H06. For more details see H03, H06 and
\protect\citet{javi07}.  Note, where velocity (vertical)
errorbars are not clearly seen, the errorbars are on the same scale
as/smaller than the plotted points.}
\label{fig:velprofiles}
\end{minipage}
\end{figure*}
\setcounter{figure}{0}
\begin{figure*}
\begin{tabular}{cc}
3C 303.1 & PKS 0023-26 \\
\hspace*{-0.6cm}\psfig{file=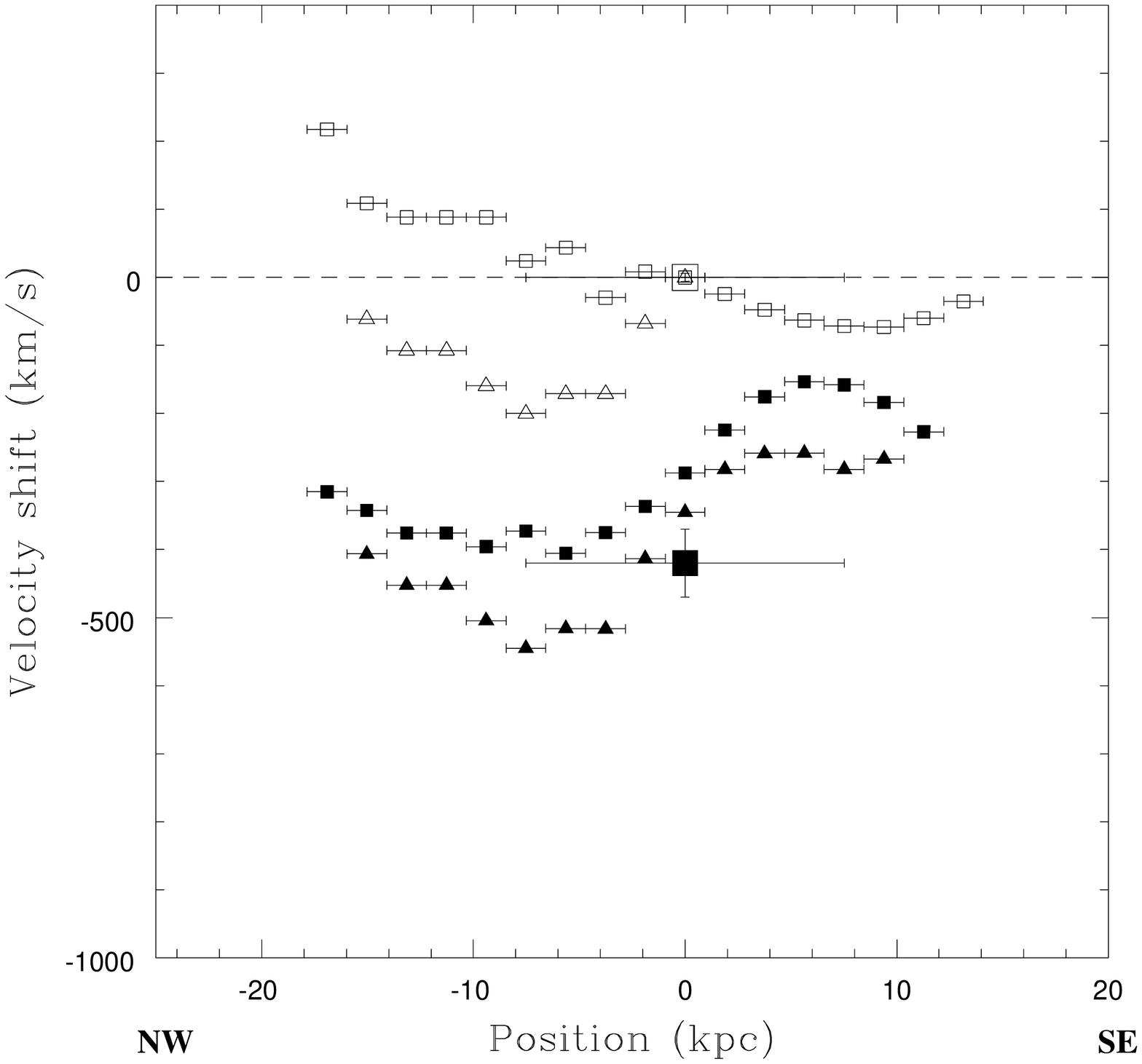,width=7.8cm,angle=0.}&
\psfig{file=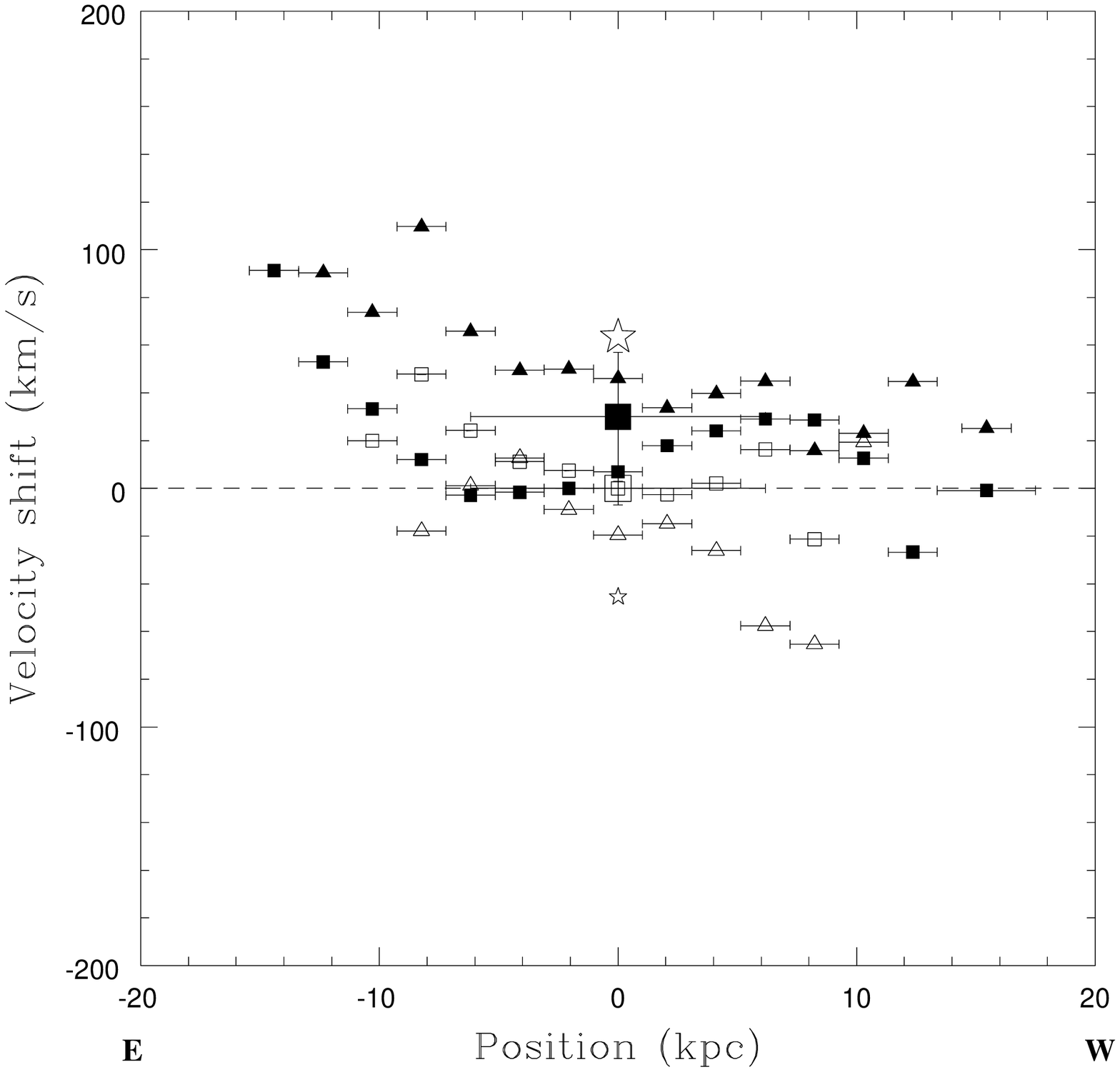,width=7.8cm,angle=0.}\\
\\
 PKS 0252-71 & PKS 1306-09 \\
\hspace*{-0.6cm}\psfig{file=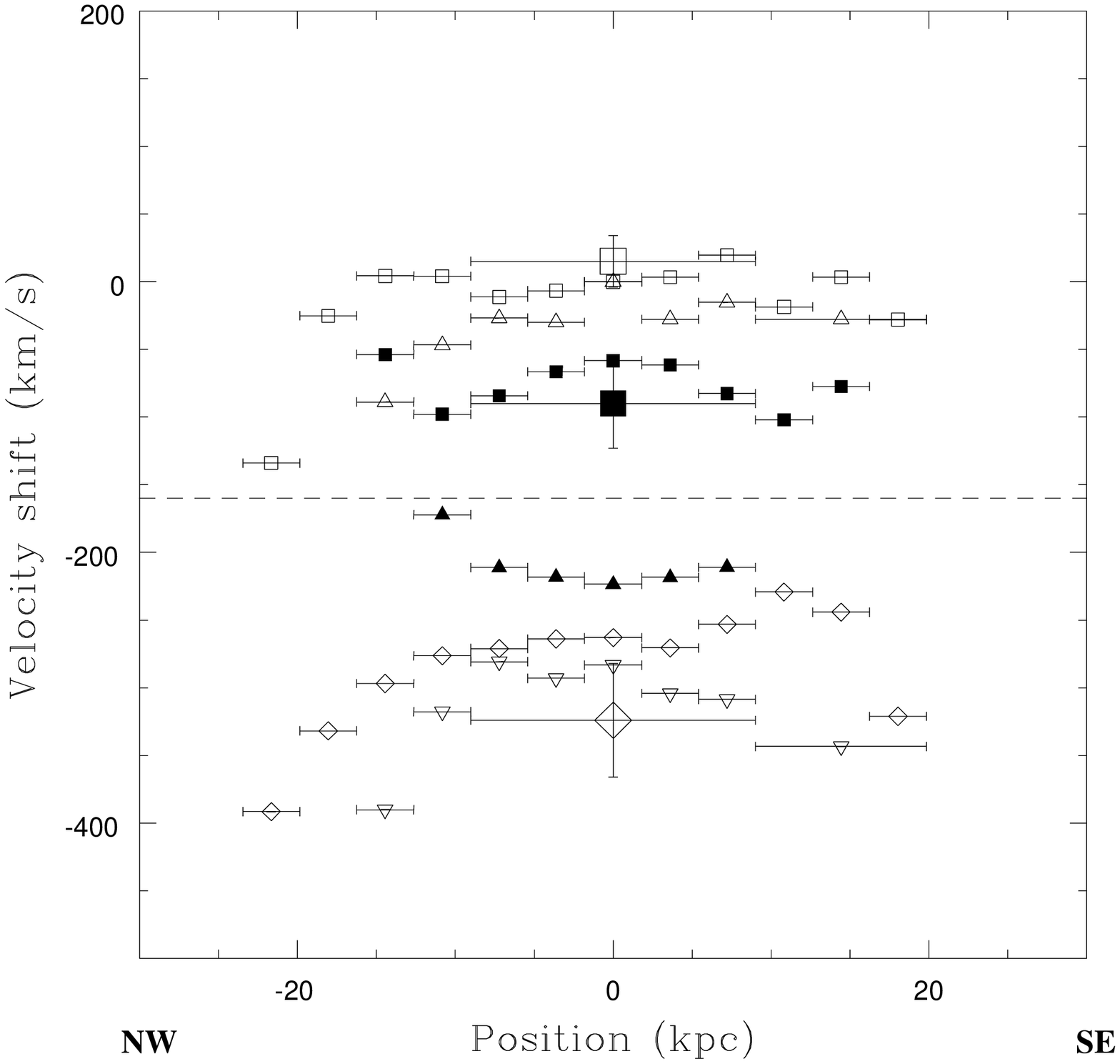,width=7.8cm,angle=0.}&
\psfig{file=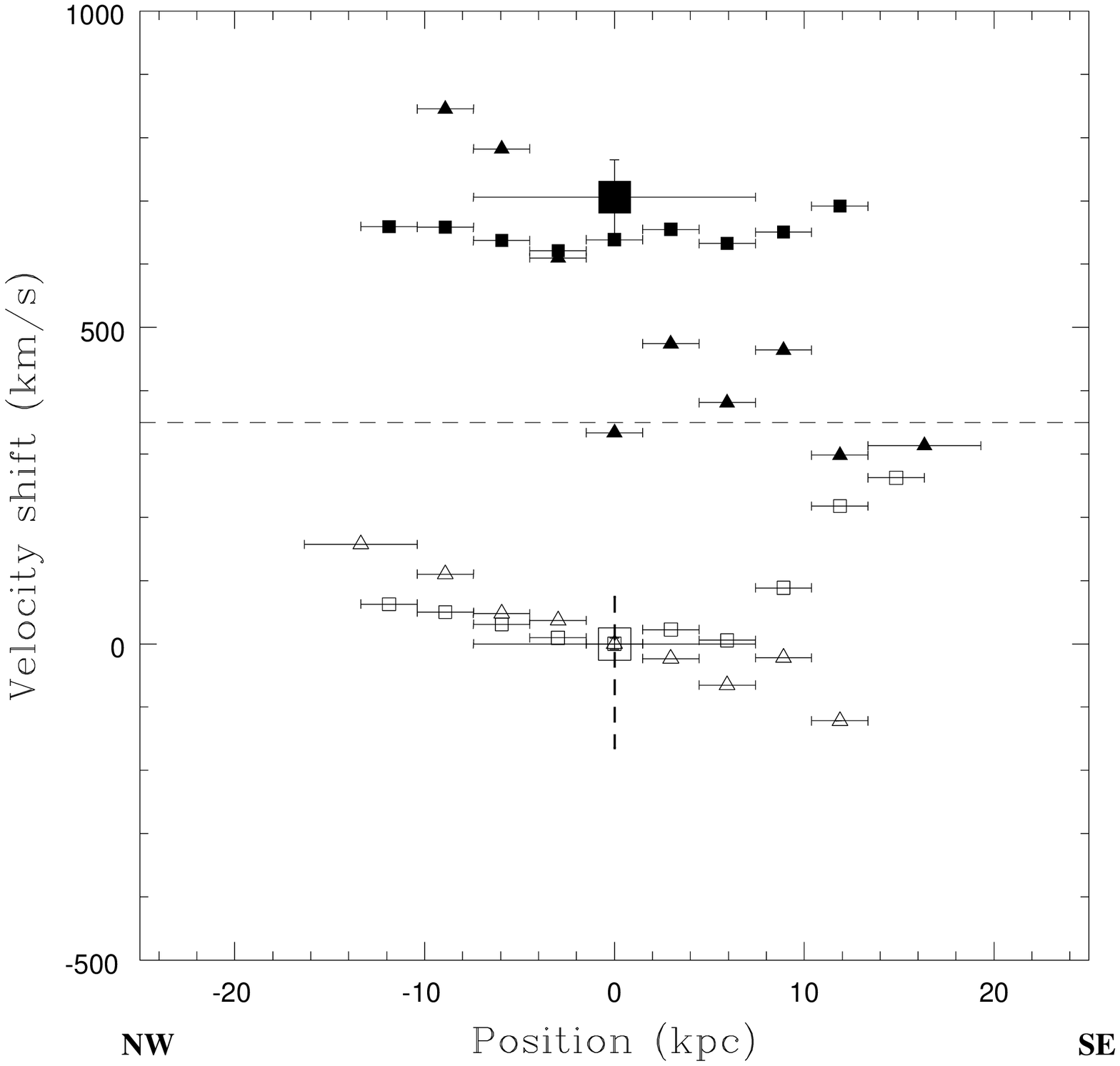,width=7.8cm,angle=0.}\\
\end{tabular}
\caption[]{Velocity profiles {\it continued}. }
\label{fig:velprofiles}
\end{figure*}
\setcounter{figure}{0}
\begin{figure*}
\begin{tabular}{cc}
PKS 1814-63: PA 65 & PKS 1814-63: PA -72 \\
\hspace*{-0.6cm}\psfig{file=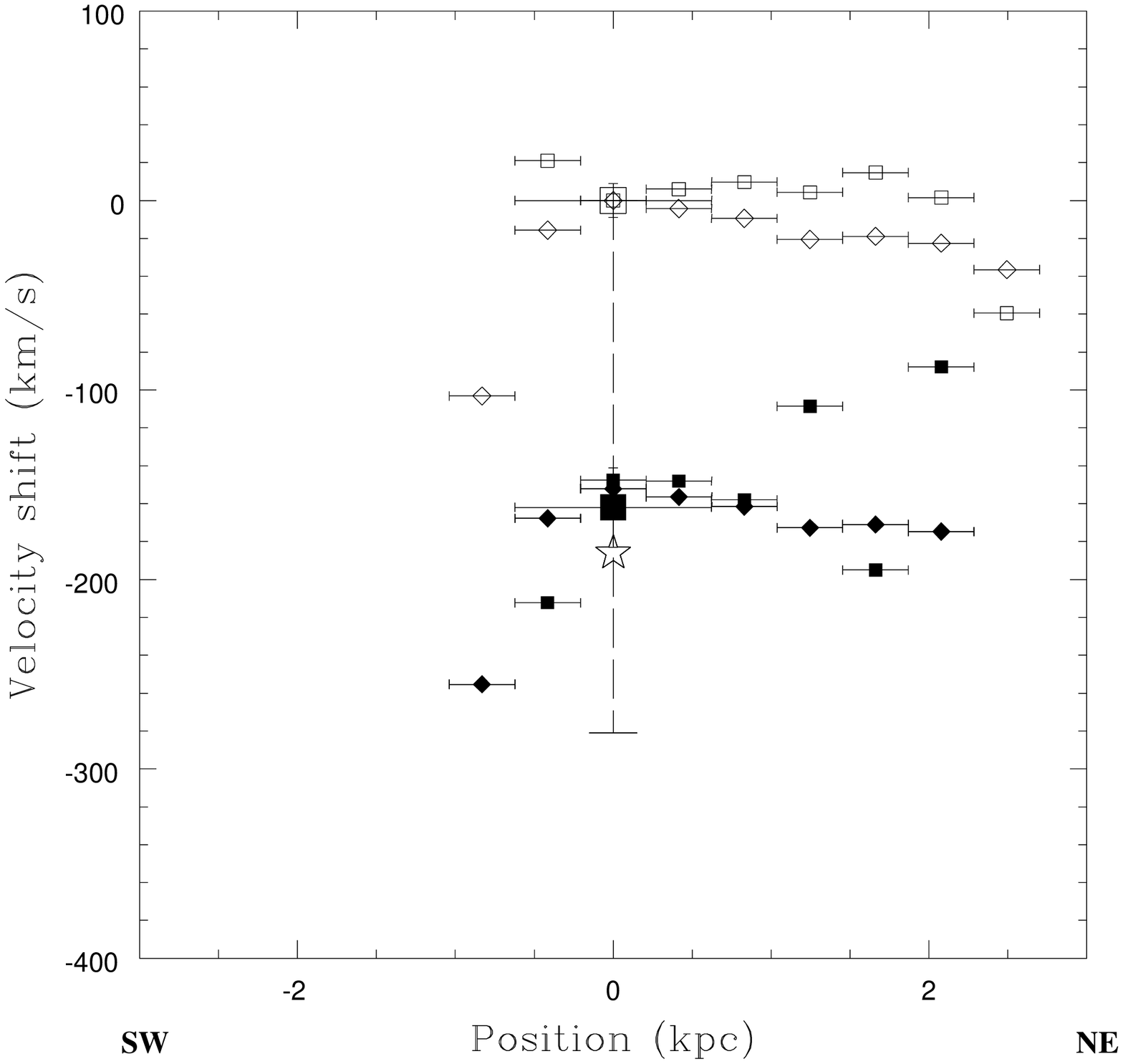,width=7.8cm,angle=0.}&
\psfig{file=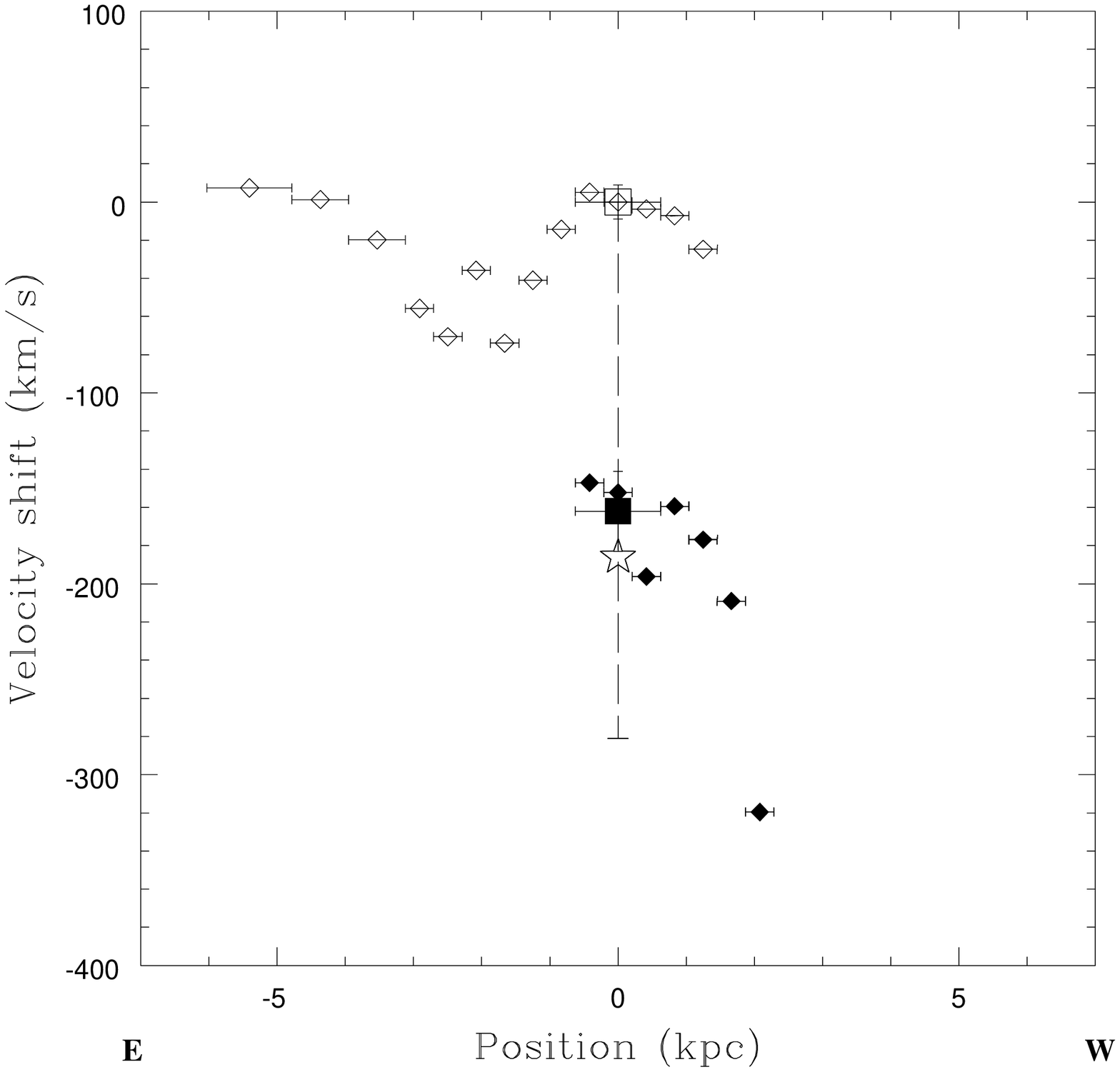,width=7.8cm,angle=0.}\\
\\
PKS 1934-63 & PKS 2135-209 \\
\hspace*{-0.6cm}\psfig{file=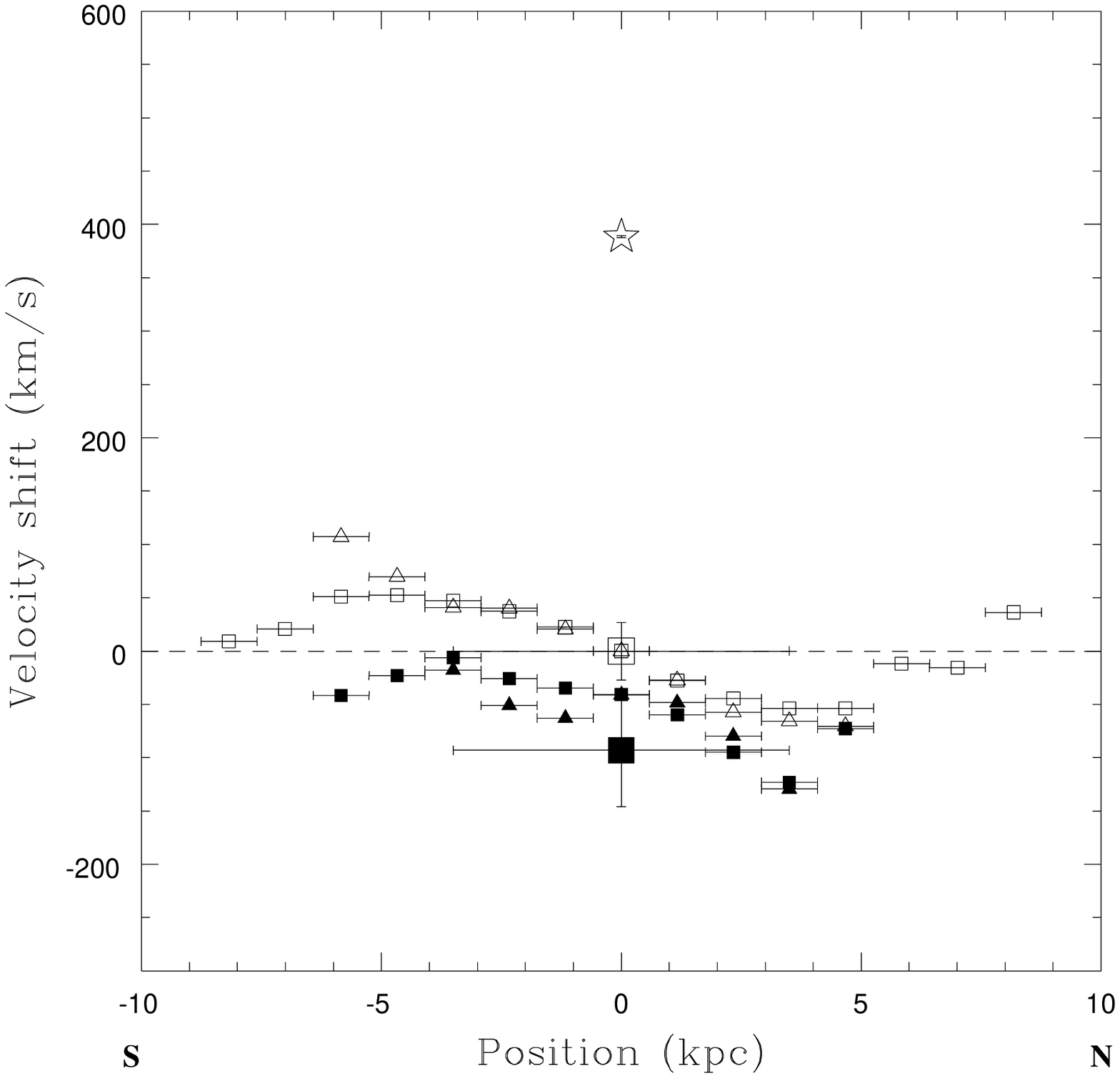,width=7.8cm,angle=0.}&
\psfig{file=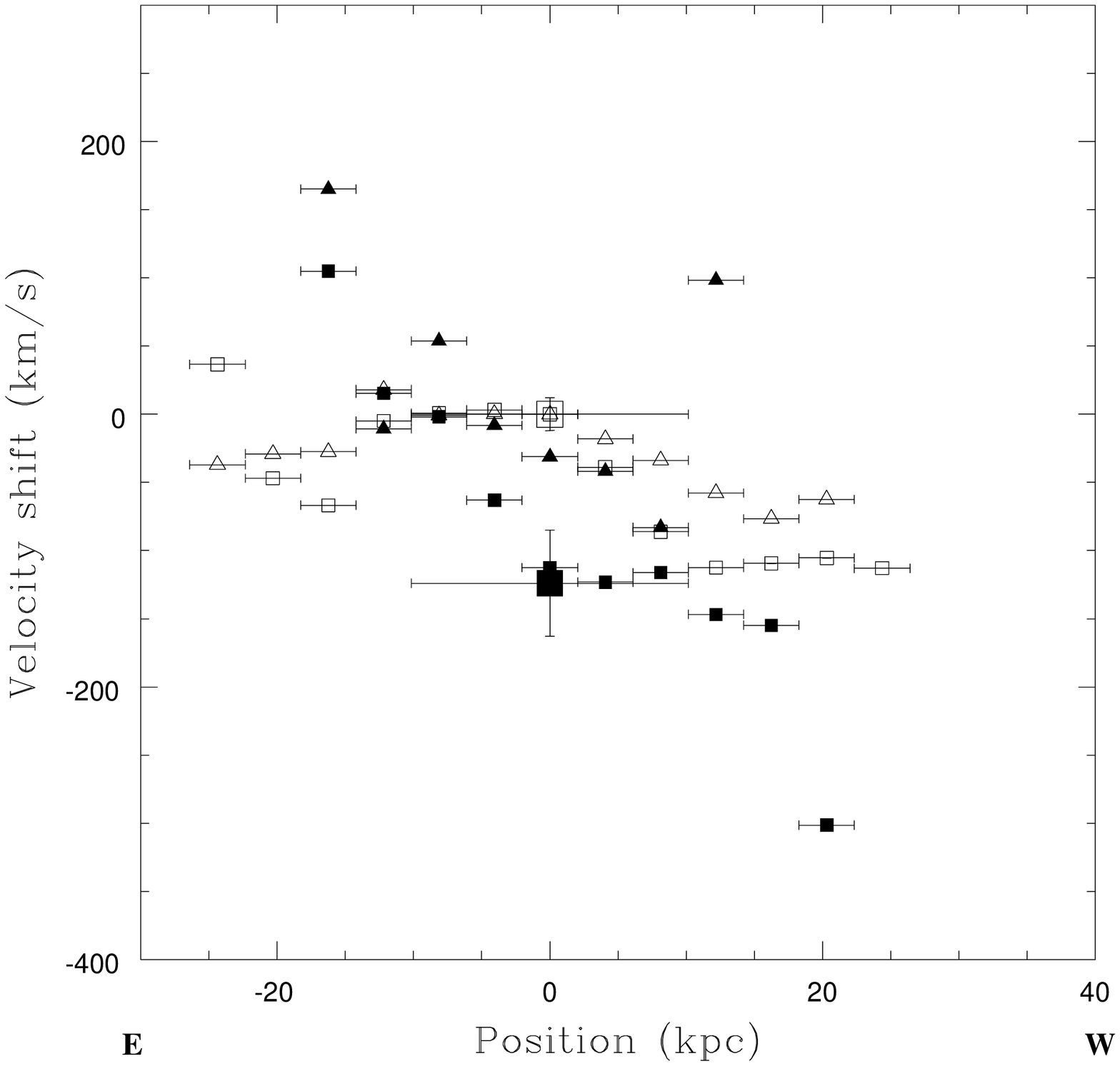,width=7.8cm,angle=0.}\\
\end{tabular}
\caption[]{Velocity profiles {\it continued}. }
\label{fig:velprofiles}
\end{figure*}
\setcounter{figure}{0}
\begin{figure*}
\begin{tabular}{cc}
3C 459: PA 175 & 3C 459: PA 95 \\
\hspace*{-0.6cm}\psfig{file=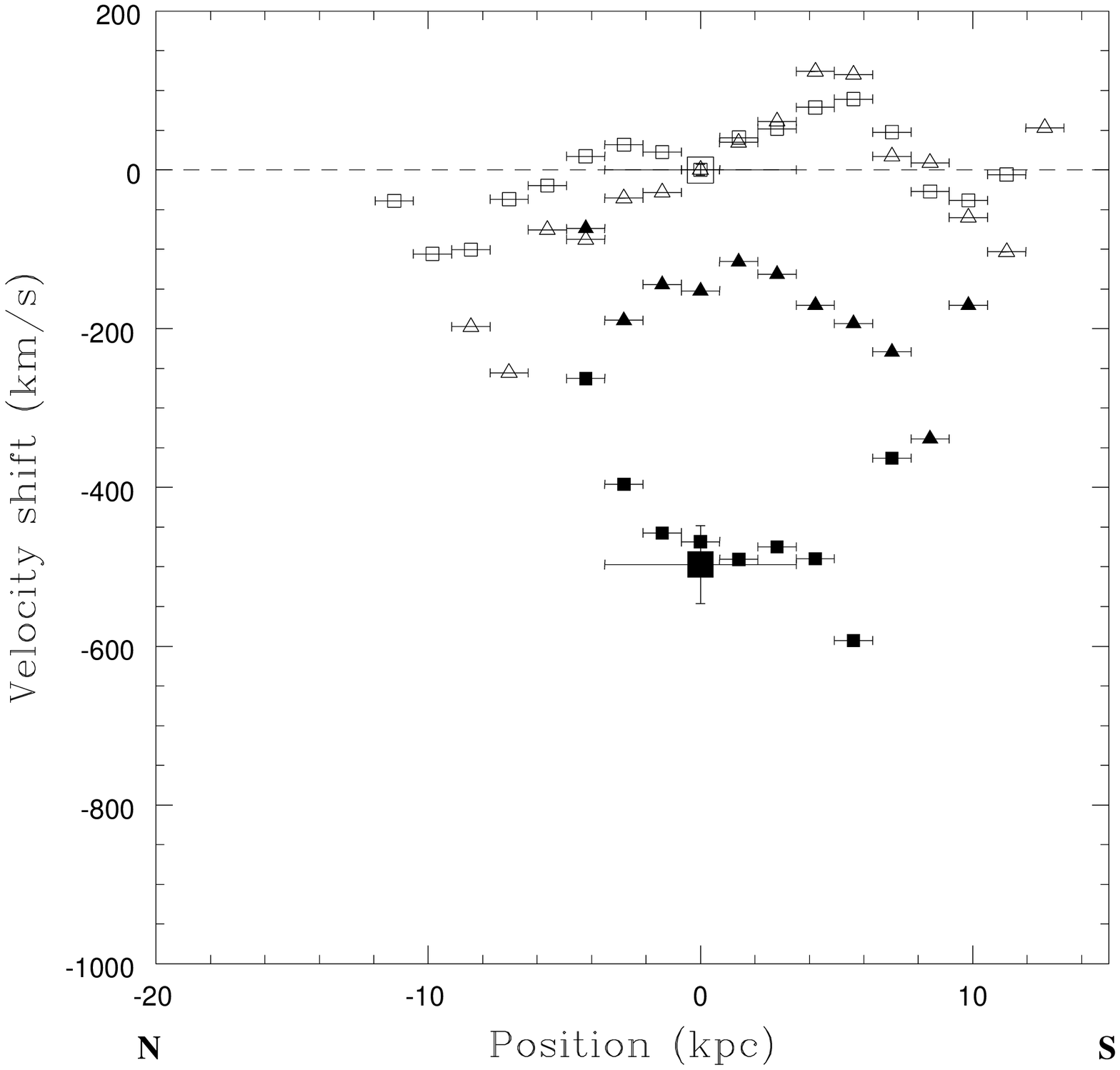,width=7.8cm,angle=0.}&
\psfig{file=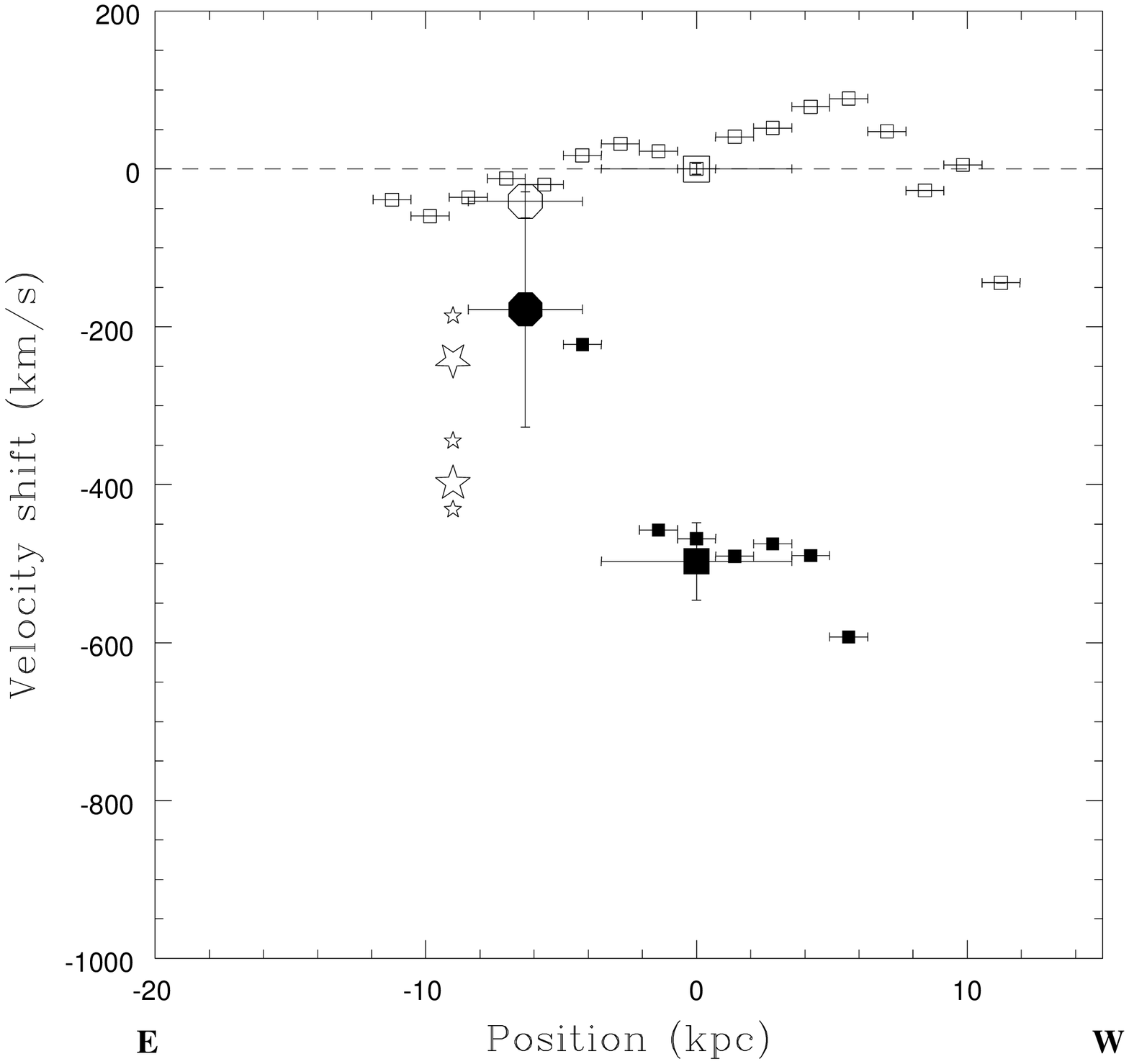,width=7.8cm,angle=0.}\\
PKS 1345+12: all PAs & PKS 1549-79: PA -5\\
\hspace*{-0.6cm}\psfig{file=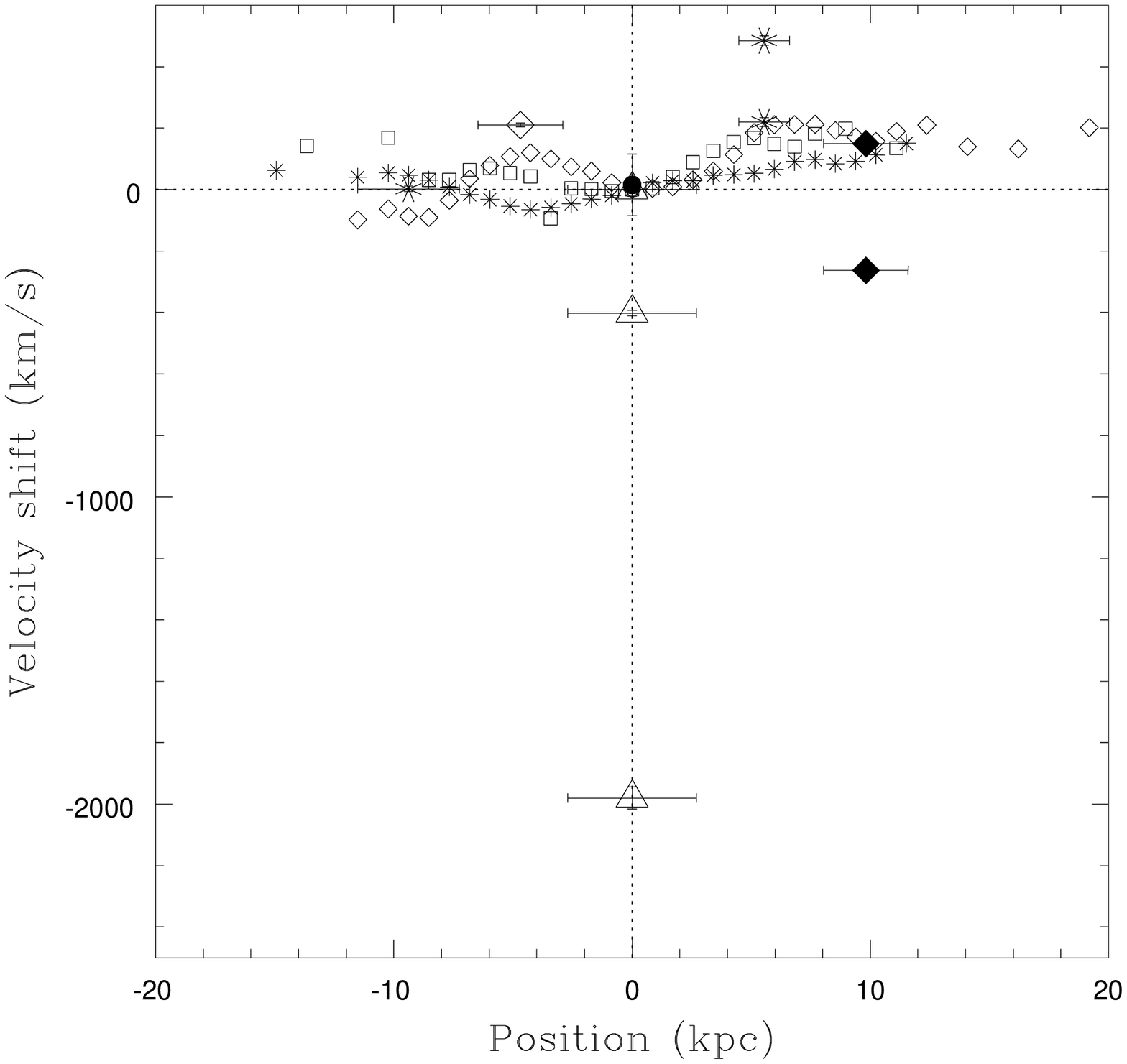,width=7.8cm,angle=0.}&
\psfig{file=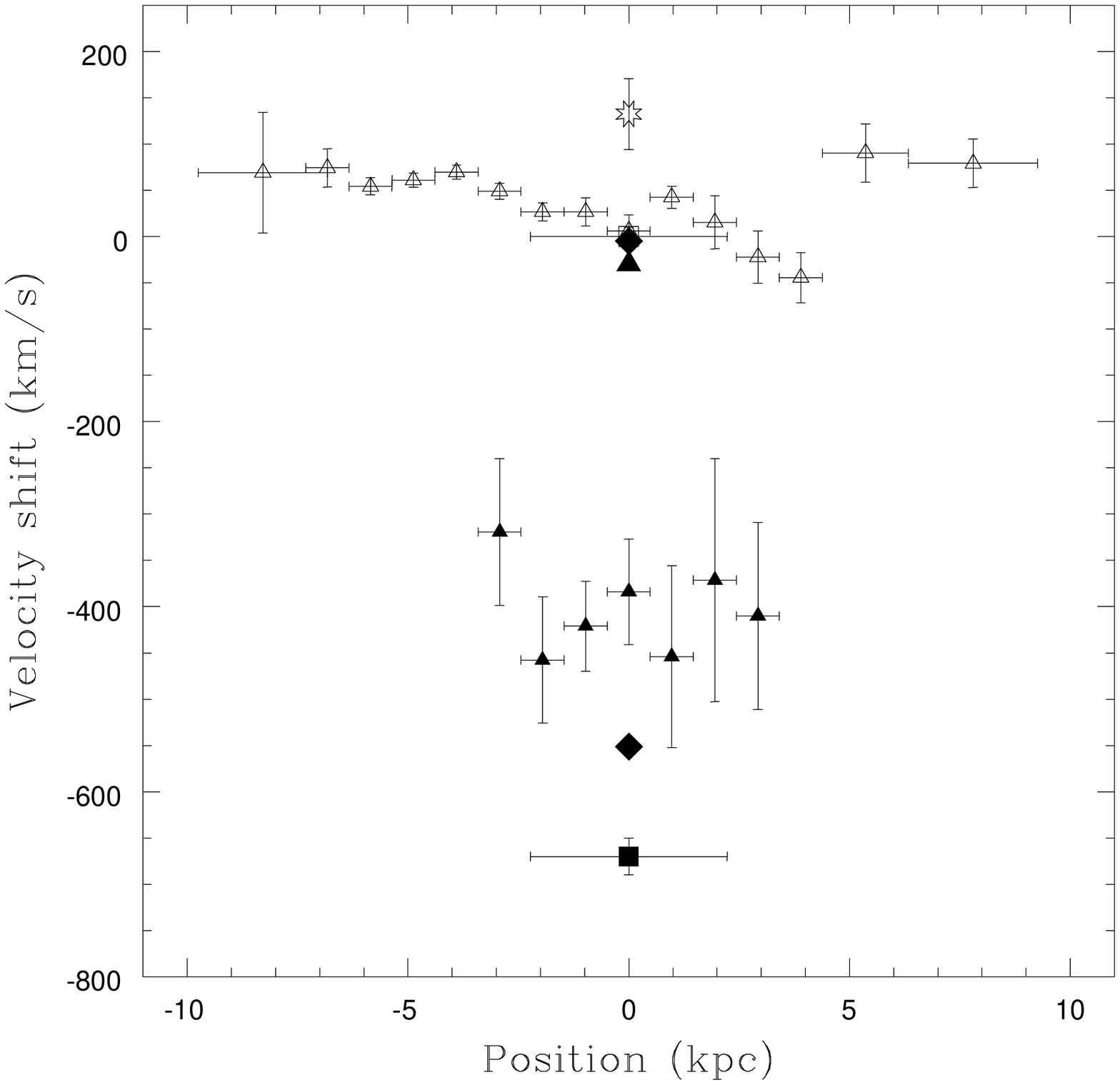,width=7.8cm,angle=0.}\\
\end{tabular}
\caption[]{Velocity profiles {\it continued}. }
\label{fig:velprofiles}
\end{figure*}
\begin{figure}
\centerline{\psfig{file=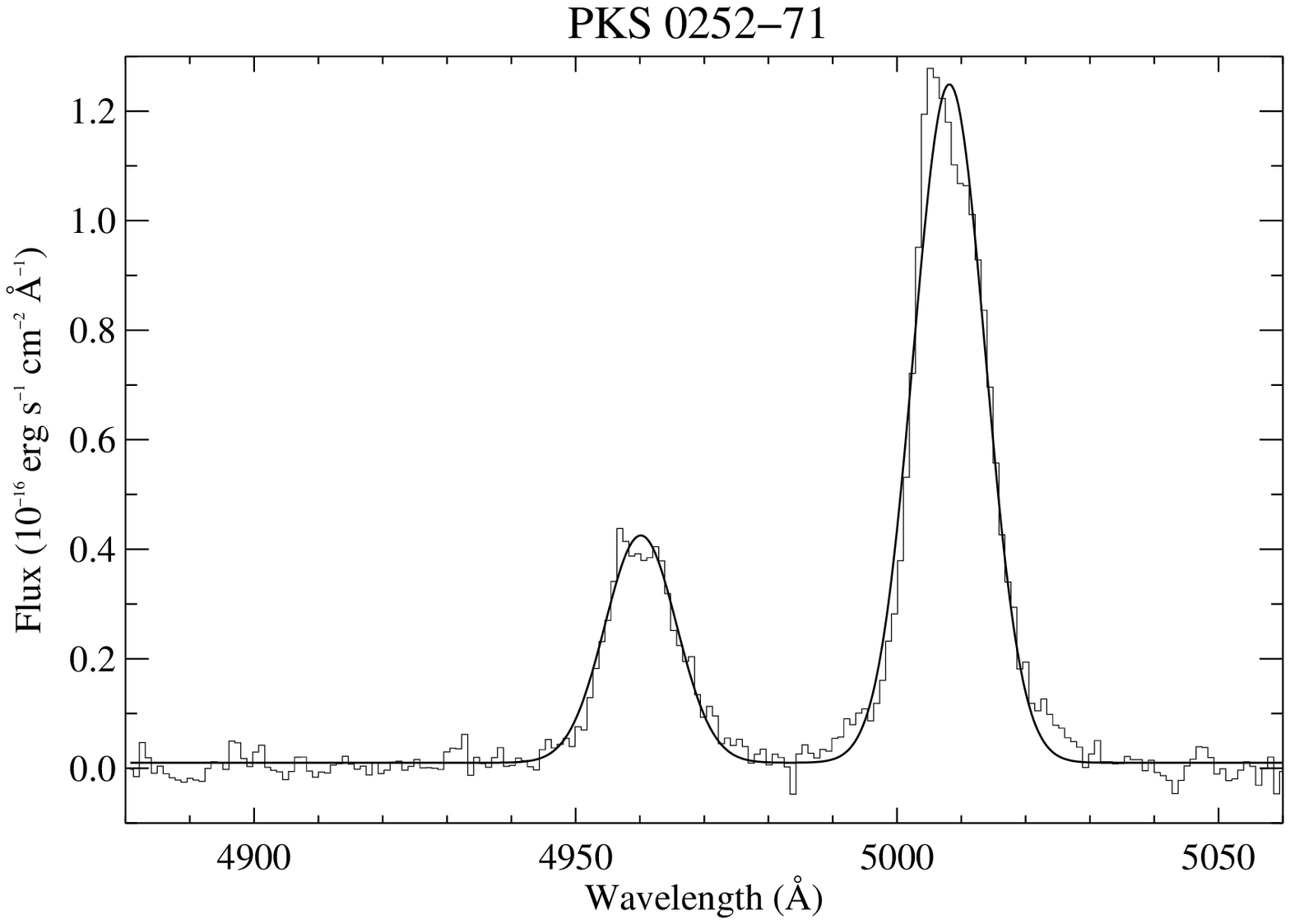,width=8cm,angle=0.}}
\centerline{\psfig{file=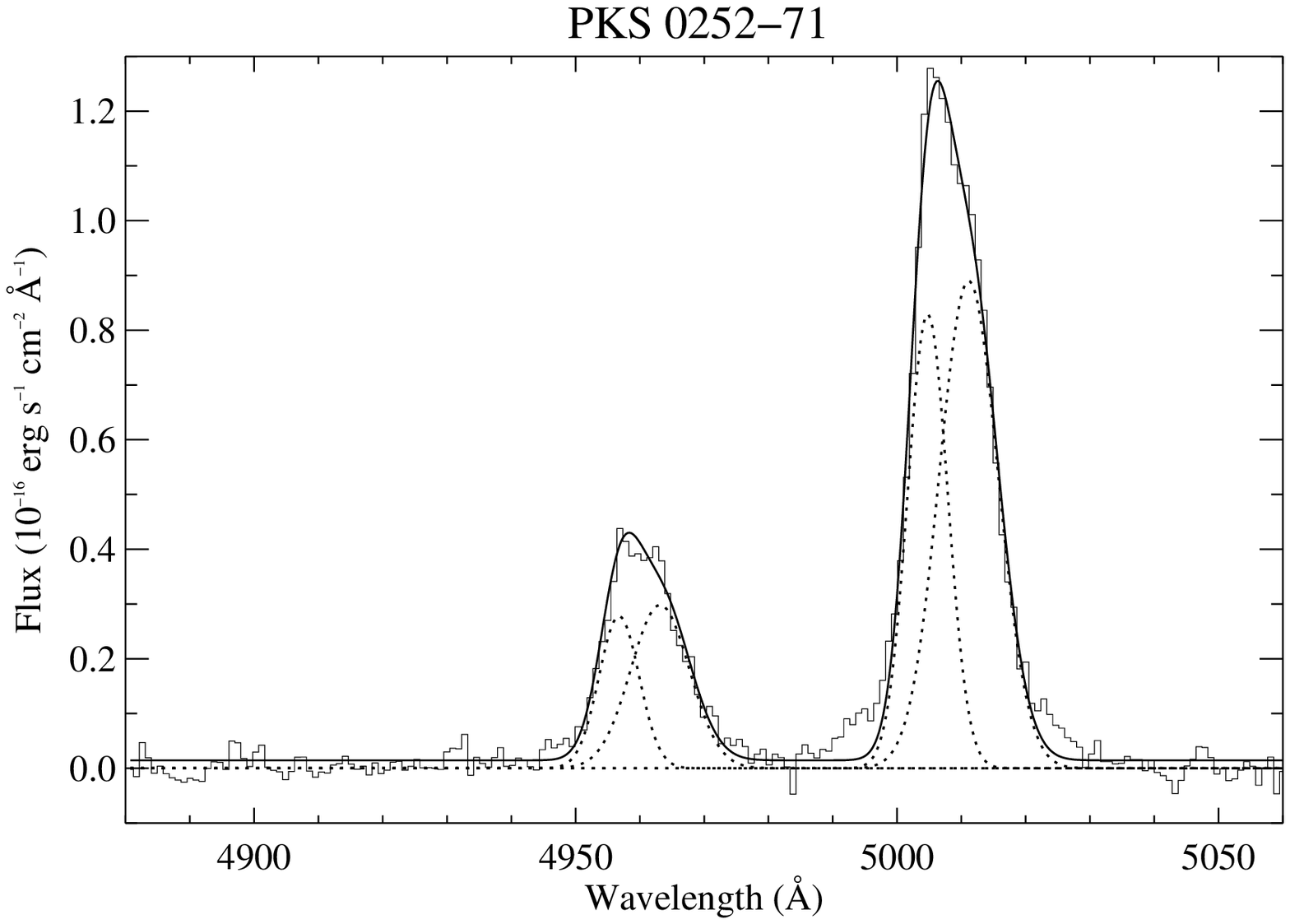,width=8cm,angle=0.}}
\centerline{\psfig{file=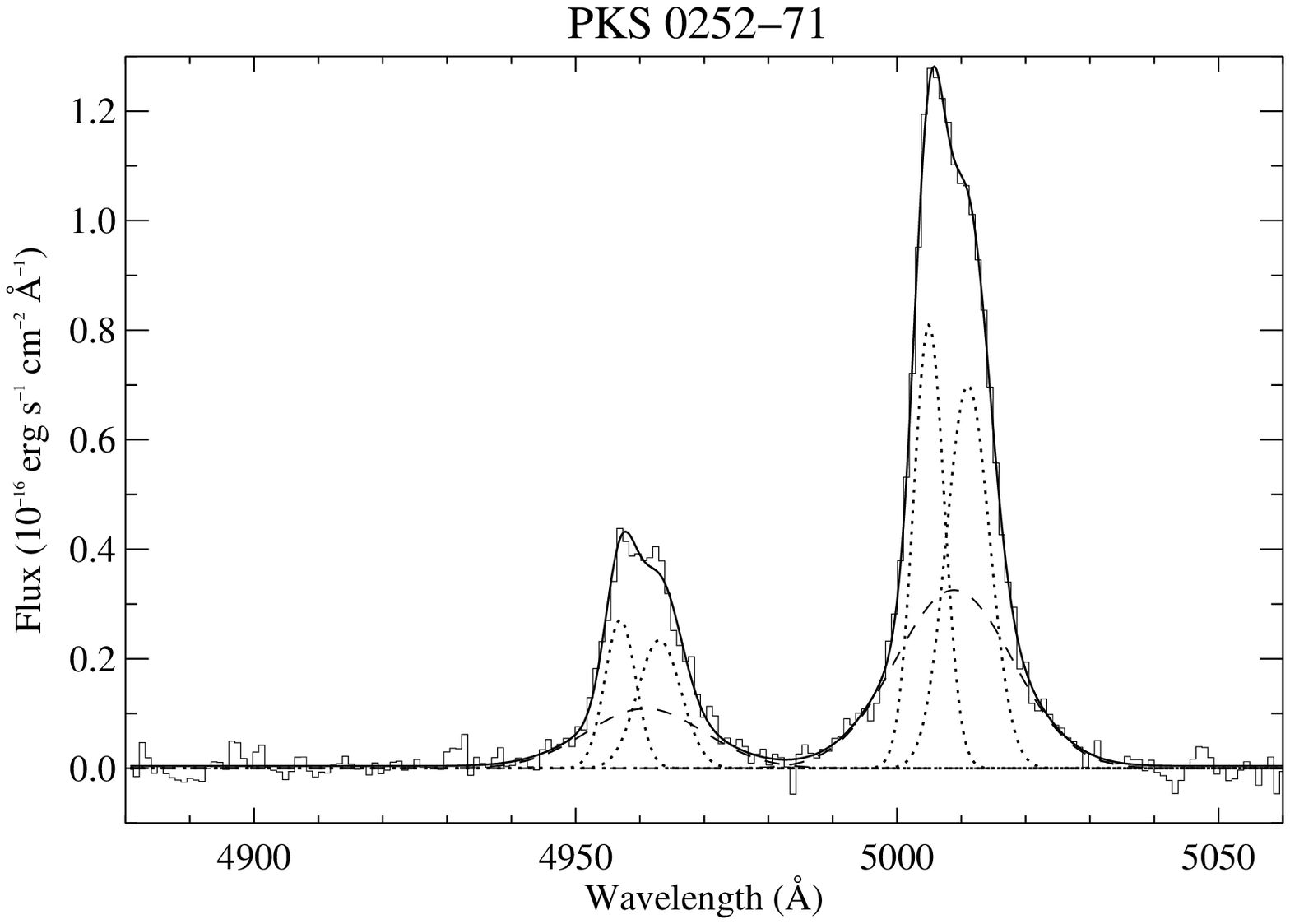,width=8cm,angle=0.}}
\caption[]{The effect oo the overall fit to the {[O
      III]}\lala4959,5007 emission line doublet in the nuclear
      aperture of PKS 0252-71 when 
fitting one      (top), two (middle) and three (bottom) Gaussian
      components to each line in the doublet. In
      each plot, the overall profile (bold line) is overplotted on the
      extracted spectrum (faint line) and the dotted, dashed and
      dot-dashed lines highlight the relative positions and fluxes of
      the contributing components. It is clear that, in the nuclear
      aperture of PKS 0252-71, a minimum of three Gaussian components is
      required to model the emission lines well.
}
\label{no-of-components}
\end{figure}

In order to investigate the impact of the activity on the
circumnuclear material, it is first necessary to establish the {\it
  exact} redshift of the galaxy rest frame.  Many of the objects in the
sample show clear evidence for extended 
emission lines, most notably the {[O II]}\lala3727 doublet and {[O
    III]}$\lambda$5007, and in some cases \ha~$\lambda$6562 (see
below). By modelling 
these, pixel by pixel, across the 
entire spatial extent, one can determine the radial velocity profile of
the host galaxy (see also e.g. H03,H06). 
All such spatial fitting was done before modelling and subtracting the
continuum and the emission lines were fitted using free fits. We
define `free-fitting' as when the only constraints used for the emission
line modelling are those set by atomic physics (i.e. the separation,
ratio of intensities and velocity widths of line components in a
doublet emitted by a single  ion).   The  radial
velocity profiles for all sources are shown in Figure 1\footnote{Note, for some
objects, the velocity fields are only marginally resolved (see Section 3
and Table 2 for details of the spatial scale and the seeing) and this is
why in some of the plots in Figure 1, the velocities do not appear to
vary significantly.}. 
\ref{fig:velprofiles}%\footnote{The radial velocity profiles for PKS
%  1345+12 and PKS 1549-79 can be found in H03, H06 and \citet{javi07}}.
We have determined the systemic velocities using a
variety of techniques: 
\begin{enumerate}
\item {\it Sources with resolved extended emission}. Where the line
   emission line  is clearly resolved we have assumed the extended
   narrow   line emission traces the quiescent halo gas as in PKS
   1345+12. For two sources (3C 277.1 and PKS 1934-63), the extended
   gas has settled into what appears to be a smooth rotation curve
    whilst for the
   remaining sources, the gas is more similar to that in PKS 1345+12
   -- consistent with gravitational motions in an elliptical galaxy
   (velocity amplitude $\leq$ 350 \kms; \citealt{tadhunter89}) but not
yet completely settled. This technique is the most reliable for
   determining the systemic velocity and we have used this for most
   sources in the sample.
\item {\it Sources with two narrow components}. In four sources (3C
  268.3, 4C 32.44, PKS 0252-71 and PKS 1306-09) we detect two narrow
  components in the nuclear aperture (for 3C 268.3, and marginally PKS
  1306-09, these components  are also extended and spatially
  resolved). Free fitting of the 
  emission lines gives similar but slightly different FWHM. Forcing
  the two narrow narrow components to have the same FWHM also gives a
  good fit, and so we assume the two narrow components are emitted by
  the same mechanism/structure. Hence, the narrow line splitting we observe
  could represent a rotation curve which is unresolved, or trace a
  bi-polar outflow in the gas. For these sources, we can therefore
  assume with reasonable confidence that 
  the systemic velocity is located between the two narrow components. 
\item {\it Other methods}. For two sources, we have little confidence
  in determining the systemic velocity. For PKS 1814-63, the 
  foreground star obscures our view of the galaxy. Using the continuum
  subtraction technique described in Section 3.2, we have been able to
model the extended line emission along two PAs. Along PA 65 the
shape  of the profile is unclear whilst along PA -72 a more defined
  shape is observed, which may be consistent with half of the rotation curve.
For PKS 2135-20, we are the least certain and have made no
  attempt to estimate the systemic velocity -- the source is
  unresolved and no narrow components are observed (the narrowest has
  FWHM $\sim$ 760 \kms). In addition to the emission line
  measurements, we have attempted to measure the stellar absorption
  lines (e.g. Ca II K and some of the higher order Balmer lines) in a
  more recently obtained VLT spectrum (see \citealt{holt07}) and
  estimate a redshift of  0.635 $\pm$ 0.004.
\end{enumerate}
Using the above methods, we have confidently determined the systemic
velocity in 12 of the 14 sources (3C 213.1, 3C 268.3, 3C 277.1, 4C
32.44, PKS 1345+12, PKS 1549-79, PKS 1934-63, 3C 459 and with some
assumptions, 3C 303.1, PKS 0023-26, PKS 1306-09 and PKS 0252-71). Due to the issues
discussed above, the velocities for 
 PKS 1814-63 and PKS 2135-20 should be used with caution. Further, for all sources,
the velocity amplitude of the narrow component is coincident with 
gravitational motions in an elliptical galaxy ($\leq$ 350 \kms;
\citealt{tadhunter89}). 

The derived/assumed systemic
velocities are marked on Figure \ref{fig:velprofiles}, and it should be
noted that zero on the y-axis of the velocity profile plots in Figure 
  {\ref{fig:velprofiles}} does not mark the assumed systemic velocity, which
  is denoted by the horizontal dashed line. The results are summarised
  % and the method(s) used for each source are highlighted
 in Table   {\ref{tab:lineparam}}. Note, whilst the extended gas was used to
   determine the location of 
the systemic velocity with respect to the various emission line
components, 
the redshifts quoted as systemic in Table {\ref{tab:lineparam}}
were derived using measurements of several emission lines (narrow
components) in the {\it 
  nuclear apertures} (as part of the modelling in Section 4.2) rather
than based on measurements of {[O III]} alone.

In addition to the velocities of the extended emission line
components in Figure {\ref{fig:velprofiles}}, the components of the
nuclear {[O III]} lines are shown (see Section 4.2). In the majority
of the radial velocity profiles, there is a clear offset in velocities
of the broader components  between the extended and nuclear apertures of order $\sim$ 100
 \kms. This is most likely due to continuum subtraction effects in the
 nuclear apertures and/or integrating over a larger aperture. 

HI 21cm absorption is detected in 10 of the 14 sources
(see Table \ref{tab:hidata}) and the velocities of the detected components are
plotted on Figure \ref{fig:velprofiles}. In 
four sources 
(3C 213.1, 4C 32.44, PKS 1345+12 and PKS 1549-79) at least one
component of HI is in agreement with the assumed systemic
velocity. However, it is striking that the majority of detected HI
components are significantly blueshifted with respect to the optically
derived systemic velocities and, in the case of PKS 1345+12 at least,
appear to trace the optical outflow. In only two sources is the HI
redshifted, consistent with  infalling gas (PKS 0023-26 and PKS
1934-63). 

\subsection{Emission line modelling}
\label{section:emlmodelling}
As discussed in Section 1, the nuclear emission lines in compact radio
sources are often broad with asymmetric profiles requiring multiple
Gaussian components to model them. In Section {\ref{sect:halo}}, we
identified the  rest frame for each source. Here, after subtracting
the continuum, we model all strong emission lines in the nucleus 
to search for outflows. In addition to multiple Gaussian fitting
following H03, we present the results of single
Gaussian fitting and asymmetry indices. These latter two steps will
allow us to make comparisons between our data and previous work.

\subsubsection{Multiple Gaussian modelling}
The nuclear emission lines in compact radio sources are often broad
with asymmetric profiles. Hence, single 
Gaussians are unable to adequately model the line profiles and
multiple components are required. As an example,  Figure 
{\ref{no-of-components}} shows how fitting 1, 2 and 3 Gaussians to {[O
    III]}\lala4959,5007 in the nucleus  of PKS 0252-71 affects the
quality of the overall fit.  

When modelling the emission line spectra of AGN, it is common to start
with the brightest emission lines, e.g. the {[O III]}\lala4959,5007
doublet (e.g. \citealt{villar99a}). The {[O III]} lines were fitted
using three constraints in accordance with atomic physics: the flux
ratio between {[O III]}$\lambda$4959 and {[O III]}$\lambda$5007 was
set at 2.99:1 (based on the transition probabilities;
\citealt{osterbrock89}); the widths of
the corresponding components of each line were forced to be equal; and
the shifts between the corresponding components of each line were
fixed to be 48.0\AA. Note that the fitting program used can only work
with a wavelength difference and not a ratio of wavelengths. However,
we find the incurred error is smaller than our estimated uncertainty. 

Following the technique of H03, we have modelled the {[O III]}
lines using the {\it minimum} number of Gaussians required to give a
good fit\footnote{Note, we define a `good fit' to the emission lines as that which, by eye,
  has acceptibly low residuals and fits all of the major features
  (peaks/wings etc.) of
  the overall line profile well.}. Hence, for all sources in the
sample the fits to {[O III]} 
required between 2 and 4 Gaussian components to model the doublet
well. The best fitting models are presented in Figure
{\ref{fig:o3models}} and the line data (velocity
widths\footnote{Line widths are corrected for instrumental width using
  the equation: 
  FWHM$_{\rm c}$ = $\sqrt{{\rm FWHM}^2_{\rm m} - {\rm IW}^{2}} $ where
  FWHM$_{\rm c}$ is the corrected Full Width at Half Maximum 
(FWHM) of the line (in \AA),  
FWHM$_{\rm m}$ is the measured FWHM of the line (in \AA) and
IW is the instrumental width or spectral resolution, calculated using
the measured widths of the night sky lines. }, relative
shifts) are presented in Table {\ref{tab:shifts}}. The errors quoted
in Table 4 are a combination of the measurement errors (taken from {\sc
  dipso}) and an estimation of the expected errors based on the
spectral resolution of the spectra. Typically, for the weaker and/or
broader components, the {\sc dipso} error is representative of the true error 
on the result. For the stronger and/or narrower components, the {\sc dipso} error is
very small ($<<$1 per cent) and is not a good indicator of the true errors. For these
sources, we have estimated a percentage error based on the spectral resolution
and signal-to-noise of the data. This is typically of order few-10 per cent.
  In addition, the
velocities  of the nuclear {[O III]} lines are overplotted on the
radial velocity profiles in Figure {\ref{fig:velprofiles}}.

From Figure {\ref{fig:o3models}} it is clear that a number of sources
show extreme line widths and shifts, particularly PKS 1345+12, 4C
32.44, PKS 1549-79 and 3C 459. It is interesting to note that these
sources account for 4 out of the 5 smallest  sources in the sample, each
having a radio source with linear size $<$1 kpc\footnote{Note, for 3C
  459, whilst the overall radio source is highly extended the compact
  core is consistent with the other compact radio sources in the
  sample.}. For the fifth source in 
this group, PKS 1814-63, non-detection of broader, more blueshifted
components may be real although difficulties in subtracting the
continuum due to the presence of a foreground star may also influence
the result.

When modelling the emission lines in AGN, it is often 
  assumed that one model will reproduce {\it 
  all} emission lines. This technique has been particularly successful
  in studies of jet-cloud interactions in powerful radio galaxies
  (e.g. Villar-Mart{\'{i}}n et al. 1999; \citealt{taylor03}).
Hence, after
  modelling {[O III]} in  nuclear aperture of each source, we
  attempted to model the other nuclear emission lines with
  the same components and the same velocity widths and shifts as the
  corresponding {[O III]} model, hereafter the {[O III]} model.

 As well as the constraints from
  the {[O III]} model, some emission line doublets required further
  constraints in accordance with atomic physics: the  shift between the
  corresponding components of each line in all doublets (e.g.  {[Ne
  III]}\lala3868,3968, {[Ne V]}\lala3346,3425, {[N II]}\lala6548,6583, 
{[O I]}\lala6300,6363 and {[S II]}\lala6716,6731) were set and for
  some doublets (e.g. {[Ne
  III]}\lala3868,3968, {[Ne V]}\lala3346,3425, {[N II]}\lala6548,6583,
  and {[O I]}\lala6300,6363) the flux ratios were set based on the transition
  probabilities. For  {[S II]}\lala6716,6731, the flux ratio was
  required  to be within the range 0.44 $<$ {[S II]}6716/6731 $<$
  1.42, the ratios corresponding to the high and 
  low density limits respectively.

With the exception of PKS 1345+12
and PKS 1934-63,  one model (the `{[O III]} model') fitted the other
emission lines well. For PKS 1345+12, only the narrow component was
consistently found in all lines whilst in PKS 1934-63, none of the
emission line components was consistent in all lines. However, it
should be noted that, due to the complexity of the 
emission line profiles in all sources, it was often
  necessary to force the position of the
narrow component to be at the systemic redshift to obtain a good fit
to the lines with the {[O III]} model. Further, in 
sources where two narrow components are detected, whilst
  free fitting gave different FWHM for the two components, it was also
  possible to obtain a good overall fit forcing the two narrow
  components to have equal FWHM (e.g. 3C 268.3, 4C 32.44, PKS 0252-71,
  PKS 1306-09, as discussed in Section {\ref{sect:halo}}). In these
  sources the derived systemic redshift was an average of the
  redshifts of the two components.

In addition to emission originating from the NLR, we have detected
broad components to the permitted lines (due to the signal-to-noise of
the data, often only detected in \ha), i.e. from the BLR, 
in three sources: 3C 268.3, 3C 277.1 and PKS 1549-79 (H06). 3C 277.1
is known to be a quasar 
and so the detection of a broad 
component to the permitted lines was not surprising. However, the
detection of a broad component in \ha~ in 3C 268.3 has led to us
re-classifying this radio source as a BLRG rather than an NLRG. The
detection of broad H$\alpha$ and Pa$\alpha$ in PKS 1549-79 was
reported in H06. 
The line data for the BLR components is presented in Table
{\ref{tab:shifts}}.

\begin{figure*}
\begin{tabular}{cc}
\psfig{file=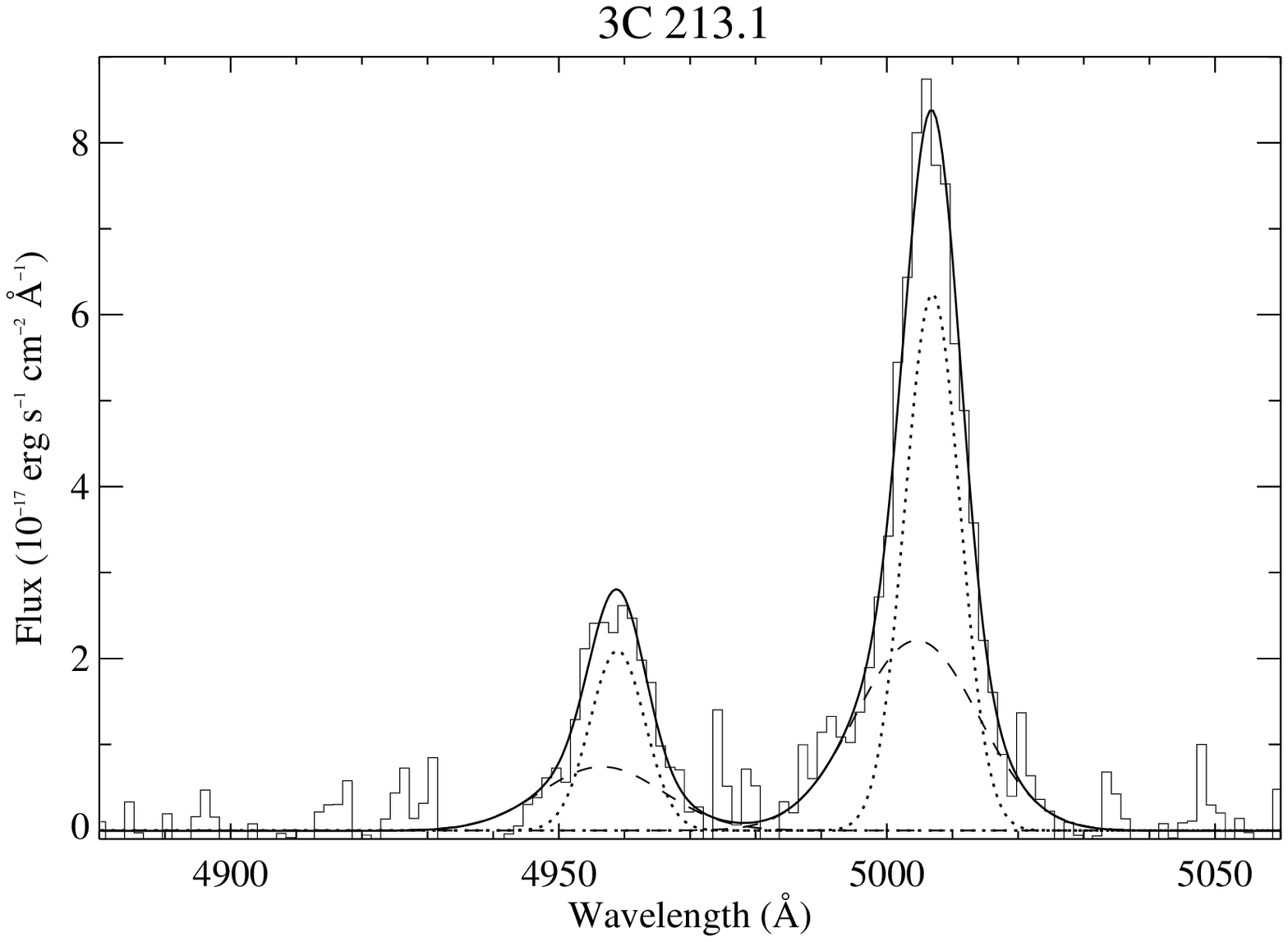,width=8cm,angle=0.}&
\psfig{file=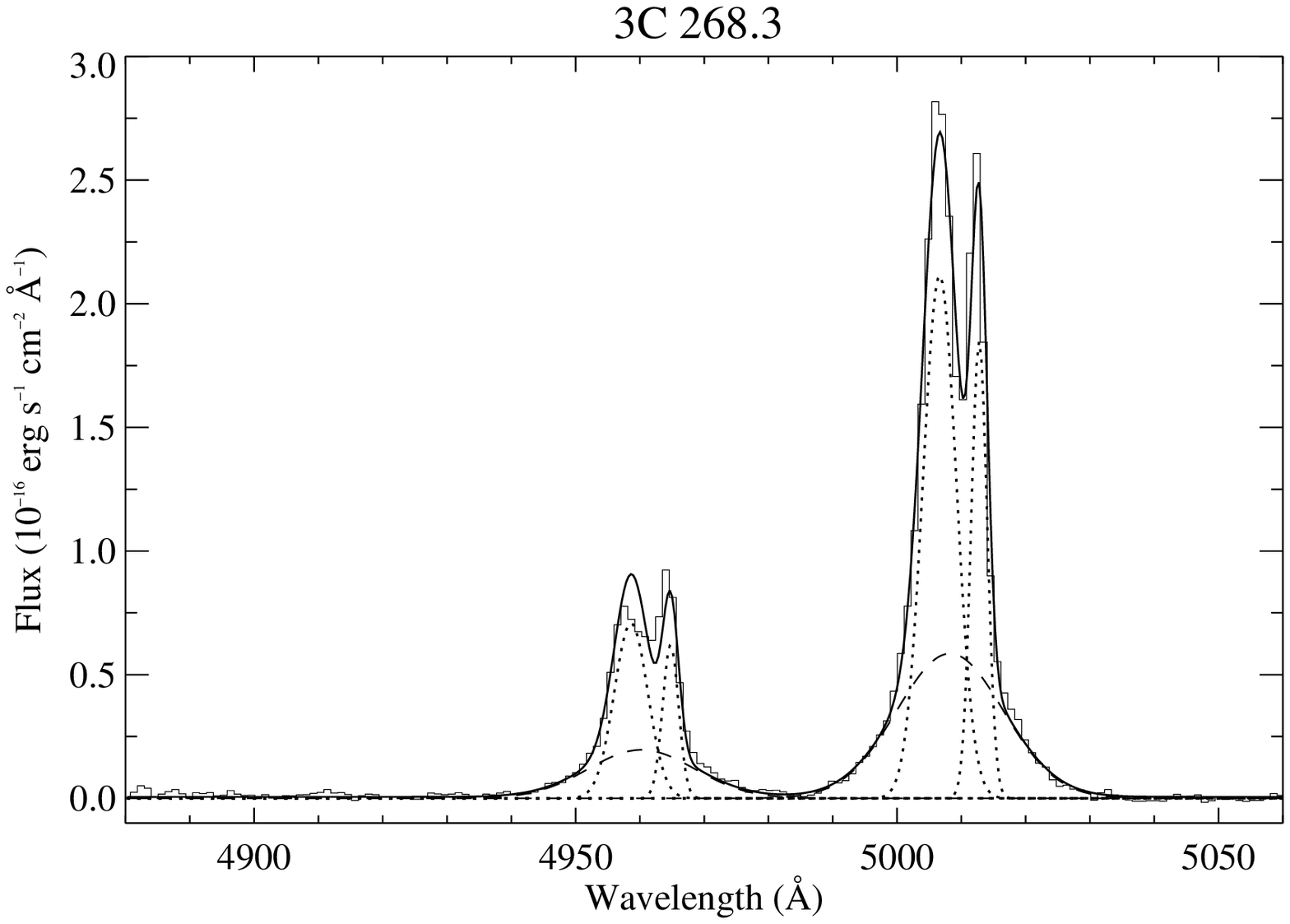,width=8cm,angle=0.}\\
\psfig{file=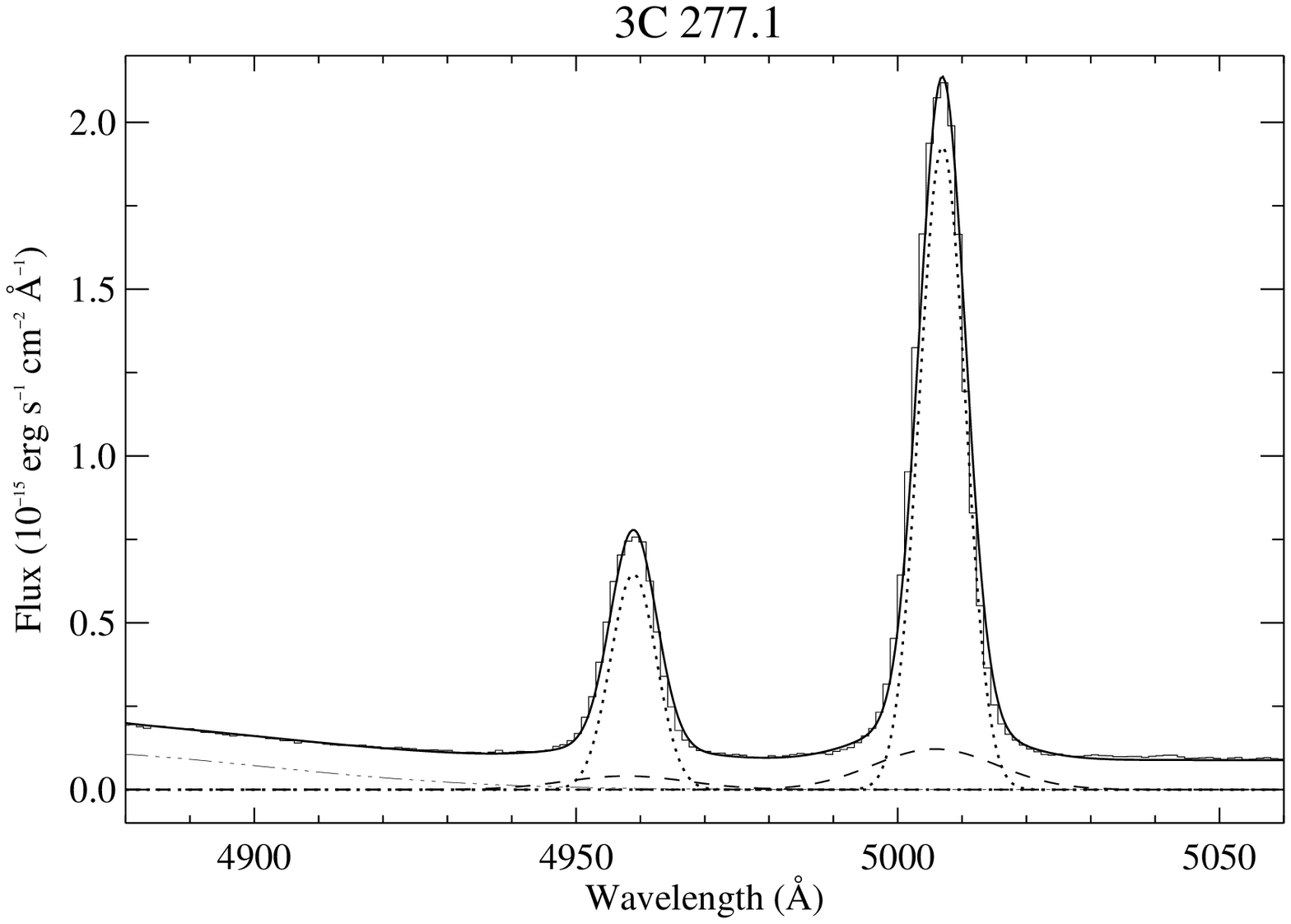,width=8cm,angle=0.}&
\psfig{file=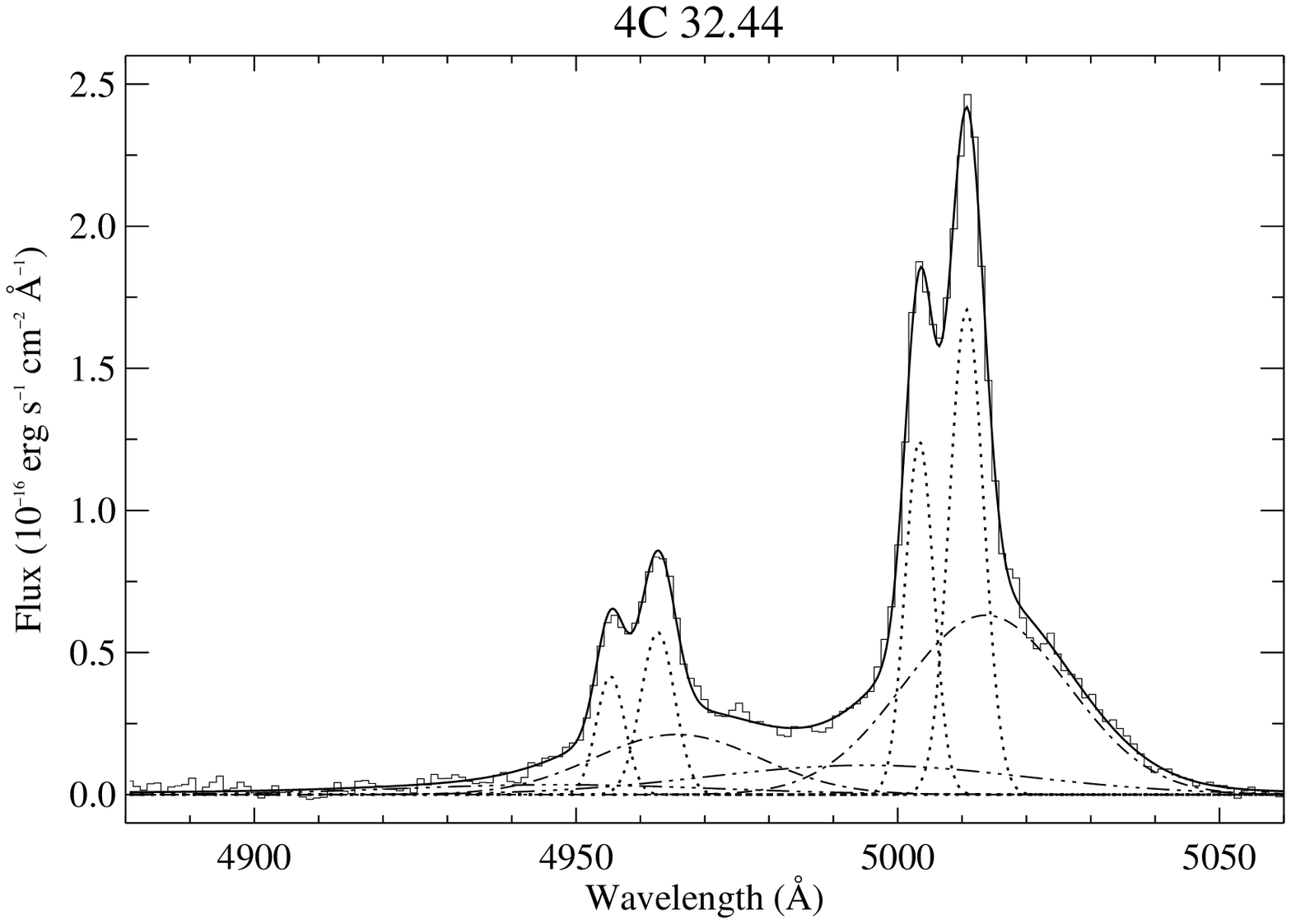,width=8cm,angle=0.}\\
\psfig{file=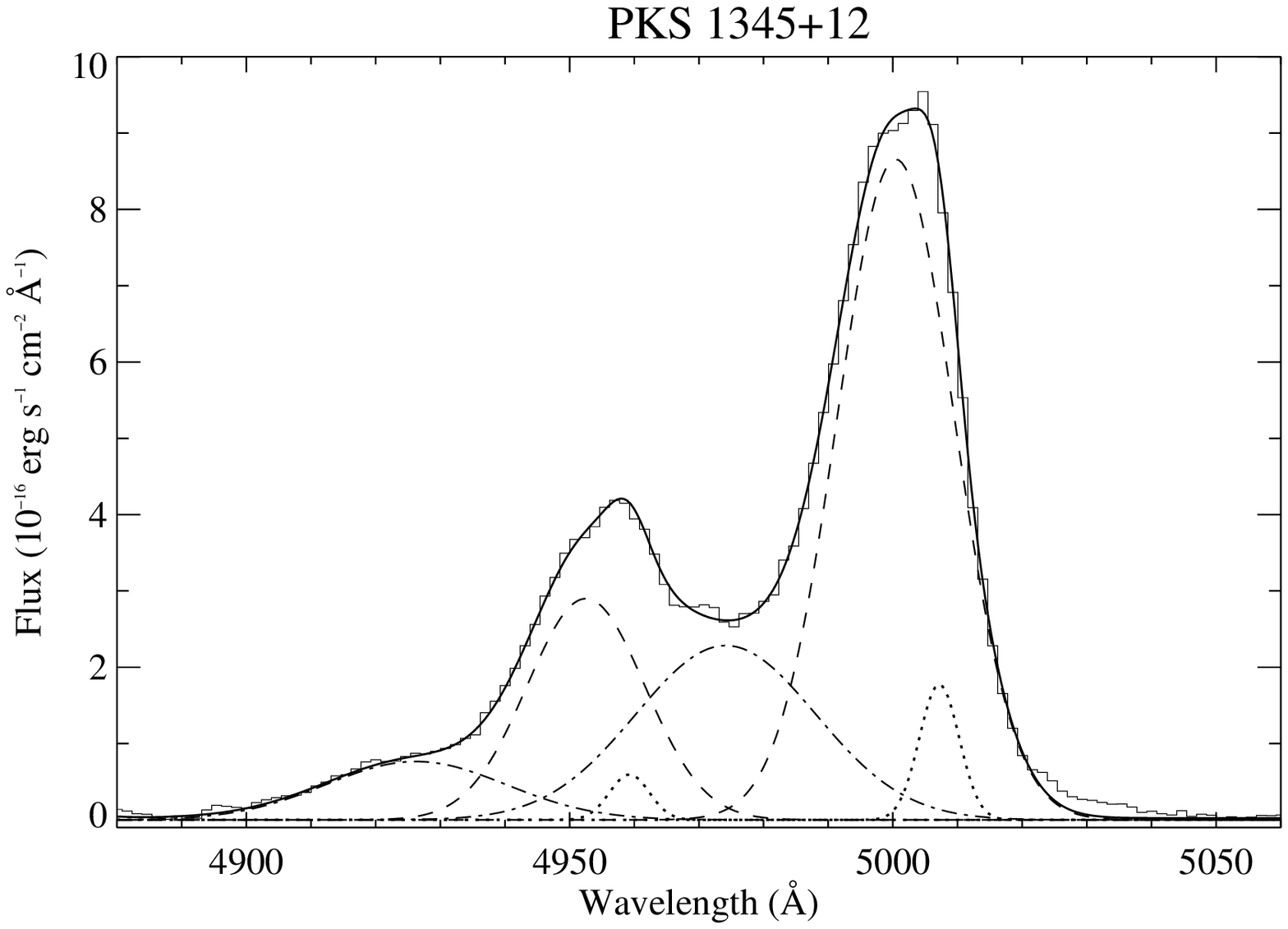,width=8cm,angle=0.}&
\psfig{file=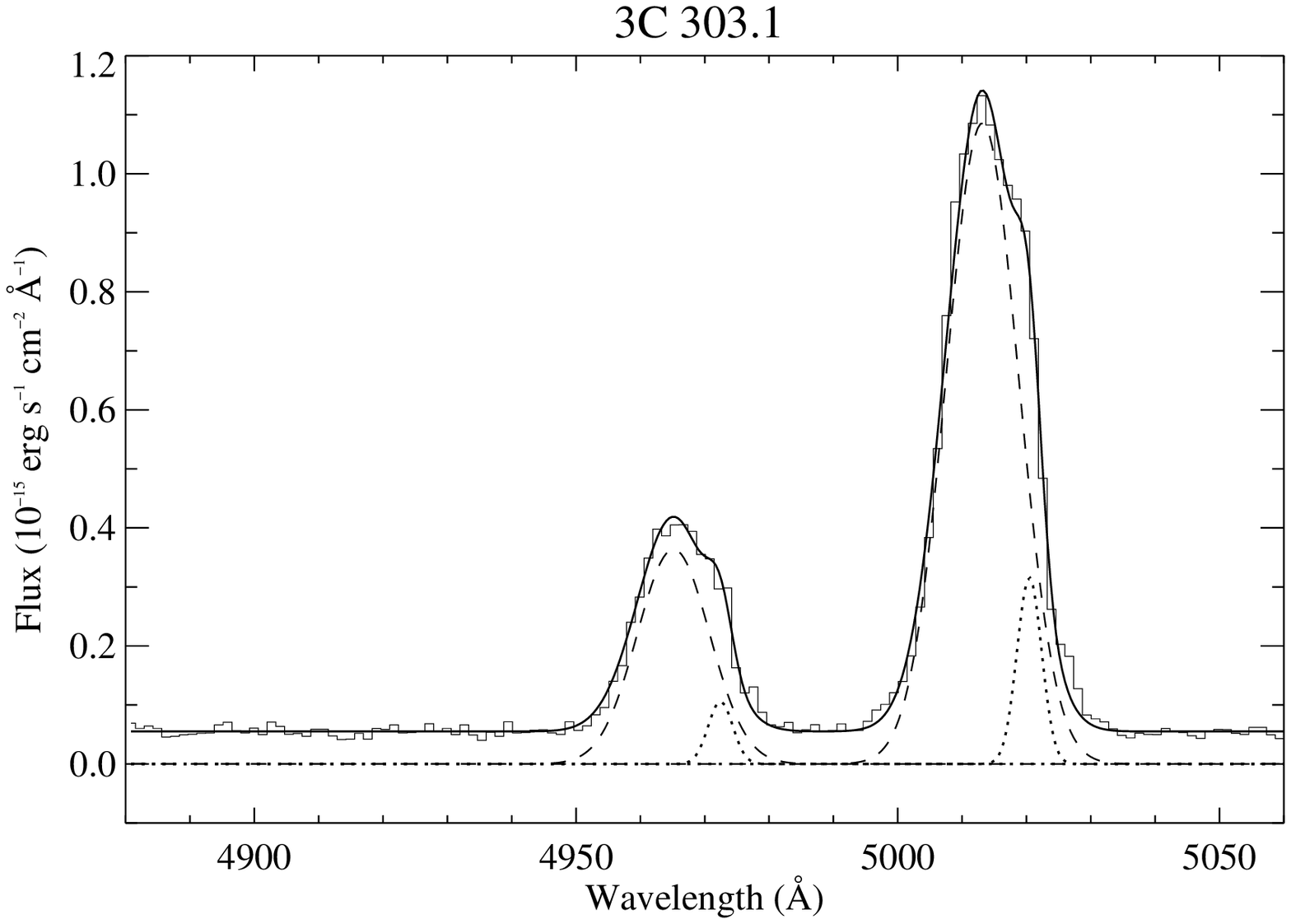,width=8cm,angle=0.}\\
\end{tabular}
\caption[]{Models to the {[O III]}\lala4959,5007 emission line doublet
in the nuclear apertures of all 14 sources in the sample. The faint
solid line represents the continuum subtracted spectrum and the bold
solid line is the overall model fit to the doublet. The components
are: narrow (dotted), intermediate (dashed), broad (dot-dashed) and
very broad (dot-dot-dashed) for all sources except 3C 277.1 where the
dot-dot-dashed line traces the BLR component of \hb~(see Section 3.3
for the kinematic definitions). All plots are on
the same wavelength (i.e. velocity) scale to allow for comparisons between the
outflows.}
\label{fig:o3models}
\end{figure*}
\setcounter{figure}{2}
\begin{figure*}
\begin{tabular}{cc}
\psfig{file=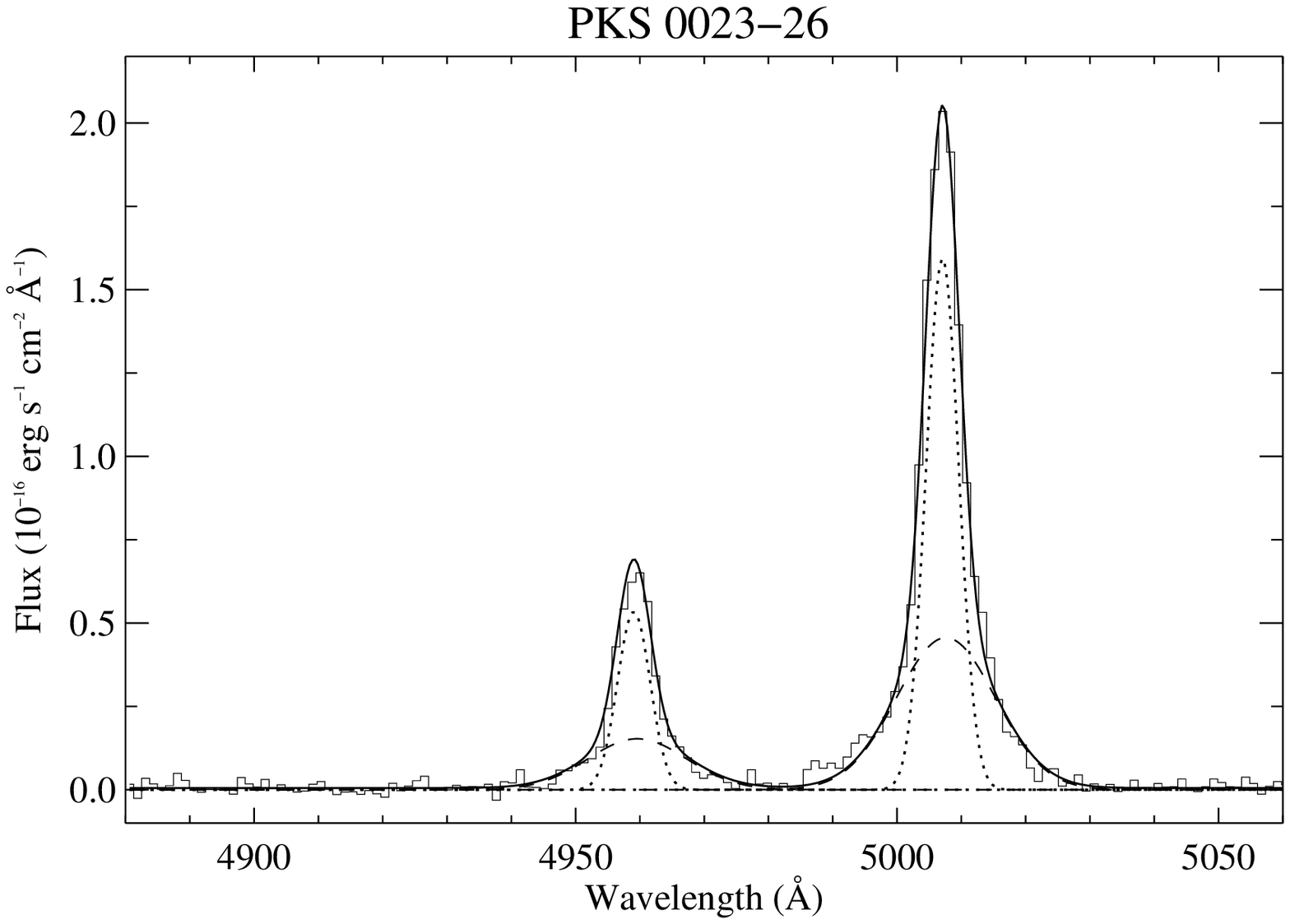,width=8cm,angle=0.}&
\psfig{file=plots/o3model/pks0252-o3model-new.ps,width=8cm,angle=0.}\\
\psfig{file=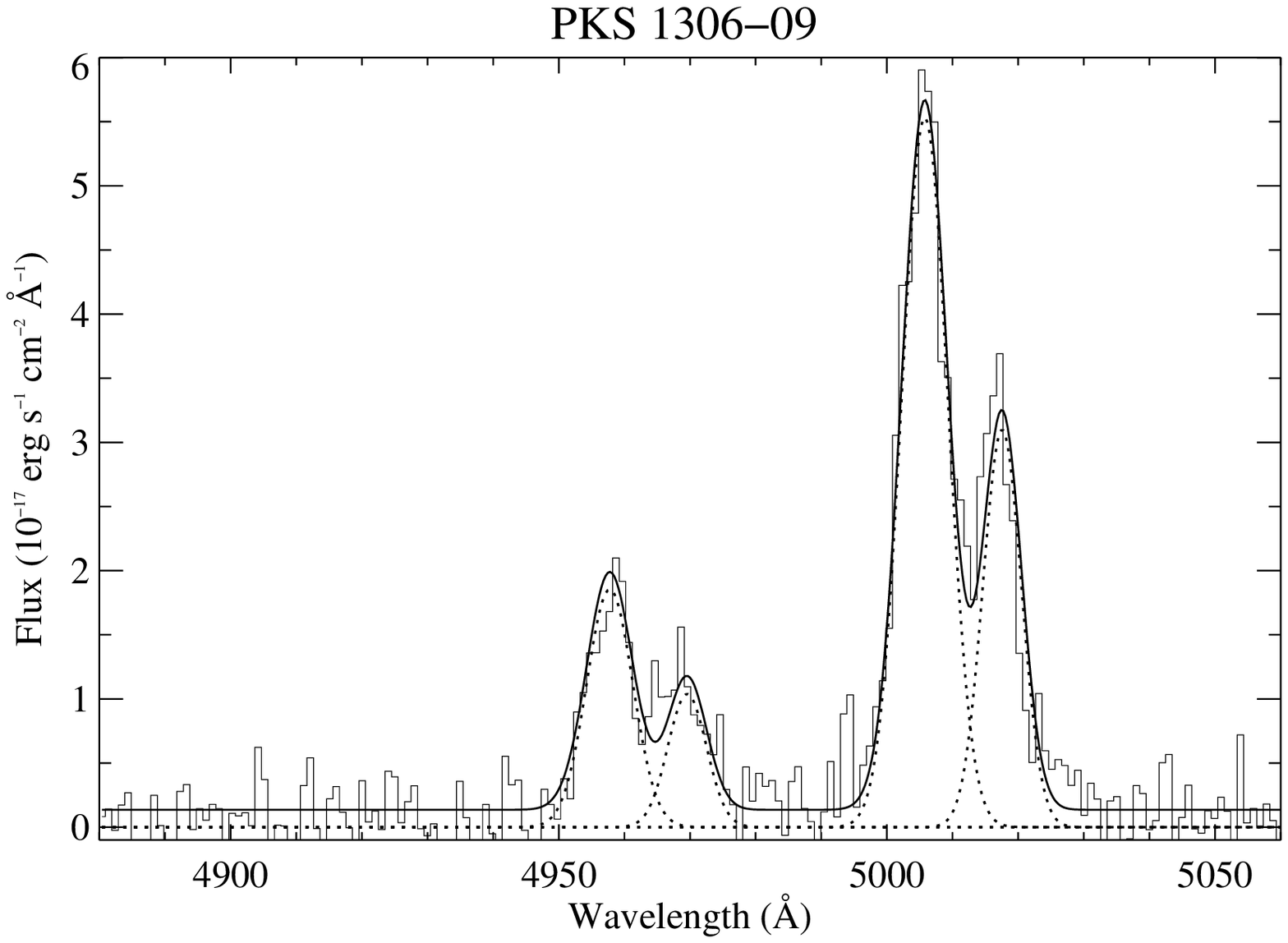,width=8cm,angle=0.}&
\psfig{file=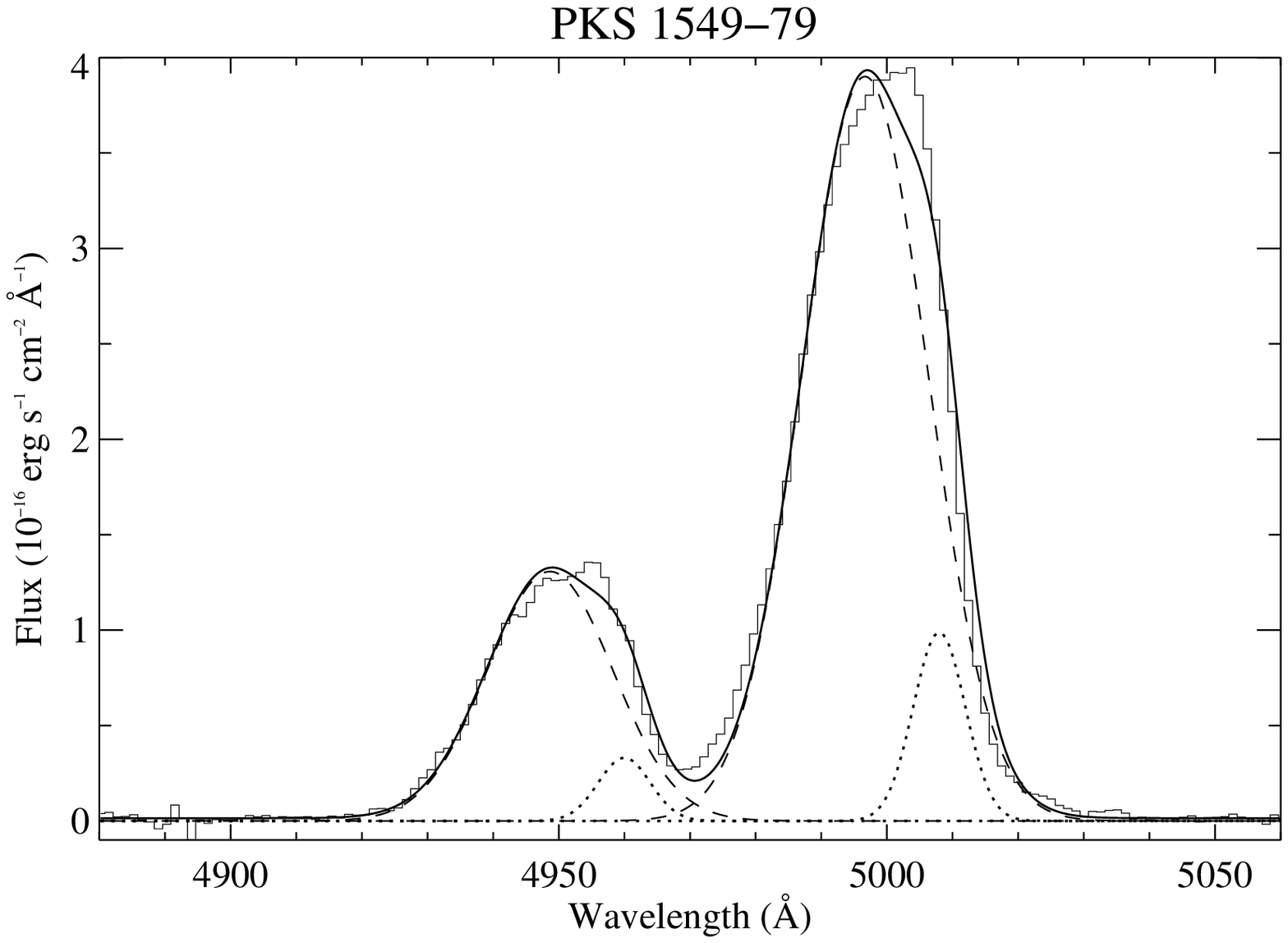,width=8cm,angle=0.}\\
\psfig{file=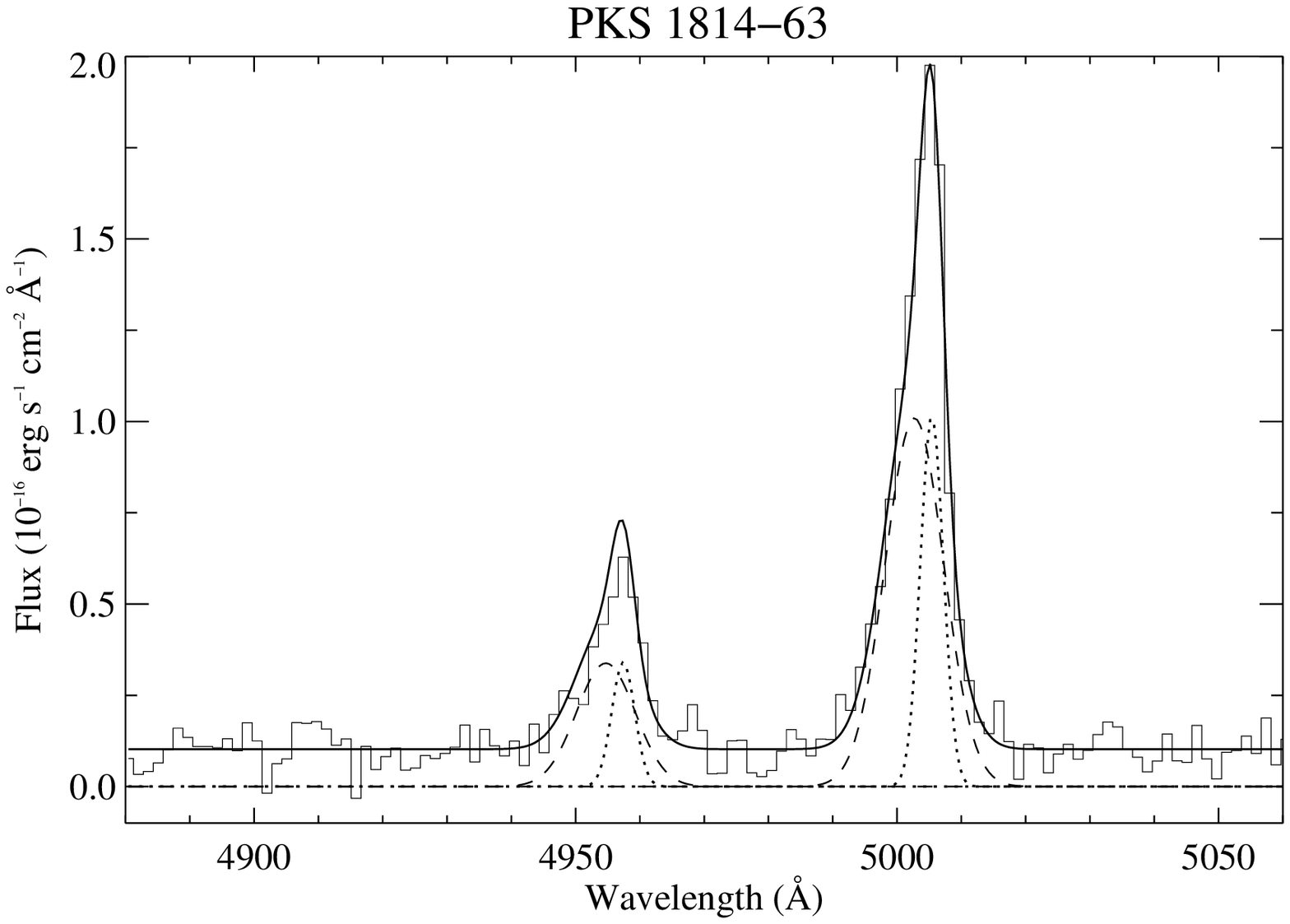,width=8cm,angle=0.}&
\psfig{file=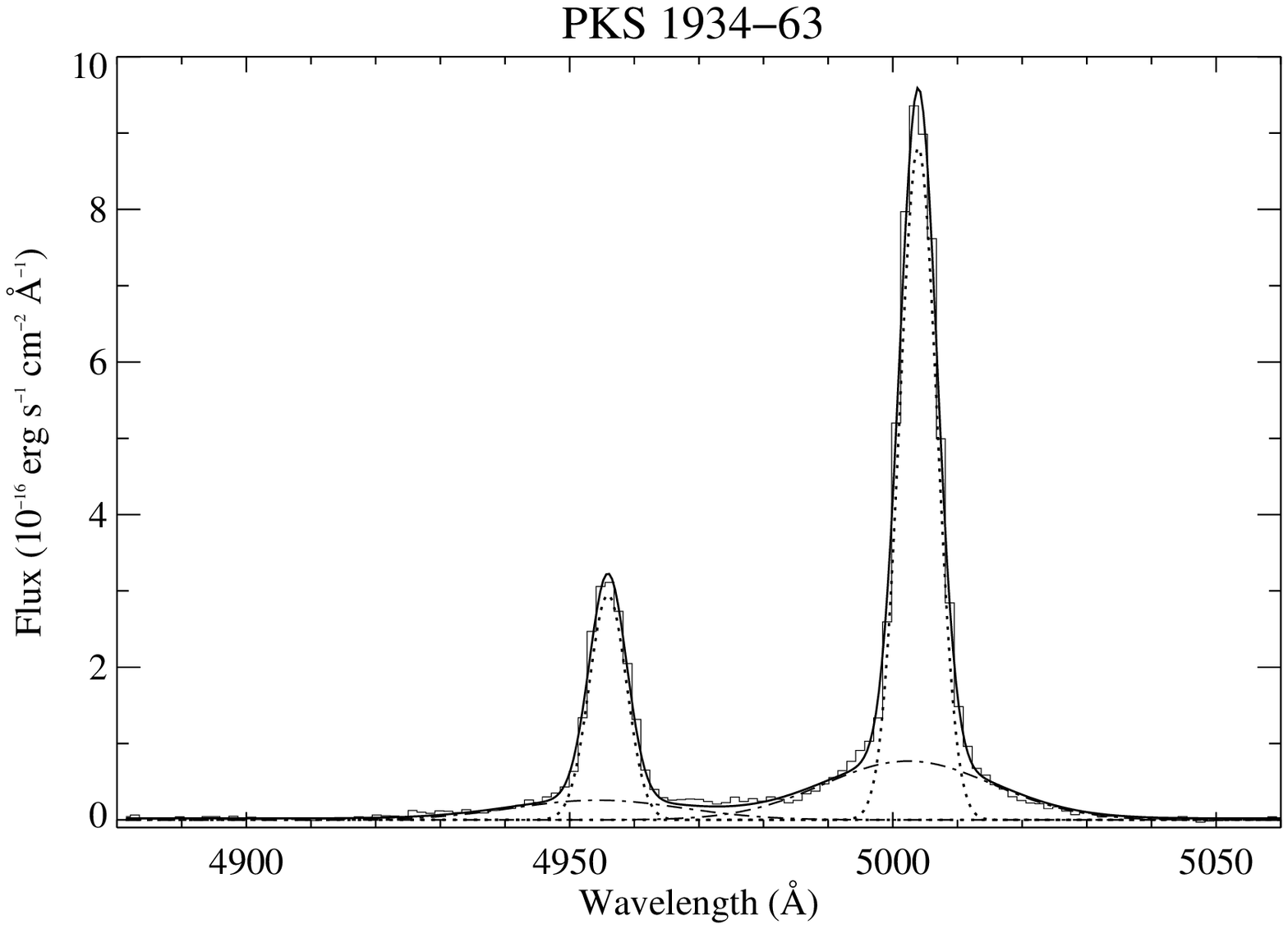,width=8cm,angle=0.}\\
\psfig{file=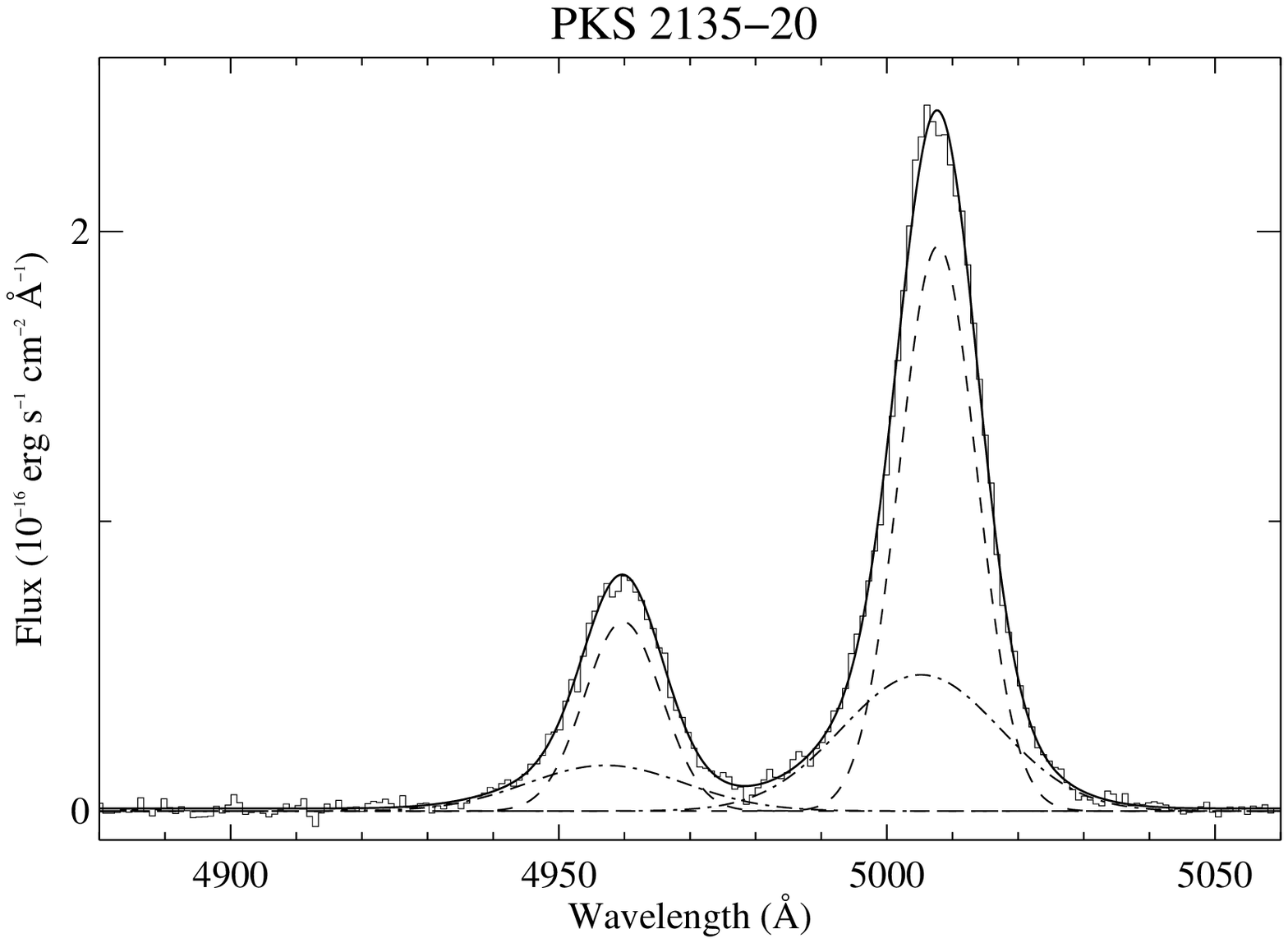,width=8cm,angle=0.}&
\psfig{file=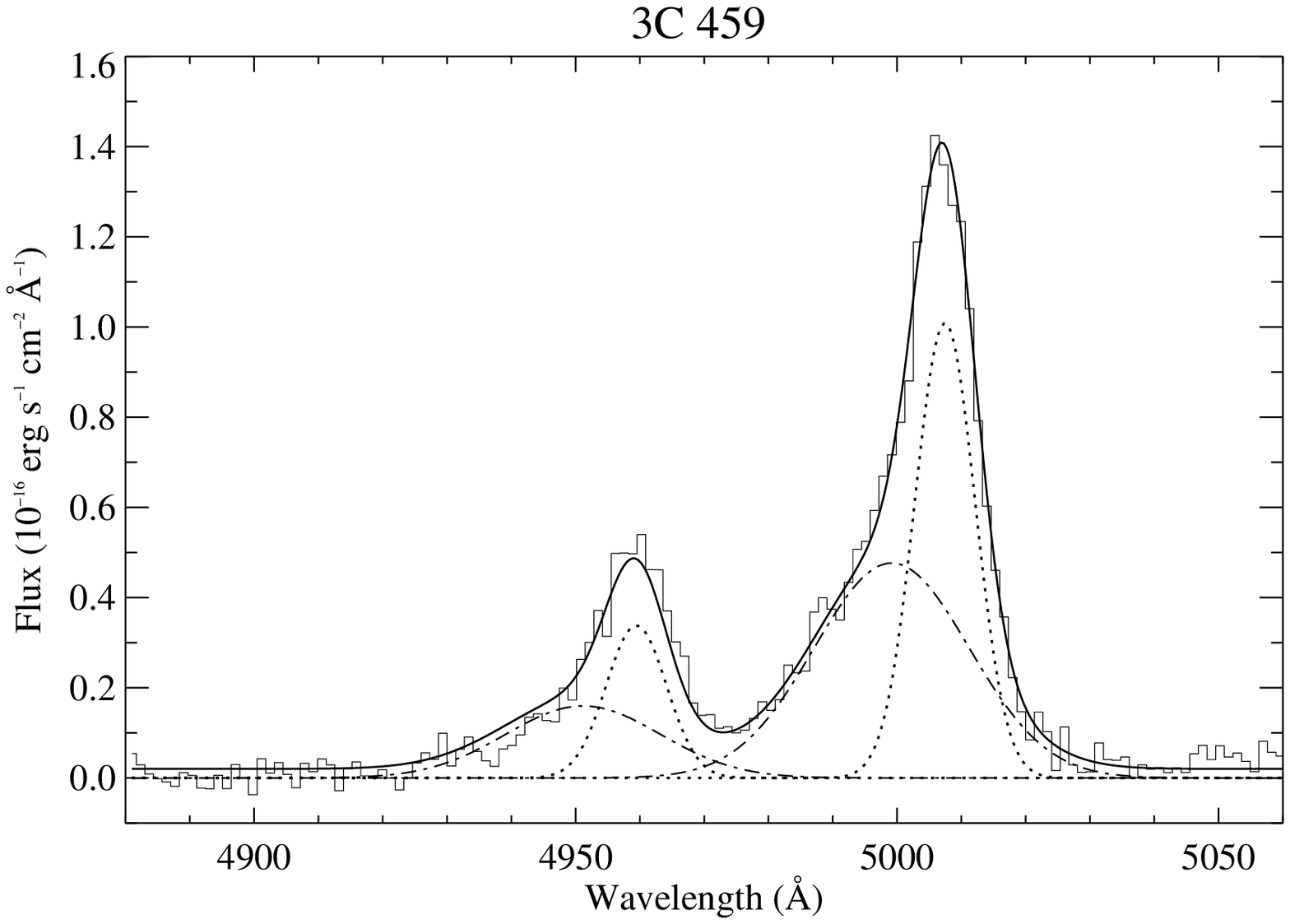,width=8cm,angle=0.}\\
\end{tabular}
\caption[]{{\it continued}. Models to the {[O III]}\lala4959,5007 emission line doublet. }
\label{fig:o3models}
\end{figure*}

\subsection{Single Gaussian Modelling}

Whilst multiple Gaussian modelling has worked well for the majority of
sources in the sample, two GPS sources (PKS 1345+12 and PKS 1934-46)
were notable exceptions. In PKS 1345+12, a common narrow component was
observed in all lines although different emission lines required
different velocity widths and shifts for the broader components to
provide good fits (H03). In PKS 1934-46, the situation is somewhat worse --
it is less clear that a common narrow component exists for all lines
when observed at this resolution/signal-to-noise.

For PKS 1345+12, H03 suggested a stratified ISM with three distinct
regions, each responsible for the emission of one of the kinematic
components. Hence, the different intermediate and broad components
could be explained by gradients in ionisation potential and/or
critical density across the regions emitting the intermediate and
broad components. Stratification of the NLR was suggested as the
reason for correlations between emission line FWHM (when fitting a
single Gaussian) and ionisation potential or critical density in
Seyfert galaxies \citep{derobertis84,derobertis86}. This is consistent
with models in which there is a continuous variation of density,
ionisation and velocity across a spatially unresolved inner narrow
line region (INLR). This technique was applied to a sample of extended
radio galaxies by \citet{taylor04}. In the latter work, no
significant correlations were found in Cygnus A 
or in the sample of NLRGs, but 3/4 BLRGs studied did show significant
correlations with ionisation potential and/or critical density, a
feature expected for an INLR on a scale less than that of the central
obscuring torus (r $<$ 100 pc). 

\begin{figure*}
\begin{tabular}{ccc}
PKS 1345+12 & PKS 1549-79 & PKS 1934-63 \\
\psfig{file=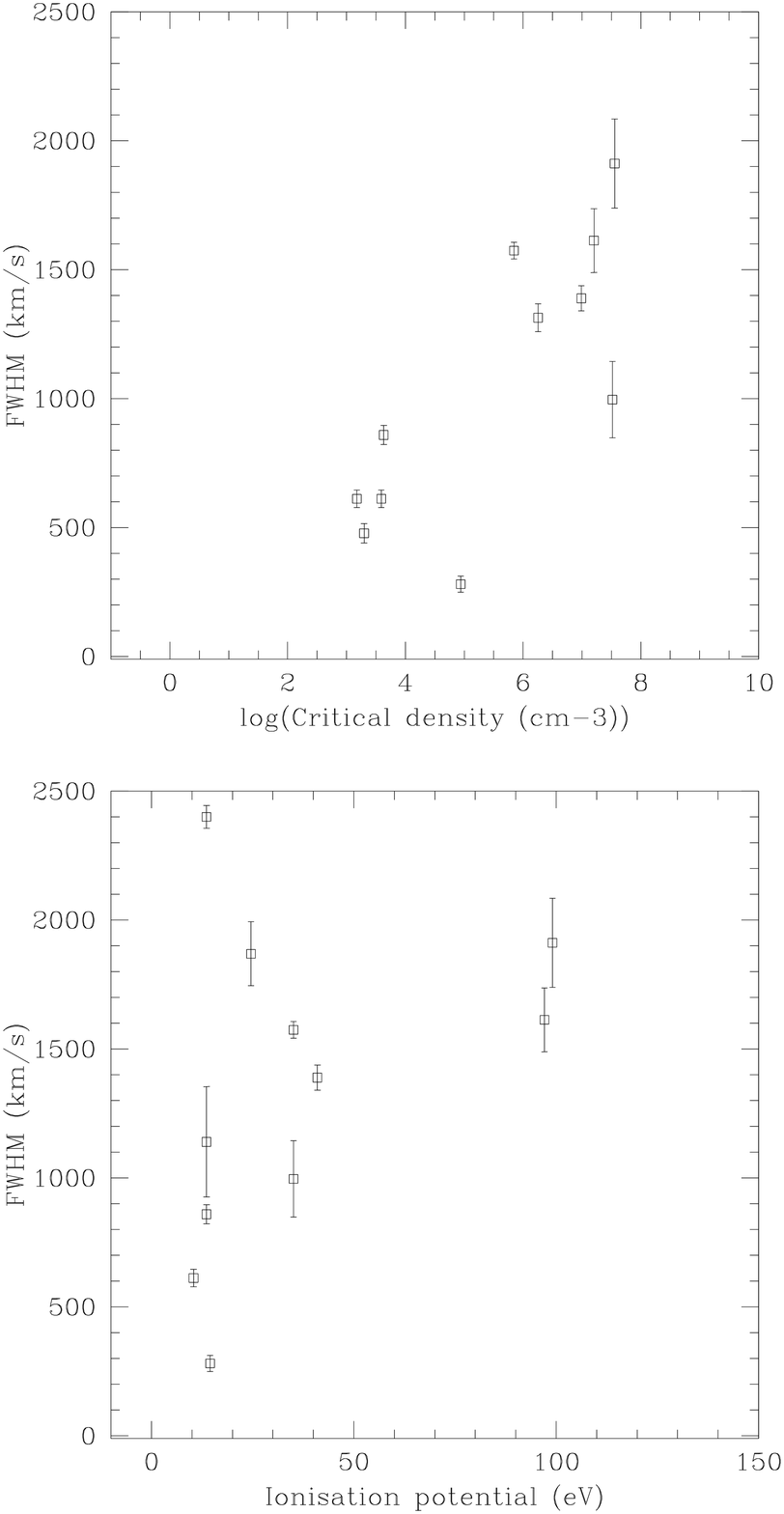,width=5.5cm,angle=0.}&
\psfig{file=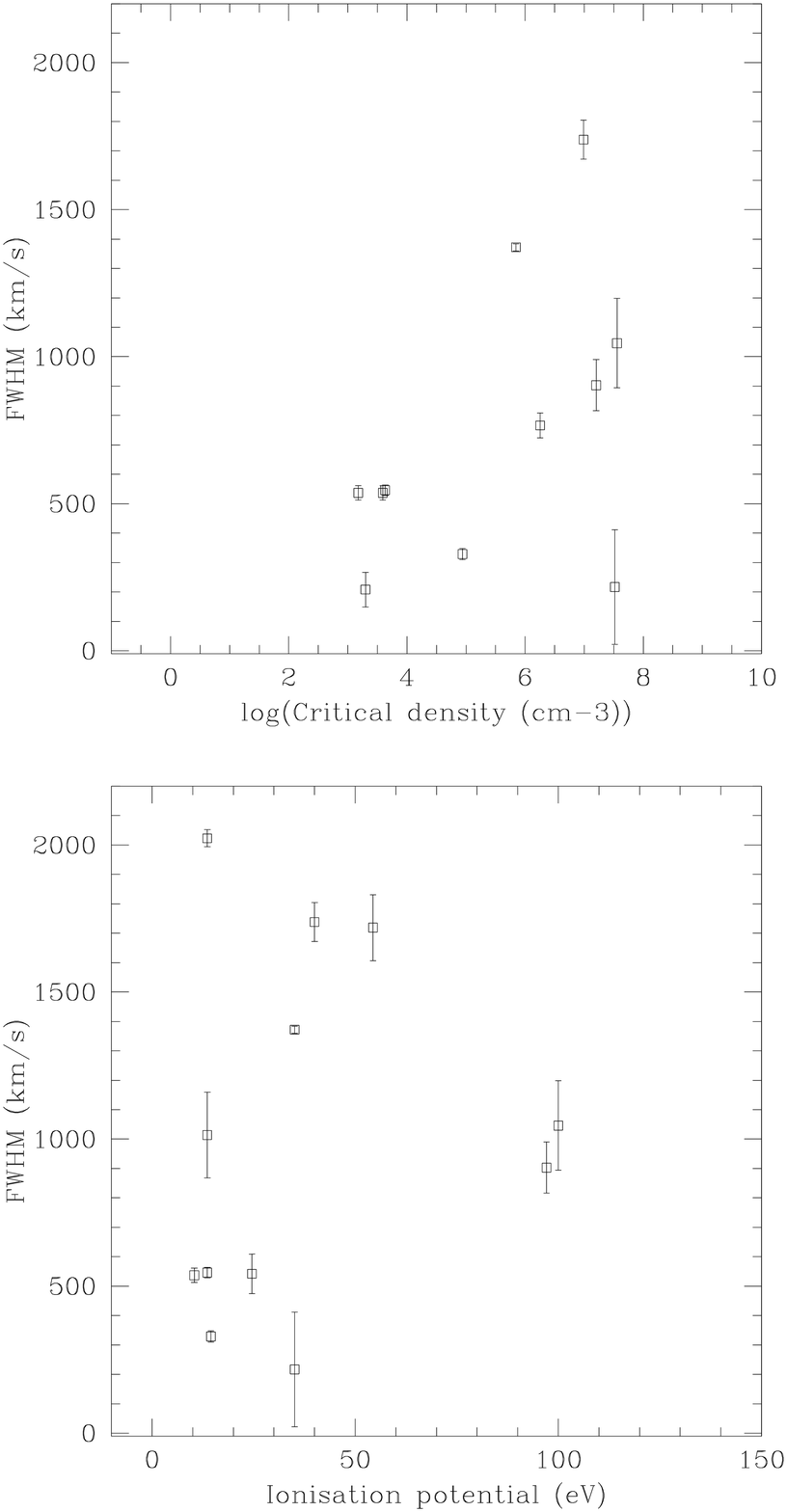,width=5.5cm,angle=0.}&
\psfig{file=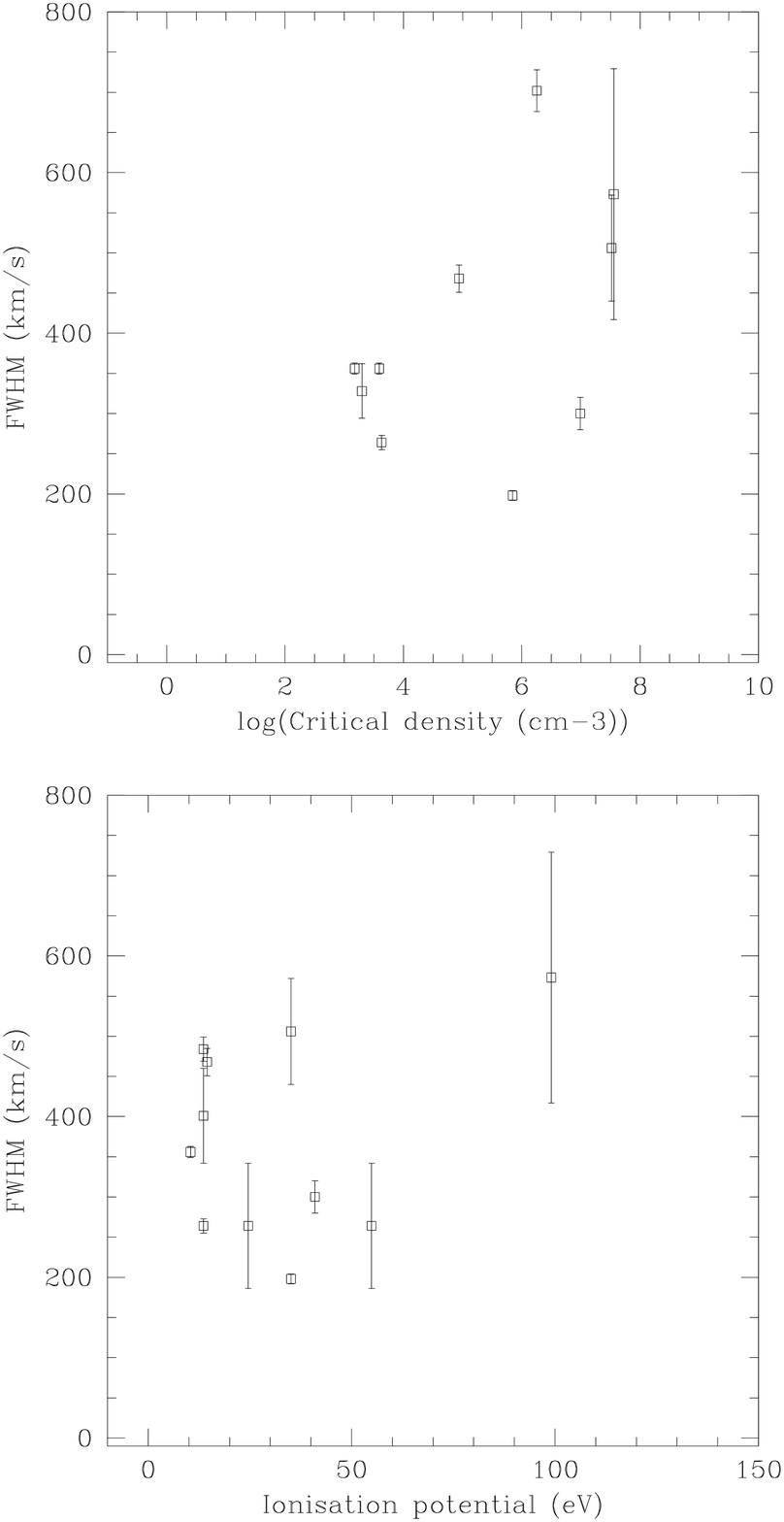,width=5.5cm,angle=0.}\\
\end{tabular}
\caption[Correlation plots]{Correlation plots for the nuclear
  apertures of PKS 1345+12, PKS 1549-79 and PKS 1934-63. For each source the
  plots are: {\it top }:   rest frame FWHM ({\kms}) versus log(critical
  density {\cmc}) and {\it bottom }: rest frame FWHM ({\kms}) versus ionisation
  potential (eV).}
\label{fig:correlations}
\end{figure*}

We have modelled all nuclear emission lines in all sources using
single Gaussian free-fits. 
No clear correlations were observed for the
majority of the sources, including the BLRGs,  in contrast to the
results of \citet{taylor04}. Figure \ref{fig:correlations} shows plots
of the single Gaussian FWHM versus both critical density and
ionisation potential for the smaller sources in the sample:
 PKS 1345+12, PKS 1549-79 and PKS
1934-63. Whilst visual inspection of the  plots suggest trends,
particularly in critical density, Spearman Rank analysis shows these
visual trends to be insignificant. 
 
\begin{table*}
\caption[]{Rest frame emission line modelling results. Columns are: 
$(a)$ object name; 
$(b)$ component following the definitions above: n (narrow; rn and bn
  denote the  `reddest' and `bluest' component where two narrow components are
  detected), i (intermediate), b (broad), vb (very broad) and BLR
  (broad line region component; i.e. a highly broadened component only
  observed in the permitted  lines); $(c)$ \& $(d)$ velocity width
  (FWHM) and error of 
  each component; $(e)$ \& $(f)$ velocity shift
  and error of the component from the systemic velocity. Note, the
  results for PKS 1345+12 and PKS 1934-63 are for {[O III]}. 
In sources for which free fitting
gave two narrow components with different widths but the lines can also be
modelled adequately by forcing the FWHM of the two narrow components
to be equal, the FWHM of this alternative model for the narrow
component is given and marked $\dagger$. }
\begin{center}
\begin{tabular}{llrrrr} \hline\hline
 & & &  \\
\multicolumn{1}{c}{Object} &   &
\multicolumn{1}{c}{Velocity} & \multicolumn{1}{c}{$\Delta$} &
\multicolumn{1}{c}{Velocity} & \multicolumn{1}{c}{$\Delta$} \\
  &   &\multicolumn{1}{c}{width} & & \multicolumn{1}{c}{shift} \\
 & & \multicolumn{1}{c}{\kms} & \multicolumn{1}{c}{\kms} & 
\multicolumn{1}{c}{\kms} & \multicolumn{1}{c}{\kms}\\
\multicolumn{1}{c}{$(a)$} & \multicolumn{1}{c}{$(b)$}  &
\multicolumn{1}{c}{$(c)$} & \multicolumn{1}{c}{$(d)$} &
\multicolumn{1}{c}{$(e)$} & \multicolumn{1}{c}{$(f)$}\\
 & & &  \\ \hline\hline
3C 213.1    & n & 523 & 44 & -- & -- \\
            & i & 1287 & 174 & -142 & 65 \\
%\\
3C 268.3    & rn& 309 & 12 & -- & -- \\
            & bn& unres & -- & -- & -- \\
            & i & 1152 & 88 & -121 & 22 \\
            &BLR& 5664 & 525 & -760 & 122 \\ 
%\\
3C 277.1    & n & 462 & 6 & +50 & -- \\
            & i & 1340 & 157 & -79 & 36 \\
            &BLR& 5177 & 185 & 335 & 31 \\
%\\
4C 32.44    & rn& 316 & 6 & 176 & 8 \\
            & bn& 242 & 6 & -264 & 6 \\
            & b & 1831 & 73 & 360 & 22 \\
            & vb& 3548 & 380 & -852 & 442 \\
            & n$\dagger$ & 281 & 5 \\
%\\
PKS 1345+12 & n & 340 & 23 & -- & --\\
            & i & 1255 & 12 & -402 & 9 \\
            & b & 1944 & 65 & -1980 & 36 \\
%\\
3C 303.1    & n & 51 & 27 \\
            & i & 747 & 17 & -438 & 20 \\
%\\
%\\
PKS 0023-26 & n & unres & -- \\
            & i & 1002 & 69 & 33 & 14 \\
%\\
PKS 0252-71 & rn$\dagger$& 335 & 30 & -- & --\\
            & bn$\dagger$& 335 & 30 & -- & --\\
            & i & 1236 & 68 & 65 & 24 \\
%\\
PKS 1306-09 & rn& 329 & 34 & 706 & 7 \\
            & bn& 425 & 20 & --  & --\\
            & n$\dagger$&365 & 13 \\%& 706 & 7 \\
%\\
PKS 1549-79
            & n & 383 & 15 &-- & --\\
            & i & 1282 & 25 & -679 & 20 \\
%\\
PKS 1814-63 & n & unres & \\          
            & i & 569 & 35 & -162 & 21 \\
%\\
PKS 1943-63 & n & unres & \\
            & b & 1785 & 103 & -93 & 43 \\
%\\
PKS 2135-20 & i & 762 & 15 \\
            & b & 1686 & 84 & -157 & 29 \\
%\\
3C 459      & n & 528 & 24 &-- & -- \\
            & b & 1647 & 48 & -497 & 49 \\
\hline\hline
\end{tabular}
\label{tab:shifts}
\end{center}
\end{table*}
\begin{table*}
\begin{center}
\caption[Emission line parameters.]{Various emission line parameters
  used in the statistical analysis. Columns are: $(a)$ object; $(b)$
  heliocentric redshift of the systemic velocity; $(c)$ maximum
  shift velocity taken as the shift between the broadest {\it NLR} component
  and the systemic velocity (\kms); ${d}$  FWHM of
  the {[O III]}$\lambda$5007 line when fitting a single Gaussian
  (\kms); $(e)$ the asymmetry index, AI$_{20}$ and $(f)$ the kurtosis
  parameter, R$_{20/50}$. 
$\star$ indicates total range of HI absorption and the PKS 1306-09 data is from 
Raffaella Morganti (priv. comm.). $\dagger$ {\it assumed} optical
  z. $^a$ The shift between the two detected narrow
  components. $\star$ Redshift derived from the stellar absorption lines. }
\begin{tabular}{llrrrr} \hline\hline
% &  & & &  \\
\multicolumn{1}{c}{Object} & \multicolumn{1}{c}{$z$} &
\multicolumn{1}{c}{Max shift} & 
\multicolumn{1}{c}{{[O III]} FWHM}& \multicolumn{1}{c}{AI$_{20}$} &
\multicolumn{1}{c}{R$_{20/50}$} \\
\multicolumn{1}{c}{ } & \multicolumn{1}{c}{ } &
\multicolumn{1}{c}{(\kms)} &  \multicolumn{1}{c}{(\kms) } &
\multicolumn{1}{c}{ } \\
\multicolumn{1}{c}{$(a)$ } & \multicolumn{1}{c}{$(b)$ } &
\multicolumn{1}{c}{$(c)$} & \multicolumn{1}{c}{$(d)$} &
\multicolumn{1}{c}{$(e)$}& \multicolumn{1}{c}{$(f)$}  \\
\hline
3C 213.1    & 0.19392 $\pm$ 0.00004          & -142   & 687 $\pm$ 22  & -0.015 $\pm$ 0.002 & 1.65 $\pm$ 0.17\\
3C 268.3    & 0.37171 $\pm$ 0.00006          & -760   & 668 $\pm$ 15  & -0.20  $\pm$ 0.02  & 1.53 $\pm$ 0.15\\
 %           &                                &       \\
3C 277.1    & 0.31978 $\pm$ 0.00008          & -79  & 483 $\pm$ 4   & -0.10  $\pm$ 0.01  & 1.79 $\pm$ 0.18\\
4C 32.44    & 0.36801 $\pm$ 0.00004          & -852 &  986 $\pm$ 28  & -0.054 $\pm$ 0.005 & 2.16 $\pm$ 0.22\\
PKS 1345+12 & 0.12351 $\pm$ 0.00008          & -1980 & 1547 $\pm$ 32 & 0.55   $\pm$ 0.06  & 2.22 $\pm$ 0.22\\
%            &                                &       &  \\
3C 303.1    & 0.27040 $\pm$ 0.00006          & -438  & 835 $\pm$ 8   & -0.067 $\pm$ 0.007 & 1.45 $\pm$ 0.15\\
%\\	    
PKS 0023-26 & 0.32162 $\pm$ 0.00003          & +33   &  262 $\pm$ 8   & -0.046 $\pm$ 0.005 & 3.79 $\pm$ 0.38\\
%            &                                &       & \\
PKS 0252-71 & 0.56288 $\pm$ 0.00009          & +65   &  714 $\pm$ 9   & -0.34  $\pm$ 0.03  & 1.68 $\pm$ 0.17\\
PKS 1306-09 & 0.46685 $\pm$ 0.00009          & 706$^a$  &1009 $\pm$ 43 & -0.46  $\pm$ 0.05  & 1.20 $\pm$ 0.12\\
PKS 1549-79 & 0.15220 $\pm$ 0.00003          & -679  &  1372 $\pm$ 12 & 0.24   $\pm$ 0.02  & 1.43 $\pm$ 0.14\\
PKS 1814-63 & 0.06466 $\pm$ 0.00007$\dagger$ & -162  & 411 $\pm$ 17  & 0.28   $\pm$ 0.03  & 2.55 $\pm$ 0.26\\
%            &                                &       & \\
PKS 1943-63 & 0.18129 $\pm$ 0.00008$\dagger$ & -93 & 198 $\pm$ 6   & -0.001 $\pm$ 0.0001& 1.69 $\pm$ 0.17\\
PKS 2135-20 & 0.63634 $\pm$ 0.00003$\dagger$ & -157 &  919 $\pm$ 7   & 0.033  $\pm$ 0.003 & 1.84 $\pm$ 0.18\\
 & 0.635 $\pm$ 0.004$\star$\\
3C 459      & 0.22012 $\pm$ 0.00003          & -497 &  995 $\pm$ 33  & 0.25   $\pm$ 0.03  & 2.40 $\pm$ 0.24
\\\hline\hline
\end{tabular}
\label{tab:lineparam}
\end{center}
\end{table*}

The line widths (FWHM) derived from the single Gaussian modelling can
also be used to compare this sample of compact radio sources to other
samples of radio sources presented in the literature.
 The measured single Gaussian FWHM of the nuclear {[O III]}
lines for this sample are 
presented in Table {\ref{tab:lineparam}} and are compared to a sample
of extended radio sources in Section 5.

\section{Discussion}

\subsection{Extreme emission line outflows}
The nuclear emission lines in all of the compact radio sources in this
sample are highly broadened, with complex emission line profiles, and
require multiple Gaussian components to model them. These components
are often shifted significantly with respect to one another and are 
therefore likely to trace flows in the nuclear emission line gas. However, in
order to distinguish between inflows and outflows, it is necessary to
accurately define the systemic velocity.

In Section 4.1, following a similar approach to that used for PKS 1345+12
(H03) and PKS 1549-79 (H06), we argued that the
extended narrow component(s) represents the ambient, quiescent ISM in
the galaxy halo. We are confident of this result in 12 of the 14
sources studied (see Section 4.1) -- kinematically the narrow 
component is the least disturbed component and is typically the only
extended  component.  In some of the sources in which HI
absorption is detected, the velocity of the narrow component is 
 also consistent with a deep, narrow component of
 HI which could be associated with a  circumnuclear
 disk or torus 
\citep{morganti01}. 

Previous studies have often struggled to
determine the relationship between the HI, the optical
kinematics and the systemic velocity (i.e. is the HI at rest, or
tracing a flow of gas?) due to the inaccuracies in
tying down the systemic velocities 
(see \citealt{morganti01} for a discussion).
 In this paper, we have
accurately established the systemic velocities  in 12 out of 14 sources
using detailed optical emission line modelling and have compared them
to the
HI data available in the literature in Figure {\ref{fig:velprofiles}} and
Table \ref{tab:hidata}. 

HI absorption is detected in 10/14 sources. Comparing the optical and
HI data, it is striking that the majority of absorption components are
significantly blueshifted with respect to the optical systemic
velocity, often by several hundreds of \kms, and represent outflows of
neutral gas. 
Only three sources have HI components consistent with the
optical systemic velocity (3C 213.1, 3C 268.3, and PKS 1345+12) which
may be consistent  with a circumnuclear disk or torus. However, in PKS
1345+12, the latest data show the component of HI at the systemic
velocity is  actually most likely associated with an off-nucleus cloud
\citep{morganti04}. 
Unlike the optical outflows, both the narrower and broader line
components have significant blueshifts. Multiple HI components are
observed in 5 sources. In PKS 1345+12, these are broadly consistent
(velocity width and shift) with the optical components. Finally, two
sources contain components consistent with infalling clouds (PKS
1934-63 and PKS 0023-26).

One of the key trends highlighted above (Section 4.2) is the detection of
significantly shifted 
emission line components observed in the majority of sources in the sample.
These are predominantly {\it blueshifted}, with the broader
components  blueshifted by the largest amounts. Figure
\ref{fig:shift-histograms}
shows the distribution of shifts\footnote{Note, for the sample
  described in this paper, the broadest (i.e. broadest NLR component)
  to systemic velocity shift is 
  used rather than the broadest to narrow velocity shift. This is
  because several sources have multiple narrow components which are
  believed to trace rotation curves in the galaxy rest frame and using
  the systemic velocity is the most consistent method for all sources
  in the sample.} for
the sample. All sources except PKS 1306-09
-- which has no `broad' component, only two narrow
components -- are included in this plot and 
the shifts used are for the NLR components only. 

For all but two sources (PKS 0023-26 and PKS 0252-71) the broad component is
blueshifted with respect to the systemic velocity with velocities up to
2000 \kms~although most sources occupy the range -900 $< v_{\rm shift}
<$ 100 \kms. 
Figure \ref{fig:shift-histograms} further sub-divides the sample into the larger
CSS sources and the smaller GPS sources. The effects of radio source size
and orientation are discussed below.

The  {[O III]} line emission is known to be resolved  in three of
the sources in  this sample: 3C 268.3, 3C 
277.1 and 3C 303.1. These sources have been imaged at high resolution with HST 
\citep{devries97,devries99,axon00} and \citet{odea02} have obtained
corresponding HST/STIS spectroscopy of 3C 277.1 and 3C 303.1. For these
sources, virtually all of the line emission detected by both
\citet{devries99} and \citet{axon00} lies  within our slit.  In our
2-D spectra, we observe clear extended line emission  in 3C 268.3 and
3C 277.1 although the line
emission imaged by HST lies within our defined nuclear aperture,
within which no obvious spatial structure is resolved.
The extended line emission  that we detect will be presented 
 in Holt et al. (2008, in prep.) along with a discussion of
extended apertures. It is difficult
for us to compare our kinematic results to those of \citet{odea02} as
whilst their spectra are of significantly higher spatial and spectral
resolution compared to ours, from the figures in their paper, the
lines are at much lower signal to noise than our spectra and they
have, for the main part, modelled 
the emission lines with single Gaussians. Depending on the dominance
of different components, this can lead to erroneous line shifts
(see e.g. H06). Comparing our single Gaussian modelling, we observe similar
line widths (FWHM$\sim$500\kms) in 3C 277.1 but significantly higher
widths in 3C 303.1 (FWHM$\sim$800\kms~compared to $\sim$500\kms~in
\citealt{odea02}). Further, the resolved {[O III]} emission may
account for the narrow line splitting in 3C 268.3.

\begin{table*}
\caption[HI data]{Summary of the HI data for the sample compiled from
  the literature. Columns are: 1) object, 2)
  \& 3) HI
  velocity shift and error with respect to the optical systemic velocity
  derived in this paper where negative velocities imply blueshifted
  HI; 4) \& 5) HI full width at half maximum (FWHM) and 
 estimated uncertainty  ($^{a}$ full width at zero intensity (FWZI) of
 the broad, shallow 
  absorption); 6) optical depth ($^{b}$ quoted fractional absorption); 7) HI 
  column density derived assuming $T_{spin}$ = 100 K and 8)
  references: 
 v0: \protect\citet{veroncetty00}; m1:
  \protect\citet{morganti01}; m3: \protect\citet{morganti03a}; v3:
  \protect\citet{vermeulen03}; g6: \protect\citet{gupta06}; H06:
\protect\citet{holt06}; l6: \protect\citet{labiano06};
m: Morganti, private communication. $\star$
  component of HI consistent with the optical systemic
  velocity. $\dagger$ \protect\citet{veroncetty00} give e.g.  FWZI  and we have
  estimated the FWHM from their figures but also give their data
  here. }
\begin{tabular}{lrrrrrrc}\hline\hline
Object & vel shift & $\Delta$ &FWHM &$\Delta$& $\tau$&N(HI) & ref\\
 & (\kms) & (\kms) & (\kms) & (\kms) & (10$^{-2}$)&(10$^{20}$ cm$^{-2}$) \\
(1)&(2)&(3)&(4)&(5)&(6)&(7)&(8) \\ \hline
3C 213.1 & -9$\star$ & 7 & 115 & & 0.05&0.11&v3\\
3C 268.3 & -63 & 4 & 19 & & 0.30&0.1 & v3\\
         & -152$\star$ & 3 & 101 & & 1.00&1.85 &v3 \\
 & -116 & 10 & 67.1 & 6 & 2.5 & 3.2 & l6\\
4C 32.44 & -128 & 20 & 229 & &0.17&0.71 &v3 \\
PKS 1345+12 & -15$\star$&&$\sim$180&&1&2&m3\\
 & $\sim$ -1600 -- $\sim$175&  & $\sim$1800$^{a}$ &  &0.2& 1 & m3\\
 & -360 & 1.7 & 23.5 & 3.9 & 0.0018$^{b}$& 0.08 & g6\\
 & -393 & 0.8 & 22.8 & 1.8 & 0.0023$^{b}$& 0.10 & g6\\
 & -450 & 0.4 & 129.9 & 0.7 & 0.0107$^{b}$& 2.68 & g6\\
%\\
%\\
PKS 0023-26 & -64 & 40 & 39 & & 0.20 & 0.14 & v3\\
& 45 & 6 & 126 & & 0.93 &2.14 &  v3\\
PKS 1306-09 &-515 -- -276 &&239$^{b}$&&0.3&&m\\
PKS 1549-79 & -74 & 10 &80&&2&400&m1,H06 \\

PKS 1814-63 & -186 &&$\sim$50&&21.3&1700& m1\\
& 0 -- -281 &&$\sim$280$^{a}$&&0.8&2000&m1 \\
& -192$\dagger$ &2&62$\dagger$&&21.7&9&v0$\dagger$\\
& $\dagger$$\sim$-370 -- $\sim$0 &&$\dagger$372$^{a}$&&&& \\
PKS 1934-63 & 389 & 2 & 20 &&0.22&0.06&v0 \\
3C 459 & -398 & 9 & 400& &0.8&270& m1 \\
 & -241 & 10 & 130 & & 0.31&0.72 & v3\\
 & -431 & 27 & 71 & 34 & 0.0012$^{b}$&0.17 &g6 \\
 & -344 & 6 & 121 & 11 &0.0034$^{b}$ &0.8 & g6\\
 & -186 & 31 & 164 & 63 &0.0012$^{b}$ &0.38 & g6\\\hline\hline
\end{tabular}
\label{tab:hidata}
\end{table*}

From the results presented here, it is clear that, with the exception
of PKS 0023-26, PKS 0252-71 and PKS 1306-09, all sources in the sample
have evidence for fast, blueshifted flows in the circumnuclear
ISM. However, from the kinematical evidence alone it is impossible to
distinguish between material in outflow on the side of the nucleus
closest to the observer and material infalling on the far size of the
nucleus. \citet{holt03} argued that, for PKS 1345+12, as the reddening
increased significantly with FWHM, the  blueshifted components were
likely to be consistent with  material being observed on the observer's
 side of the nucleus and therefore tracing an outflow. We will discuss
 the issue of reddening in a future paper (Holt et al. 2008, in
 prep.).

\subsection{The role of orientation}
The original work on the compact flat-spectrum radio source PKS
1549-79 by \citet{tadhunter01} suggested that the direction of jet
propagation was oriented close to the observer's line of sight (LOS) as:
\begin{enumerate}
\item PKS 1549-79 has a flat radio spectrum, often associated
  with radio loud quasars whose axes are close to the observer's LOS;
\item PKS 1549-79 has  core-jet radio morphology. As radio galaxies
  have two-sided radio jets, observing one-sided radio structure is
  usually interpreted as an orientation effect with the observer's LOS
  aligned close to the direction of jet propagation (see H06, Figure
  10)\footnote{It should be noted that highly asymmetric radio
  structures are also consistent with jet-cloud interactions in a
  non-homogeneous ISM.}.
\end{enumerate}
Similarly,  radio maps of PKS 1345+12 (e.g. \citealt{lister03})
show highly asymmetric radio
jets, again suggesting the radio source is pointing close to the
observer's LOS.  This argument is supported 
by the detection of a broad (FWHM $\sim$2600
\kms)  component Pa$\alpha$  \citep{veilleux97} and the detecton of a
point source component in high-resolution near-IR images
\citep{evans99}.  However, this orientation is disputed by 
\citet{lister02,lister03} -- see H03 for a discussion. 

Interestingly, PKS 1345+12 and PKS 1549-79 contain two of the most
extreme outflows in the  sample discussed in this paper ($\sim$2000
\kms~and $\sim$680 \kms~respectively). Perhaps, then, the particular
orientation of these objects with respect to our LOS is the
reason why we observe such extreme kinematics in these sources. In
contrast, for example, PKS 1934-63 has a weak core and  is a symmetric
double, as 
well as being one of the most compact sources in the sample, and has
one of the smallest observed outflow velocities.

Due to the angular scales of the compact radio sources, 
compact flat-spectrum radio cores for the sources in this sample are 
not generally detected, even 
with VLBI (e.g. \citealt{tzioumis02}) -- the core is only detected in
5/14 sources. 
It is therefore not possible to
determine the radio source orientation accurately for most of the
sample using the standard 
methods, for example, the $R$ (core-dominance) parameter
(e.g. \citealt{orr82,giovannini01}), radio flux density 
comparison of the jet and counter-jet (e.g. \citealt{giovannini01}),
jet proper motion 
measurements (e.g. \citealt{giovannini01})  or
jet-motion modelling (e.g. \citealt{lister03}). Three sources have
quoted orientation information in the literature: PKS 1549-79
($R_{\rmn 2.3 GHz}$ = 1.310;
\citealt{morganti01}), 3C 459 ($R_{\rmn 4.8 GHz}$ = 5.65;
\citealt{morganti93}; R$_{\rmn 2.3 GHz}$ = 0.394; \citealt{morganti01})
 and
PKS 1345+12 (jet motion modelling by \citealt{lister03}: 82$^{\circ}$
to LOS with opening angle of 46$^{\circ}$, although this result is
inconsistent with the arguments presented in  H03). 

Despite the lack of radio information to accurately determine the
radio source orientations, it is still interesting to make a rough
estimate of the importance of orientation with respect to the observed
outflow velocities. For the remainder of the sample, we have therefore
adopted the rather crude approach of comparing the relative extents of
the radio jets on either side of the nuclei in the radio
maps. Assuming both jets are generated by 
the same central 
source, if they expand through empty space they should have similar
intrinsic extents on either side of the core and any observed differences will
be due to orientation effects. It should be noted that,
should the jets expand through a dense ISM, interactions between the
radio source and this ISM could significantly alter the path of at
least one of the jets. Hence, a highly asymmetric ISM can alter the
relative extents of the jets and mimic the effects of orientation with
respect to the observer and these results should therefore be used
with caution. However, \citet{best95} studied the angular
asymmetries in a sample of extended FR II 3CR sources and found the
distribution of asymmetry angles to be consistent with the predictions
of unified schemes. 

\begin{figure}
\centerline{\psfig{file=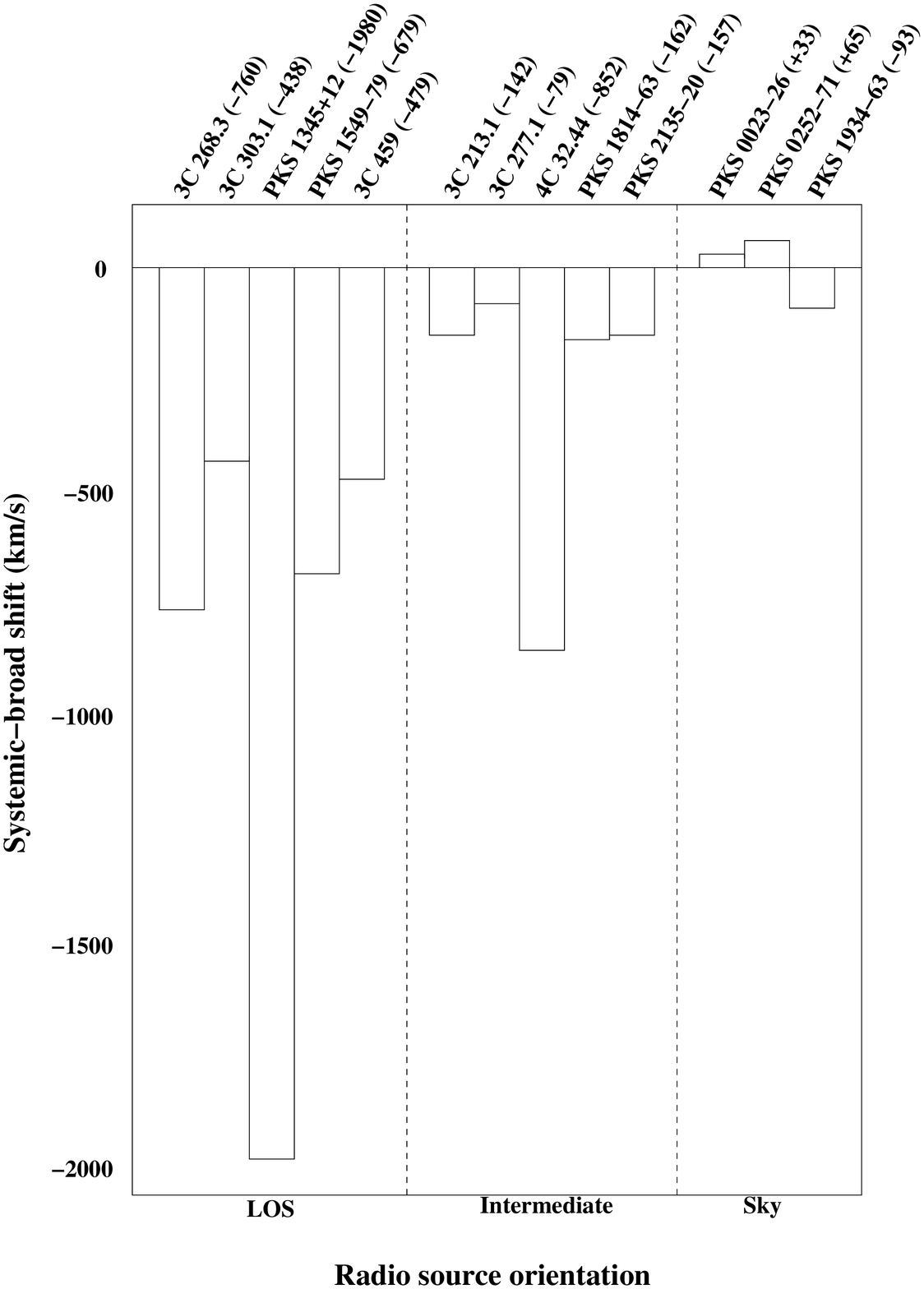,width=7cm,angle=0}}
\caption[Orientation of radio sources in the sample
  histogram]{Histogram  showing the putative 
orientation of the sources with respect to their outflowing velocities.}
\label{fig:orientation}
\end{figure}

We have therefore classified the sources into three broad categories: 
\begin{enumerate}
\item {\it Close to the LOS}. All sources with an
  obvious core-jet and a large core/extended radio flux ratio e.g. PKS
  1549-79, PKS 1345+12 and 3C 303.1
\item {\it Close to the plane of the sky}. Symmetric radio morphology
  (e.g. similar jet extents and/or fluxes): e.g. PKS 1934-62, PKS 0252-71 and
  PKS 0023-26. 
\item {\it Intermediate}. Sources not consistent with the other two categories.
\end{enumerate} 
 Figure {\ref{fig:orientation}} shows
 the grouping with respect to the largest outflow velocity.

\begin{table}
\begin{center}
\caption[Largest projected linear sizes.]{Comparison of the maximum
  outflow velocity with the largest projected linear size of the   
  sources in this sample. $^a$ compact core.}
\begin{tabular}{lr|lr} \hline\hline
 & & &  \\
\multicolumn{2}{c|}{D $<$ 1 kpc} &
\multicolumn{2}{c}{D $>$ 1 kpc} \\
 & & &  \\ \hline\hline
4C 32.44    & (-852)  & 3C 213.1    & (-142)\\
PKS 1345+12 & (-1980) & 3C 268.3    & (-760)\\
PKS 1549-79 & (-679)  & 3C 277.1    & (-79)\\
PKS 1814-63 & (-162)  & 3C 303.1    & (-438)\\
3C 459$^a$      & (-497)  & PKS 0023-26 & (+33)\\
            &         & PKS 0252-71 & (+65)\\
            &         & PKS 1306-09 & (0)\\
            &         & PKS 1934-63 & (-93)\\
            &         & PKS 2135-20 & (-157)\\
 & & &\\ \hline\hline
\end{tabular}
\label{tab:linsize}
\end{center}
\end{table}

Despite the crudeness of the method, the data suggest
that the orientation of the radio source to the observer's line of sight/the
asymmetry of the radio morphology may be
  an important factor in determining the maximum outflow
velocity  observed. Sources with jets most likely pointing towards us
contain the most extreme outflow velocities, and 
sources likely to be close the plane of the sky exhibit the least evidence for
fast outflows in the NLR.  
Only two sources in the sample appear to be
`mis-fits' in this classification scheme -- 3C 277.1 and
4C 32.44. This is most likely due to both the over-simplification of
the method and inherent scatter in the data. Note also, whilst the
flat-spectrum core appears strong in the radio map of 3C 277.1 and we observe strong
optical signatures of the quasar in our optical spectra (broad
permitted lines and a strong blue power-law continuum),   the radio emission
is resolved into two distinct lobes which suggest the radio axis is
not closely aligned with our line of sight. We have therefore placed
3C 277.1 in the  `intermediate' category.

Such a strong trend in orientation--outflow velocity has been
suggested for other classes of AGN. The most extreme
outflows  observed in any type of AGN are those in  BAL quasars. Whilst the
BAL outflows are on a different scale to those observed in compact
radio sources (on the scale of the BLR rather than the NLR), it has
been suggested that the extreme velocities observed may be partly due
to an orientation effect  (e.g. \citealt{weymann91,elvis00}).

\subsection{Outflow driving mechanism}
It is clear that extreme emission line kinematics exist in all of the
compact radio sources in this sample. The emission line flows are
likely to trace outflows in the emission line gas, similar to the
results presented for PKS 1345+12 (reddening; H03) and PKS 1549-79 
(source orientation; H06). 

If compact radio sources are truly young AGN in which the outflows
have not yet swept aside the natal cocoon, then because of the large
gas and dust concentrations in the central regions, the effects of {\it all}
types of outflow (quasar winds, jets, starburst superwinds) are likely
to be more visible as there is more warm/cool gas around for the winds
and/or jets to interact with.  Whilst jet-cloud interactions are
expected to produce extreme kinematics, all of these scenarios can
potentially explain broad line widths and large velocity shifts and
so, from the kinematical evidence alone, it is not possible to
distinguish between the different outflow driving mechanisms.

Some of the best evidence for jet-driven outflows is provided by the
co-alignment and similar scales of the radio source and the optical
emission line structures. More than 30 CSS sources have been observed
with HST in both broad and narrow bands (to isolate {[O II]} and/or
{[O III]} emission) by \citet{devries97,devries99}
and \citet{axon00}. The de Vries et al. samples include 3C 213.1, 3C 268.3,
3C 277.1 and 3C 303.1 and the Axon et al. sample includes 3C 268.3, 3C
277.1, 3C 303.1 and PKS 1345+12. These imaging studies reveal that the
optical and radio emission is both:
\begin{enumerate}
\item  on similar scales (30-90\% of the optical line
emission is concentrated within a few kpc and the radio sources are
either completely embedded within this optical line emitting gas or
extend just beyond it; \citealt{axon00} sample) and;
\item  strongly aligned (typically $\lesssim$
10$^{\circ}$ in all sources, including the sources also found in our
sample,  in which
the data could be accurately registered in the
de Vries et al. sample, and in 6/11 sources in the
Axon et al. sample, often elongated into jet-like structures). 
\end{enumerate}
The
radio-optical alignment is also observed across the entire redshift
range probed by these samples (0.1 $\lesssim$ z $\lesssim$ 1.5), rather
than confined to higher redshifts   as in more
extended radio sources (z 
$\gtrsim$ 0.6; e.g. \citealt{mccarthy87,dekoff96}). Such close
association between the optical emission line gas and the radio source
suggest that the radio source is strongly interacting with the ambient
medium as it expands through it \citep{odea02}. Follow-up
HST/STIS spectroscopy of three CSS sources with resolved {[O III]}
line emission (including 3C 277.1 and 3C 303.1) provides further evidence
that the outflows are likely to be driven by the expanding radio jets;
the emission lines have complex, broad profiles (FWHM $\sim$ 500\kms) 
which are  offset with respect to the systemic velocity by
300-500\kms~in the region(s) of the radio hotspots \citep{odea02} and
the line ratios are consistent with a mixture of fast shocks (500-1000\kms)
and photoionisation/precursor \citep{labiano05}.  

Finally, PKS 1345+12 and PKS 1549-79 were recently observed at
higher resolution with HST/ACS \citep{batcheldor07}. In these sources, the
optical line emission is clearly concentrated in the central regions,
ruling out galaxy wide starbursts as the dominant  outflow driving
mechanism. Further, the radio emission in both sources is on similar
scales to the optical line emission. 
However, the region of line emitting gas is only marginally resolved
and, whilst there is a suggestion that the isophotes
may be elongated in the direction of the radio axis, this evidence is
far from conclusive. 

Despite the strong evidence for  radio/optical alignments and the
co-spatial scales of the optical and radio emission, we find no
evidence for anti-correlations between the emission line widths and
the ionisation potential (Figure \ref{fig:correlations}) that are a
clear signature of shocks in some extended radio sources.
Hence, to confidently distinguish between the different ionisation
mechanisms, it is necessary to combine the kinematical and  imaging
data with a detailed study of the line ratios (Holt et al., 2008, in prep.).

\subsection{Are the kinematics in compact radio sources more extreme
  than in extended radio sources?}
\label{sect:extendedvscompact}
\begin{table*}
{\small
\begin{center}
\caption[Comparison sample parameters.]{Parameters for the comparison
samples used in both this paper and in Paper II. Columns are: $(a)$
sample name; $(b)$ redshift range; 
$(c)$ radio power range at 5 GHz except $^a$ at 178 MHz, $^b$ at 1.4 GHz and
 $^c$ no radio data was given for this sample; $(d)$ number of sources
used; $(e)$ completeness of the sample. $\dagger$ see Section 2 for details. Note, for sources
which are included in more than one sample, data are included only
once and are taken from the sample with the highest quality
data. $\star$ This comparison sample is only used in Holt et al.,
2008, in prep.}
\begin{tabular}{l|r|r|r|r|r} \hline\hline
 & & & & & \\
\multicolumn{1}{c|}{Sample} &  \multicolumn{1}{c|}{z}& 
\multicolumn{1}{c|}{Radio power}&  \multicolumn{1}{c|}{N}& 
\multicolumn{1}{c|}{completeness}& \\
\multicolumn{1}{c|}{ } &  \multicolumn{1}{c|}{}& 
\multicolumn{1}{c|}{log P$_{\rm 5 GHz}$}&  
\multicolumn{1}{c|}{}& & \\
\multicolumn{1}{c|}{ } &  \multicolumn{1}{c|}{}& 
\multicolumn{1}{c|}{(W Hz$^{-1}$) } &  \multicolumn{1}{c|}{}& 
 & \\
\multicolumn{1}{c|}{$(a)$}& \multicolumn{1}{c|}{$(b)$}&
\multicolumn{1}{c|}{$(c)$}& \multicolumn{1}{c|}{$(d)$}& 
\multicolumn{1}{c|}{$(e)$} \\\hline 
\multicolumn{1}{l|}{\bf Compact radio sources} & & & &\\
This paper                & $<$ 0.7 & 26-28 & 14 & complete$\dagger$ & \\
\protect\citet{gelderman94}& $<$ 0.9 & 22-29 & 16 & not complete \\
\protect\citet{hirst03}    & 0.5-3.6 & $>$ 26.5$^a$ & 9 & complete \\
PKS 1151-34                & 0.258   & 26.8  &  &\\
PKS 1718-643               & 0.014   & 24.3  &  &\\
 & & & & &\\
\multicolumn{1}{l|}{\bf Extended radio sources} & & & &\\
2 Jy                        & $<$ 0.7  & 25-29 & 36 & complete $\star$\\
radio loud objects from \protect\citet{heckman84}    & $<$ 0.7  & $>$ 24.5$^b$ & 45 & not complete  \\
radio loud objects from \protect\citet{brotherton96}& $<$ 0.95 & $^c$      & 60 & not complete\\
\protect\citet{taylor04}    & $<$ 0.2  & 26-29 & 12 & complete\\
\hline\hline
\end{tabular}
\label{tab:sample-param}
\end{center}
}
\end{table*}
As discussed in Section 1, previous data have suggested that the 
kinematical properties of the {\it nuclear} emission line gas are
 different in compact radio sources compared to more extended
radio sources. For example, comparison of the emission line profiles
in samples of compact (e.g. \citealt{gelderman94}) and extended
(e.g. \citealt{brotherton96})  
radio sources in the literature suggest that the line profiles of the
compact radio sources  
are often broader and more asymmetric. However, to date, no attempt
has been made 
to properly quantify the differences between the emission line profiles of 
compact and extended radio sources.

Many of the spectra in the literature are not of sufficient
signal-to-noise and/or resolution to model the emission line profiles
in detail and so we have used three line profile parameters on the
strong {[O III]}$\lambda$5007 line to compare our sample to the data
in the literature, for example, i) the line width (FWHM) derived  
from single Gaussian modelling ;
ii) the asymmetry index, AI$_{20}$, and iii) the kurtosis parameter,
R$_{20/50}$. 
The latter two are defined in \citet{heckman81}. All the results for
our sample are 
presented in Table \ref{tab:lineparam}. 
As our sample of compact radio sources is small,
we have formed a larger sample including data from the literature (see
Table \ref{tab:sample-param}).  
Our comparison
sample of extended radio sources for these three parameters comprises
all radio loud objects in the  
samples of \citet{heckman84} and \citet{brotherton96} (see Table
\ref{tab:sample-param}). The latter 
sample comprises solely quasars whereas the sample of
\citet{heckman84} includes both radio galaxies and quasars.
\begin{figure}
\centerline{\psfig{file=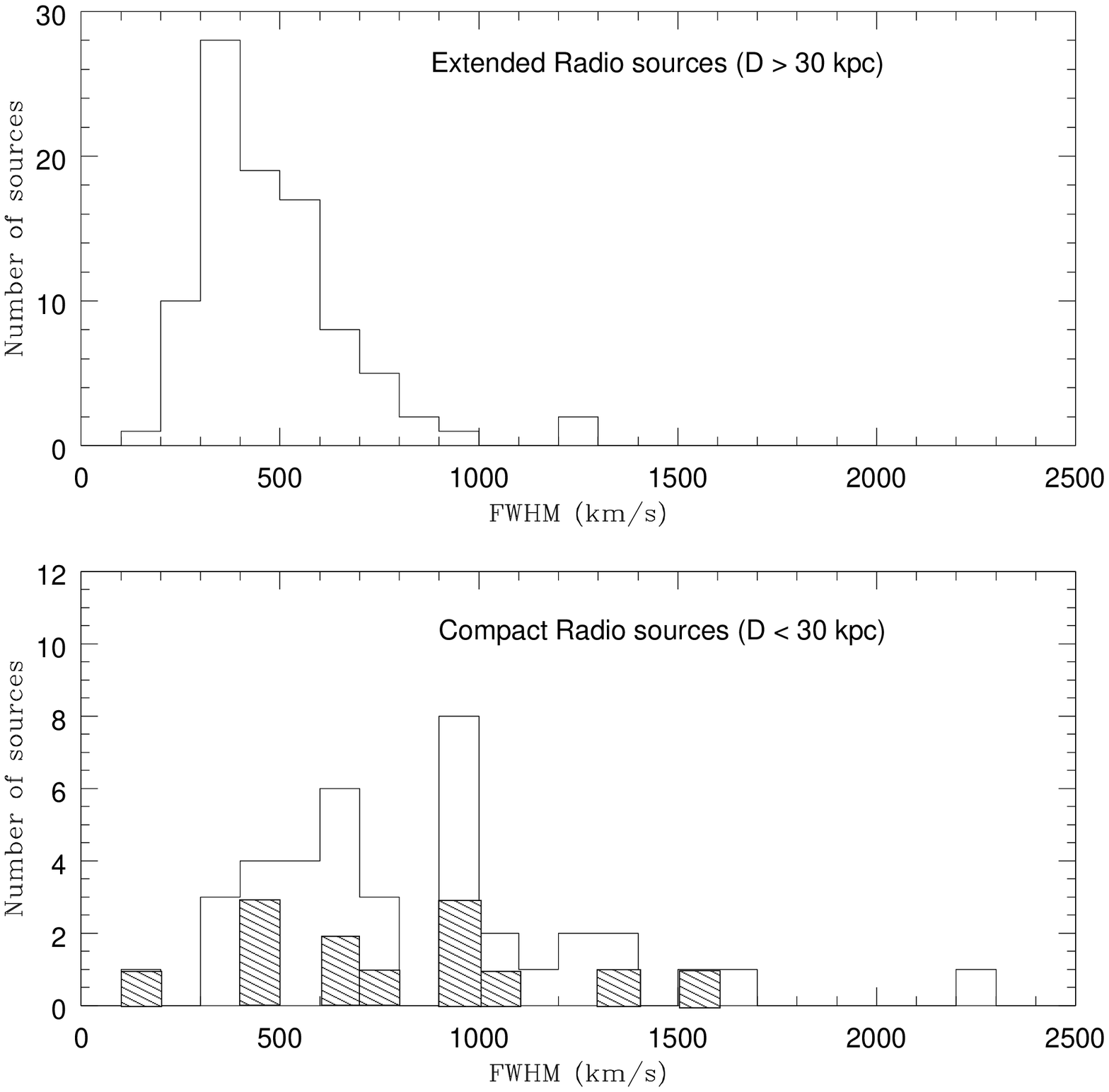,width=9cm,angle=0.}}
\caption[Histograms for emission line FWHM.]{Histograms showing the
  distribution in observed optical emission line  FWHM of {[O III]}
  in samples of extended (D $>$ 30 kpc;
  from \protect\citealt{heckman84} and \protect\citealt{brotherton96})
  and compact (D $<$ 30 kpc; this sample plus sources from 
\protect\citet{gelderman94}, four sources from the sample of
  \protect\citet{hirst03} and PKS 1718-49 from
  \protect\citet{fosbury77}). The shaded part of the bottom plot
  highlights the sources from this sample.
}
\label{fig:histfwhmnew}
\psfig{file=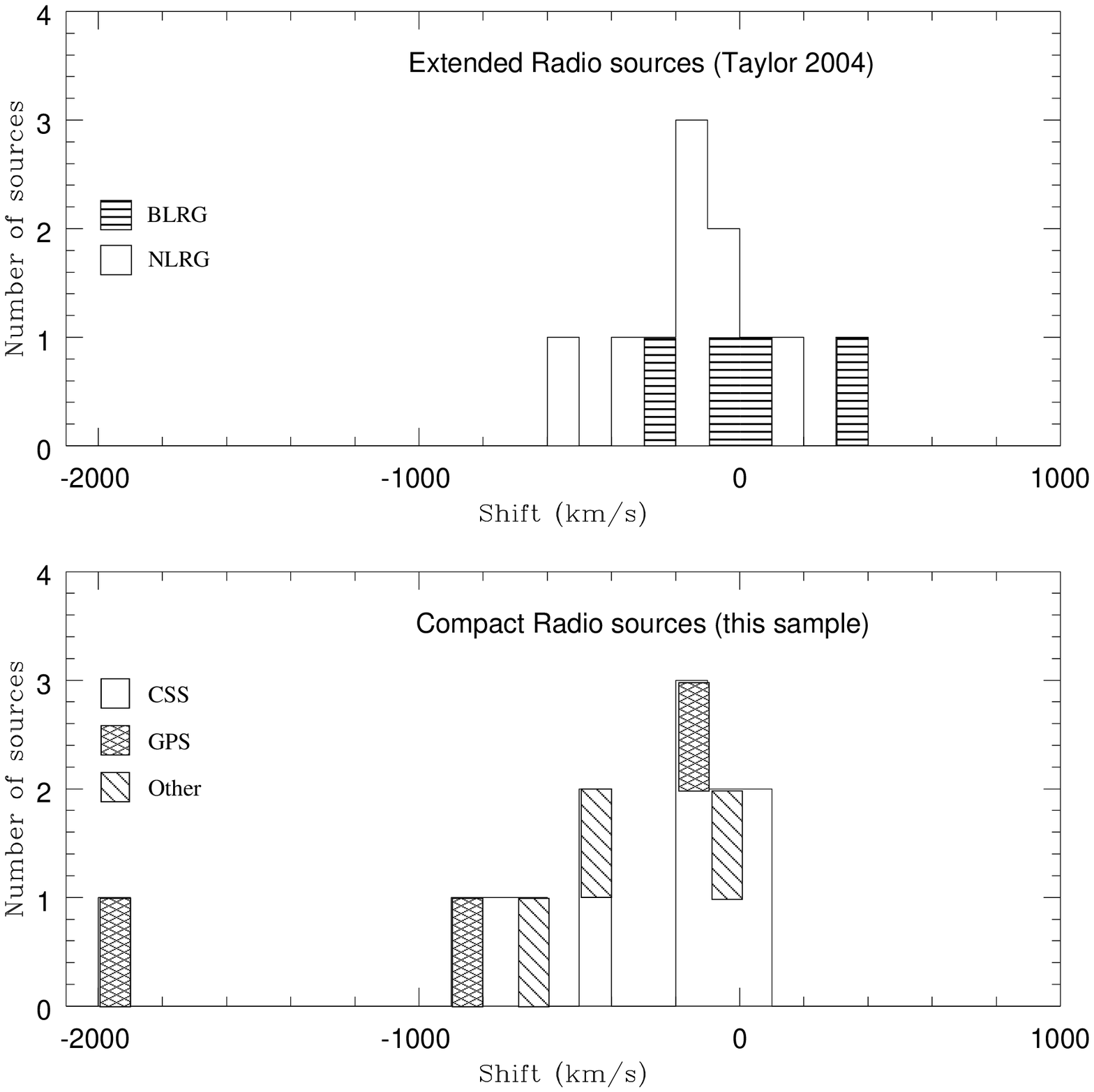,width=9cm,angle=0.}
\caption[Histogram of velocity shifts.]{Histograms showing the
  maximum nuclear outflow velocity for this sample of compact radio sources
  (bottom) and a comparison sample of extended radio sources taken
  from \protect\citet{taylor04}. Note, for the extended sources the
  shift plotted is broadest--narrowest component whilst for the compact
  sources, the shift between the broadest--systemic velocity is used
  as in a number of sources, two narrow components are observed (see
  Section 4.1). For the compact radio sources, the plot is further
  divided by scale -- CSS, GPS and `other' comprising compact-core and
  compact flat spectrum radio sources. For the extended sources, the
  sample is divided into BLRG and NLRG. All shifts are derived from the
  nuclear apertures only.}
\label{fig:shift-histograms}
\end{figure}

Figure \ref{fig:histfwhmnew} shows that the distribution of FWHM of {[O
    III]}$\lambda$5007 is markedly different for the compact and
    extended radio source samples.  
For the extended sources, the distribution peaks at $\sim$
200-300 \kms~which is consistent with gravitational motions in an
elliptical galaxy \citep{tadhunter89}. However, more extreme kinematics
are observed in the sample of compact radio sources -- the spread is larger
and it peaks at FWHM $\sim$1000 \kms~-- velocities which are too large
to be explained by gravitational motions unless the gas was located close to
the central black hole (i.e. in the BLR or an INLR). A one-tailed
Kolmogorov-Smirnov  
(K-S)  test shows the distributions are different at the 99.9 per cent
    confidence level\footnote{Whilst the K-S test reveals a strong
    trend, the reader should also remember that the various samples
    have different selection criteria which may affect the test to
    some degree.}. 
Whilst this effect might be due, in part, to
  orientation effects in both compact radio sources (see Figure 5) and/or
  foreshortening of larger radio sources, it has 
  been argued that the 
  majority of CSS sources can not be larger radio sources
  foreshortened by projection effects \citep{fanti90}.

Whilst the line widths show a clear difference between the compact and
extended 
sources, a K-S test shows that neither the asymmetry index or the kurtosis 
parameters are significantly different. Whilst we may initially expect the
broad, blueshifted components to give highly asymmetric line profiles,
examination of Figure {\ref{fig:o3models}} shows in many sources,
either the peak flux 
of the broadest components is often below the 20\% flux level (i.e. 
asymmetries would only be evident in parameters using e.g. the 10\% flux level
as in PKS 1345+12), or the emission lines appear symmetric although
the dominant component 
may be highly shifted (e.g. PKS 1549-79). 

In addition to the line widths/profiles comparison to older data in
the literature, 
we have compared our sample to a recent study of a sample of extended
3C radio sources
taken with similar observational set-up \citep{taylor04} enabling
comparisons between, for  
example, the shifts between the different components of a line (see
Table \ref{tab:sample-param}).  
A histogram
of the broadest-narrow velocity shifts for both samples in shown in Figure 
\ref{fig:shift-histograms}. The distributions are markedly
different (99.9\%: K-S test) with the compact radio sources spanning a
much larger range in velocity (out to $\sim$2000\kms~compared to
$\sim$600\kms). It is interesting to note that this trend 
of more extreme velocities in smaller
sources is also evident within the sample of compact radio sources,
with the two most extreme velocities being detected in  smaller GPS
sources (PKS 1345+12 and 4C 32.44). Indeed, all but one of the most extreme
velocities (v $>$ 500 \kms) are detected in sources with 
projected linear size D $<$ 1kpc whilst 
 the cluster of sources in the low outflow velocity region (-100
$<$ $v_{\rmn shift} <$ 100 \kms) are predominantly the larger CSS
 sources (Figure \ref{fig:shift-histograms} and Table \ref{tab:linsize}). 
Such velocity evolution with source size is consistent with the idea
that compact radio sources are young, starting as GPS sources and
expanding to become CSS then extended radio sources.

Whilst the nuclear regions of extended radio sources appear to be less
extreme than in more compact radio sources, 
extreme velocities ($v$ $\gtrsim$ 1000 \kms;
e.g. \citealt{solorzano01}) are commonly observed in 
extended radio sources, but in the {\it extended} emission line regions
(EELRs). EELRs   are typically on similar scales as, and are aligned with, the
radio axis and are therefore associated with jet-cloud
interactions. Hence, in extended radio sources, the 
regions emitting the broader components are often
significantly displaced from the nuclear aperture, coincident with the
radio emission. 

With this in 
mind, it is not surprising that extreme kinematics are observed in the
{\it nuclear} apertures of compact radio sources. Compact radio sources
are on the scale of the host galaxy ($<$15 kpc), often contained
within the nuclear regions ($\lesssim$1 kpc), 
and so the signatures of interactions between the radio jets and the
ISM will occur in the circumnuclear regions. Indeed, 
it is in the
nuclear apertures of compact radio sources that we observe both broad,
blueshifted components, which are generally spatially unresolved, as
well as any quiescent components similar to those observed in the
nuclear regions of extended radio sourcces. 
 The consistency between the spatial scales of the emission
line regions and the radio emission has  been confirmed in six sources
in our sample (3C 213.1, 3C 268.3, 3C 277.1, 3C 303.1, PKS 1345+12 and
PKS 1549-79) using HST broad- and narrow-band imaging
\citep{devries97,axon00,batcheldor07}.  
 Hence, 
whilst the kinematics observed in the nuclear regions of
compact radio sources are significantly different to those observed in
the nuclei of more extended radio sources, they are entirely
consistent with compact radio sources being a young, small scale
version of the jet-cloud interactions observed n the extended sources
with aligned radio and optical emission.

\section{Summary, conclusions and future work}
It is clear that all compact radio sources show evidence for disturbed
kinematics, with large line widths and shifts with respect to the
galaxy rest frame in the optical emission lines. The main conclusions
of this paper are:
\begin{itemize}
\item {\bf The extended emission line halo}. For the majority  of
  sources (12/14), we were able to determine the systemic velocity
  with  confidence. For consistency, all techniques
  focused on the extended narrow component emission. In two sources, a
  smooth `rotation curve' was observed, but in most sources, the
  extended narrow component emission was either not settled, as in PKS
  1345+12, or split into two narrow components, assumed to represent
  a  rotation curve in which the central regions were unresolved
  although this may also be a signature of bi-polar outflows.
\item {\bf Extreme emission line outflows}. All but three sources show
  evidence for  outflows in the circumnuclear ISM. The most extreme
  outflow is in the GPS source PKS 1345+12 ($\sim$2000
  \kms). Interestingly, the second most extreme outflow was observed
  in another GPS source, 4C 32.44.  As well as radio source size (CSS
  or GPS), the orientation of the radio source to the observer's line
  of sight may also be important, with higher outflow velocities observed
  in sources pointing towards the observer.
\item {\bf Blueshifted HI}. HI absorption is detected in 10/14
  sources, with multiple components observed in 5 sources. The majority
  of HI components (narrow and broad) are significantly blueshifted
  with respect to the systemic velocity and trace outflows in the
  neutral gas. In PKS 1345+12, (and others?) the outflowing HI
  components are broadly consistent with the emission line
  components. In only two sources (PKS 0023-26 and PKS 1934-63) is the HI
   redshifted, consistent with infalling gas.
\item {\bf Kinematical evidence for shocks}. As well as the extreme
  line splitting/outflows, the emission line components are also
  highly broadened. Again, the two most extreme FWHM in the NLR gas
  are observed in GPS sources: PKS 1345+12 ($\sim$2000 \kms) and 4C
  32.44 ($\sim$3500 \kms). Highly broadened components are almost
  exclusively confined to the nuclear apertures and therefore on the
  scale of the radio source. Higher resolution HST studies have
  resolved the {[O III]} emission line gas in several sources in this
  sample, revealing that the
  emission line gas is both on the same scale as, and is strongly aligned
  with, the radio source. This is consistent with observations of
  high redshift extended radio sources in which the broadest
  components are observed in regions coincident with the radio
  emission (e.g. \citealt{solorzano01}). The suggested importance of
  orientation 
  on the  observed outflow velocity also suggests the acceleration is
  confined to a spatially small region (i.e. coincident with radio
  jets) rather than ocurring across the entire nuclear region
  (i.e. due to   a quasar- or starburst wind).
\item {\bf Smaller source -- more extreme kinematics}. These results
  provide further evidence that the radio source size is also
  important in determining the outflow velocity and the dominance of
  shocks. Shocks, and the kinematics associated with them, are
  predominantly confined to smaller {\it extended} sources (D
  $\lesssim$ 120 kpc: \citealt{best00}). The statistical tests
  presented in this paper show that the nuclear kinematics, in particular
  the emission line shifts, are more extreme in compact radio sources
  than in the nuclear 
  regions of their extended 
  counterparts at the 99.9\% confidence level. Similarly, the most
  extreme outflows in the compact radio sources occur in sources with
  the smallest projected linear sizes, generally with D $<$ 1 kpc, the
  GPS sources.  Whilst this effect might be due, in part, to
  foreshortening, it has been argued that the
  majority of CSS sources can not be larger radio sources
  foreshortened by projection effects \citep{fanti90}.
\end{itemize}

\section*{\sc Acknowledgements}
JH acknowledges financial support from PPARC.
We thank the referee for useful comments on the paper. 
The William Herschel Telescope is operated on the
island of La Palma by the Isaac Newton Group in the Spanish
Observatorio del Roque de los Muchachos of the Instituto de
Astrofisica de Canarias. This research has
made use of the NASA/IPAC Extragalactic Database (NED) which is
operated by the Jet Propulsion Laboratory, California Institute of
Technology, under contract with the National Aeronautics and 
Space Administration. Based on observations made with ESO Telescopes
at the La Silla and  Paranal Observatory under programmes 69.B-0548(A)
and 71.B-0616(A).

\bibliographystyle{mn2e}
\bibliography{abbrev,refs}

\end{document}